\documentclass[reprint,
superscriptaddress,
 amsmath,amssymb,
 aps,
doi=true,
floatfix,
]{revtex4-1}
\pdfoutput=1

\usepackage{natbib}
\usepackage[colorlinks=true,linkcolor=blue,urlcolor=blue,citecolor=blue]{hyperref}

\usepackage{graphicx}
\usepackage{dcolumn}
\usepackage{bm,upgreek}

\usepackage{overpic}
\usepackage[caption=false]{subfig}
\usepackage{xfrac}
\usepackage{xcolor}
\usepackage{dblfloatfix}
\usepackage{mathtools}
\usepackage[mathscr]{euscript}

\usepackage[export]{adjustbox}

\usepackage{amssymb}
\usepackage{relsize}

\definecolor{dandelion}{rgb}{0.94, 0.88, 0.19}
\definecolor{thepurple}{rgb}{0.65, 0.24, 0.59}
\definecolor{theorange}{rgb}{0.95, 0.40, 0.13}
\definecolor{thegreen}{rgb}{0.05, 0.50, 0.25}
\definecolor{mathematicagreen}{rgb}{0.0, 0.80, 0.0}
\definecolor{igoraquamarine}{rgb}{0.0, 0.6, 0.6}
\definecolor{igorgreen}{rgb}{0.0, 0.6, 0.0}
\definecolor{springgreen}{rgb}{0.0, 0.8, 0.0}
\definecolor{skyblue}{rgb}{0.0, 0.6, 1.0}
\definecolor{black}{rgb}{0.0 0.0 0.0}
\definecolor{mhs}{rgb}{0.2,0,0.9}
\definecolor{CaptAlsid}{rgb}{1.0,0.4,0}
\definecolor{john}{rgb}{0.8,0.0, 0.2}
\definecolor{mhsnew}{rgb}{0,0,0}

\begin{document}

\preprint{APS/123-QED}

\title{Sensitive AC and DC Magnetometry with Nitrogen-Vacancy Center Ensembles in Diamond}

\author{John F. Barry}
\email[Corresponding author.\\]{john.barry@ll.mit.edu}
\affiliation{Lincoln Laboratory, Massachusetts Institute of Technology, Lexington, Massachusetts 02421, USA}
\author{Matthew H. Steinecker}
\affiliation{Lincoln Laboratory, Massachusetts Institute of Technology, Lexington, Massachusetts 02421, USA}
\author{Scott T. Alsid}
\affiliation{Lincoln Laboratory, Massachusetts Institute of Technology, Lexington, Massachusetts 02421, USA}
\affiliation{Department of Nuclear Science and Engineering, Massachusetts Institute of Technology, Cambridge, Massachusetts 02139, USA}
\author{Jonah Majumder}
\affiliation{Lincoln Laboratory, Massachusetts Institute of Technology, Lexington, Massachusetts 02421, USA}
\author{Linh M. Pham}
\affiliation{Lincoln Laboratory, Massachusetts Institute of Technology, Lexington, Massachusetts 02421, USA}
\author{Michael F. O'Keefe}
\affiliation{Lincoln Laboratory, Massachusetts Institute of Technology, Lexington, Massachusetts 02421, USA}
\author{Danielle A. Braje}
\affiliation{Lincoln Laboratory, Massachusetts Institute of Technology, Lexington, Massachusetts 02421, USA}

\date{May 10, 2023}

\begin{abstract}
Quantum sensing with solid-state spins offers the promise of  high spatial resolution, bandwidth, and dynamic range at sensitivities comparable to more mature quantum sensing technologies, such as atomic vapor cells and superconducting devices. However, despite comparable theoretical sensitivity limits, the performance of bulk solid-state quantum sensors has so far lagged behind these more mature alternatives. A recent review~\cite{barry2020sensitivity} suggests several paths to improve performance of magnetometers employing nitrogen-vacancy defects in diamond, the most-studied solid-state quantum sensing platform. Implementing several suggested techniques, we demonstrate the most sensitive nitrogen-vacancy-based bulk magnetometer reported to date. Our approach combines tailored diamond growth to achieve low strain and long intrinsic dephasing times, the use of double-quantum Ramsey and Hahn echo magnetometry sequences for broadband and narrowband magnetometry respectively, and P1 driving to further extend dephasing time. Notably, the device does not include a flux concentrator, preserving the fixed response of the NVs to magnetic field. The magnetometer realizes a broadband \textcolor{mhsnew}{near-}DC sensitivity $\sim 460$~fT$\cdot$s$^{1/2}$ and a narrowband AC sensitivity $\sim 210$~fT$\cdot$s$^{1/2}$. We describe the experimental setup in detail and highlight potential paths for future improvement. 
\end{abstract}

\maketitle

\tableofcontents

\clearpage

\section{\label{sec:intro}Introduction}
In recent years, quantum sensing using solid-state spin-defect systems has received increasing interest as an alternative to existing atomic gas and superconducting quantum sensors. Solid-state spin-defect systems offer the promise of comparable sensitivity to their atomic and superconducting counterparts but with important additional capabilities, including fixed sensing axes provided by a rigid crystal lattice and compatibility with a wide range of environments, as well as high spatial resolution, bandwidth, and dynamic range.

The most-studied solid-state spin-defect system for quantum sensing is the negatively-charged nitrogen-vacancy (NV$^\text{-}$) defect in diamond, primarily utilized for magnetometry applications. The NV$^\text{-}$ center in diamond can offer extraordinarily long coherence times compared to most other solid-state defects, but high-fidelity optical readout of the spin state of the system remains a challenge due to both the typically low efficiency of fluorescence collection and the inherently low contrast of the spin-state-dependent fluorescence. In addition, despite the theoretical availability of long coherence times in diamond~\cite{Balasubramanian2009ultralong}, in practice the dephasing time for broadband low-frequency ($\lesssim$ 10 kHz) magnetometry is limited to much shorter timescales due to strain and dipolar broadening from the surrounding bath of $^{13}$C and substitutional nitrogen, also known as P1 centers.

A recent review~\cite{barry2020sensitivity} identified certain promising methods to enhance bulk NV diamond magnetometer sensitivity. Several of these advances have been employed individually over the past decade, including high $^{12}$C purity diamonds~\cite{Balasubramanian2009ultralong}, double-quantum magnetometry to eliminate common-mode sources of dephasing~\cite{Fang2013high,Mamin2014multipulse,moussa2014preparing,Angerer2015subnanotesla,bauch2018ultralong}, spin-bath driving to reduce P1-induced dephasing~\cite{DeLange2012controlling,Knowles2014observing,bauch2018ultralong}, phase-modulated noise subtraction schemes to eliminate low-frequency noise in Ramsey sequences~\cite{Schloss2018Simultaneous,hart2021nv}, and the light-trapping diamond waveguide geometry to improve excitation efficiency~\cite{Clevenson2015broadband}. Other developments have also been reported, such as the use of a dielectric resonator to improve MW strength and uniformity~\cite{Kapitanova2018efficient,Eisenach2018broadband}. However, to date these advances have not all been successfully combined, and the best reported sensitivities have relied on specialized techniques such as a flux concentrator~\cite{Fescenko2020diamond} or microwave cavity readout~\cite{Eisenach2021cavity}, enabling 0.9~pT$\cdot$s$^{1/2}$ and 3~pT$\cdot$s$^{1/2}$ sensitivities respectively.  For conventional readout without a flux concentrator, both narrowband AC and broadband DC-sensitive magnetometry using NV ensembles have been limited to sensitivities near 10~pT$\cdot$s$^{1/2}$~\cite{Wolf2015subpicotesla,Barry2016optical,Glenn2018high,Schloss2018Simultaneous}.

Here, we report an NV ensemble magnetometer that achieves the best sensitivity reported to date for both broadband and narrowband sensing without a flux concentrator. The device combines the advances described above with additional, novel improvements, including a low-gradient magnetic bias field design, near-unity collection efficiency optics, balanced photodetection to minimize laser intensity noise, and a custom non-metal mechanical support structure to avoid AC field attenuation. Together, these developments produce a magnetometer that demonstrates $\sim 460$~fT$\cdot$s$^{1/2}$ broadband sensitivity using a Ramsey sequence and $\sim 210$~fT$\cdot$s$^{1/2}$ narrowband AC sensitivity using a Hahn echo sequence. The device does not employ a flux concentrator and thus retains the known response of the NV$^\text{-}$ centers to magnetic field.

This paper is organized as follows: Section~\ref{sec:advancesoverview} describes the systematic improvement of each aspect of a typical NV magnetometer setup, in particular the diamond, fluorescence collection optics, bias magnetic field, microwave delivery, and pulse sequences. Section~\ref{sec:expresults} details the complete experimental setup and discusses measurement of the dephasing time $T_2^*$, decoherence time $T_2$, and magnetic sensitivity of Ramsey and Hahn echo magnetometry. Section~\ref{sec:magdemo} reports the results obtained by employing the device as a magnetometer. Finally, Section~\ref{sec:outlook} provides a discussion of future prospects and next steps. Additional details on each component of the experimental setup, the measurements, and sensitivity calculations are contained in the Supplemental Material.

\section{Advances Overview}
\label{sec:advancesoverview}
The theoretical sensitivity limit for an NV$^\text{-}$ broadband ensemble magnetometer employing Ramsey interferometry is~\cite{barry2020sensitivity}
\small
\begin{align}\label{eqn:ramseyshotexact}
\eta_\text{Ram} ^\text{ens} & \approx \underbrace{\frac{\hbar}{\Delta m_s g_e \mu_B} \frac{1}{\sqrt{N\tau}}}_{\text{Spin projection limit}} \;   \underbrace{\frac{1}{e^{-\left(\tau/T_2^*\right)^p}}}_{\text{Spin dephasing}} \; \underbrace{\sqrt{1\!+\!\frac{1}{C^2 n_\text{avg}}}}_{\text{Readout}} \;\nonumber \\ & \quad \times \underbrace{\sqrt{\frac{t_\text{I}\! +\! \tau\! +\! t_\text{R}\!+\!t_\text{D}}{\tau}}}_{\text{Overhead time}}
\end{align}
\normalsize
where $\hbar$ denotes the reduced Planck constant, $\Delta m_s$ denotes the difference in spin quantum number between the interferometry states, $g_e\approx 2$ is the NV$^\text{-}$ center's electronic g factor \cite{doherty2013nitrogen}, $\mu_B$ is the Bohr magneton, $N$ is the number of NV$^\text{-}$ centers interrogated, $\tau$ is the free-precession time per measurement, $T_2^*$ is the dephasing time, $p$ is the stretched exponential parameter~\cite{bauch2018ultralong}, $C$ is the measurement contrast~\cite{taylor2008high}, $n_\text{avg}$ is the average number of photons collected per NV$^\text{-}$ center in a measurement, $t_\text{I}$ is the initialization time, $t_\text{R}$ is the NV$^\text{-}$ ensemble fluorescence readout time, and $t_\text{D}$ represents any additional dead time in the measurement sequence. Equation~\ref{eqn:ramseyshotexact} implicitly treats the case of a single NV class under the assumption the magnetic field is parallel to the NV axis; for realistic measurements of NV ensembles, order-unity geometric corrections are required to account for the reduced response of NVs whose axes are not aligned with the magnetic field; see SM Sec.~\ref{sec:ramseytheorysensitivity}.  We note the functional form of sensitivity in a Hahn echo scheme is similar, with the decoherence time $T_2$ replacing the dephasing time~\cite{barry2020sensitivity}.

The factor $\sigma_R = \sqrt{1+1/[C^2 n_\text{avg}]}$ in Eq.~\ref{eqn:ramseyshotexact} quantifies the imperfect readout of NV spins due to limited contrast and photon shot noise, so that $\sigma_R = 1$ corresponds to spin-projection-limited readout~\cite{itano1993quantum}. Typically $\sigma_R$ is expressed in terms of the readout fidelity $\mathcal{F}$, with $\sigma_R = 1/\mathcal{F}$. In the limit where $C^2n_{\textrm{avg}} \ll 1$, the readout fidelity $\mathcal{F}$ can be approximated using $\mathcal{F} \!\approx\! C\sqrt{n_\text{avg}}$.  

Equation~\ref{eqn:ramseyshotexact} makes clear that sensitivity optimization of a shot-noise-limited device requires maximizing the dephasing time, the measurement contrast, the number of interrogated spins, and the average number of photons detected per NV$^\text{-}$ per measurement. In practice, the latter two quantities are often difficult to measure independently, and we define the total number of photons detected per measurement, $\mathscr{N} \equiv N n_\text{avg}$, as an experimentally accessible proxy to optimize instead. Then, using the readout fidelity approximation above, the shot-noise-limited sensitivity equation for a Ramsey scheme becomes
\begin{align}\label{eqn:ramseyshot}
& \eta^\text{ens,sho}_\text{Ram} \nonumber \\ &\approx \frac{\hbar}{\Delta m_sg_e \mu_B} \frac{1}{C e^{-\left(\tau/T_2^*\right)^p}\sqrt{\mathscr{N}}} \frac{\sqrt{t_\text{I} + \tau + t_\text{R}+t_\text{D}}}{\tau}.
\end{align}

While there are several techniques to optimize each of the interconnected parameters noted above, the overarching strategy used in this work follows Ref.~\cite{barry2020sensitivity}, which suggests that extending dephasing time should be a central focus when making a high sensitivity broadband NV magnetometer. Upon extending the dephasing time, however, each aspect of an NV magnetometer employing conventional optical readout must be systematically engineered for improved performance: the diamond substrate and mounting, the NV fluorescence collection optics, the static bias field, the microwave delivery system, and the magnetometry pulse sequences. The remainder of this section summarizes these design decisions, including the combination of several previously-demonstrated advances as well as the introduction of novel and non-standard techniques. The main design choices and advances are summarized in Table~\ref{tab:expmethodsummary}.

\begin{table*}[thp]  
\centering
\caption{Overview of implemented techniques to increase NV$^\text{-}$ ensemble sensitivity} %
\centering %
\begin{tabular}{p{1.8cm} p{3.0cm} p{2.1cm} p{11cm}}   
\hline\hline   
Subsystem & Method & Parameter optimized & Method description  \\
\hline
Diamond & $\bullet\;$ $^{15}$NV isotope & $C$ & $\bullet$ Reduces number of MW tones required and undesirable cross-talk.  \\
 & $\bullet\;$ $^{12}$C isotopic purity & $T_2^*$ & $\bullet$ 99.998$\%$ $^{12}$C isotopic purity mitigates dipolar coupling to $^{13}$C nuclei~\cite{Balasubramanian2009ultralong}.\\
 & $\bullet\;$ Low-strain growth & $T_2^*$ & $\bullet$ Growth on low-strain substrates translates to a low-strain $^{15}$NV layer; strained edges are removed after growth. \\
 & $\bullet\;$ Thermal stability & \mbox{$T_2^*$, $C$,} stability & $\bullet$ 4H SiC acts as combined heatsink and heat spreader for diamond, preventing high temperatures and associated loss of contrast~\cite{Toyli2012measurement}. \\
 \hline
Optical & $\bullet\;$ Light-trapping \;\;\;\; diamond waveguide & $\mathscr{N}$, laser power & $\bullet$ Fixed 532 nm light is totally-internally reflected within diamond until absorbed, making best use of fixed laser power~\cite{Clevenson2015broadband}.
\\
 & $\bullet\;$ \!\!\!\!\!\!\!\!\mbox{TIR lens $\&$} \;\;\; reflector & $\mathscr{N}$ &  $\bullet$ Combination of total-internal-reflection (TIR) lens and dielectric reflector allows nearly all light exiting diamond to be delivered to photodiode. 
\\
 & $\bullet\;$ Balancing circuit & laser noise & $\bullet$ Balancing circuit reduces wide-band laser-intensity-induced noise to 5$\%$ above shot noise on fluorescence photocurrent.
\\
\hline
Bias \;\;\;\;\;\; magnetic \;\;\; field & $\bullet\;$ $[100]$ field  &\mbox{$T_2^*,C$} & $\bullet$ Static field along $[100]$ axis projects equally onto all four NV classes, allowing all NVs to contribute to magnetometry signal.
\\
 & $\bullet\;$ Ring geometry & $T_2^*$&  $\bullet$ Large, distant circular magnet array creates uniform field, reducing inhomogeneous broadening.
\\
\hline
Microwave  & $\bullet\;$ \raggedright Dielectric\;\;\;\;\;\;\;\;\;\;\; resonator\;\;\;\;\;\;\;\; & \raggedright MW power and uniformity & $\bullet$ Decreases required microwave power relative to broadband alternatives and increases MW uniformity over NV ensemble volume.
\\
& $\bullet\;$ Gaussian pulses & $C$ & $\bullet$ Gaussian-enveloped MW pulses mitigates  cross-talk~\cite{vandersypen2005nmr,Fuchs2009gigahertz}.
\\
\hline
Pulse \;\; sequences & $\bullet\;$ Double quantum & $T_2^*$, noise & $\bullet$ Mitigates dephasing from longitudinal-strain and temperature-induced resonance shifts while doubling the effective gyromagnetic ratio~\cite{Mamin2014multipulse,Schloss2018Simultaneous}.  
\\
 $\&$ \;\;\;\;\;\;\;\;\;\;Noise & $\bullet\;$ P1 Driving & $T_2^*$ &  
 $\bullet$ Suppresses dephasing from substitutional nitrogen~\cite{DeLange2012controlling,Knowles2014observing,bauch2018ultralong}. 
\\
subtraction & $\bullet\;$ Noise subtraction schemes & Noise & $\bullet$ Phase shifting of final microwave pulse in measurement sequences isolates magnetic signals from noise~\cite{hart2021nv}.
\\
 & $\bullet\;$ \raggedright Digital phase modulation & $C$, noise & $\bullet$ Avoids noise typically introduced by varactor or PIN diode phase shifters while allowing implementation of noise subtraction and observation of fringes in resonant excitation.
\\

\hline 
\end{tabular}\label{tab:expmethodsummary}
\end{table*}

\subsection{Diamond}

\subsubsection{Tailored diamond growth for low strain and high $T_2^*$}

\vspace{5mm}
For NV-diamond magnetometers, the spin ensemble's characteristics set fundamental limits to device sensitivity. If the magnetometer uses Ramsey-type interrogation with conventional optical readout, the sensitivity of an optimized device depends on the number of NV$^\text{-}$ defects interrogated $N$, the dephasing time $T_2^*$, and the readout fidelity $\mathcal{F}$ as detailed in Eqn.~\ref{eqn:ramseyshotexact}. All three parameters are influenced by the diamond's properties and should be maximized. %

To optimize properties for magnetometry, diamonds are grown in-house via chemical vapor deposition (CVD). The diamond employed in this work consists of a 70-$\upmu$m-thick $^{15}$N-doped layer grown on a type IIa diamond substrate. Since substrate strain tends to propagate upwards during CVD growth~\cite{friel2009control}, only low strain substrates are used~\cite{DHaenensJohansson2015large,Tallaire2017thick}. The nitrogen isotope $^{15}$N is chosen rather than $^{14}$N to reduce the number of required microwave frequencies. Following CVD growth, about 1 mm of each of the $\langle 110 \rangle$ diamond sides is removed by laser cutting, as those regions are found to exhibit higher strain than the interior. The resulting doped layer is estimated to contain on the order of $3 \times 10^{11}$ NV$^\text{-}$ defects. See Supplementary Sect.~\ref{sec:diamondappendix} for further diamond details.

The dephasing time $T_2^*$ can be limited by nuclear or electronic spins near each NV$^\text{-}$ center in the diamond. To mitigate dipolar broadening from $^{13}$C, the $^{15}$N-doped layer is grown using methane specified to $99.999\%$ $^{12}$C isotopic purity. Secondary ion mass spectrometry on a sister sample finds the layer's carbon isotopic purity to be 99.998$\%$ $^{12}$C, similar to the results in Refs.~\cite{Teraji2013effective,Teraji2015homoepitaxial}. Following the estimates in Ref.~\cite{barry2020sensitivity}, the dephasing time due to $^{13}$C in the diamond is estimated to be $T_{2,\text{DQ}}^*\{^{13}\text{C}\}$ = 250 $\upmu$s for a double-quantum (DQ) scheme.  In a DQ scheme, we measure $T_{2,\text{DQ}}^* = 14$ $\upmu$s and $T_\text{2,DQ}=136$, which corresponds to $[^{15}\text{N}] = 0.4\pm0.2$~ppm using the scaling provided in Refs.~\cite{barry2020sensitivity,bauch2020decoherence}.

\subsubsection{Thermal, mechanical, and electromagnetic considerations in sensor construction}

In mounting a diamond for use as a magnetometer, the mechanical construction must balance three primary design goals: high thermal conductivity so that elevated diamond temperatures do not reduce contrast, stable mechanical mounting to limit vibration-induced noise, and minimal use of conductive materials so that AC magnetic fields are not needlessly attenuated.

Time-varying or excessively high temperatures can be problematic for NV-based magnetometers. First, the measurement contrast $C$~\cite{taylor2008high} declines for diamond temperatures above $\sim 300$ K~\cite{Toyli2012measurement}, thereby degrading device sensitivity. Second, the NV$^\text{-}$ center's zero-field splitting $D\sim 2.87$ GHz shifts at \mbox{$\approx$ -74 kHz/K~\cite{Acosta2010temperature,Kucsko2013nanoscale,Toyli2013fluorescence}}. As a result, unwanted temperature variation can detune the applied MWs from the intended Zeeman resonances, again degrading sensitivity.

Primary thermal loads on the diamond arise from optical excitation, MW pulses, and P1 driving, with the latter two contributions predominantly arising from heating of the ambient environment around the diamond. We note that the time-averaged optical power applied to the diamond can be as high as 4 W, sufficient to produce substantial heating of the diamond in the absence of measures to remove this heat. Moreover, maintaining the diamond at a fixed temperature is not trivial. For example, the various pulses sequences used in this work (with varying optical, MW, and P1 driving duty cycles) produce differing heat loads. In addition, small changes in the optical power delivered to the diamond can produce significant changes in diamond temperature. Further measures to ensure consistent optical heat load are discussed in Supplementary Sect.~\ref{sec:laserandAOM}.

To address the above thermal challenges, the diamond is adhered to a 50.8 mm diameter, 330-$\mu$m-thick semi-insulating 4H SiC wafer, which acts as a combined  heat-sink/heat-spreader. The thermal conductivity is approximately 490 W/(m$\cdot$K) and 390 W/(m$\cdot$K) normal and parallel to the wafer surface, respectively~\cite{cree2021silicon,Protik2017phonon}. In addition, 4H SiC is stiff (with a Young's modulus of approximately 500 GPa~\cite{chen2019study}), exhibits a loss tangent $\lesssim 10^{-4}$ at 2.87 GHz~\cite{Hartnett2011microwave}, is widely commercially available, and is compatible with deposition of dielectric coatings. A stiff mounting structure is helpful to avoid low-frequency mechanical resonances that could cause vibration-induced translation of the diamond. The low loss tangent ensures minimal microwave energy is absorbed by the SiC wafer. The SiC's optical transparency facilitates troubleshooting and alignment; some degree of transparency is maintained in the blue visible region even with the reflective dielectric coating (see Section~\ref{sec:lightcollection}). In addition, the laser light and NV$^\text{-}$ fluorescence are sufficiently intense to allow their observation despite attenuation through the dielectric coating.

With a Young's modulus of approximately 1150 GPa~\cite{Klein1992anisotropy}, diamond is an exceptionally stiff material. Nevertheless, mechanical forces related to the adhesive adhering the diamond to the SiC wafer were observed to impart sufficient stress to the diamond such that the Zeeman resonance linewidths were broadened by a factor of two or more. This adhesive-induced resonance broadening was observed when the diamond was adhered with cyanoacrylate, 5-minute epoxy, and polyurethane, but not with polydimethylsiloxane (PDMS). The latter adhesive was therefore employed for all data shown in this work. See Supplementary Sect.~\ref{sec:adhesion} for more information on adhesive selection. 

Measurement of kHz-frequency-scale AC magnetic fields is a primary use for the sensor. As magnetic fields at such frequencies are attenuated by the presence of conductive materials~\cite{hoburg1995principles}, the sensor employs non-metallic materials where possible. The mechanical support structure of the SiC wafer is made of a 3D-printed ceramic-loaded plastic (Somos PerFORM), which is affixed to a supporting breadboard using a 25.4 mm diameter glass-filled PEEK rod. Both these materials have low coefficients of thermal expansion and are extraordinarily stiff compared to standard plastics. The mounting structure also allows the SiC wafer and diamond to be rotated together as described in Supplementary Section~\ref{App:LTDWDetails}. The light pipe collecting the NV fluorescence, detailed in Section~\ref{sec:optics}, is supported by nylon flexure mounts, and the dielectric resonator, described in Section~\ref{sec:DRandMWdelivery}, is held by a 1" OD alumina tube. Four long quartz rods are attached to the supports for the diamond, light pipe, photodiodes, and dielectric resonator, so that these components are rigidly held in line with the diamond while allowing easy translation away from the diamond when desired. A diagram of the sensor head is provided in the Supplemental Material;  see SM Fig.~\ref{fig:MasterSetupDiagram}.

Finally, in addition to avoiding conductive materials, care is taken to incorporate only non-magnetic materials in the rest of the sensor head. Thus, virtually all screws are made of nylon or brass, and sensor head components are predominantly fabricated from a 3D-printed ceramic-filled plastic (Somos PerFORM), glass-filled PEEK, quartz, alumina, silicon carbide, and various other plastics. The sensor head is mounted to an aluminum breadboard stood off from the sensor head by $\sim\!100$ mm.

\subsection{Optics}\label{sec:optics}

\subsubsection{Light-trapping diamond waveguide}\label{sec:LTDW}

To use the available 532 nm excitation light most efficiently, the sensor employs the light-trapping diamond waveguide technique~\cite{Clevenson2015broadband}, where excitation light is totally-internally-reflected within the diamond until completely absorbed. To achieve this effect, the corner of an otherwise square-cuboid diamond is faceted at 45 degrees relative to the two adjacent sides, creating an additional surface through which the 532 nm laser light can be focused into the diamond. With appropriate alignment through this input facet, the light is thereafter confined by total internal reflection for a number of reflections off the four vertical sides (and possibly the top and bottom facets as well). While all paths eventually exit the diamond, in practice paths can be found where nearly all the excitation light is absorbed within the diamond. As a result, the light trapping diamond waveguide technique can make better use of fixed available laser power than conventional single-pass approaches and is particularly useful for diamonds with sufficiently low NV concentration that the diamond is optically thin for a single pass. Supplemental Section~\ref{App:LTDWDetails} has additional details on the light-trapping diamond waveguide technique.

\subsubsection{TIR and dielectric reflector light collection}\label{sec:lightcollection}

For conventional optical readout of NV centers~\cite{barry2020sensitivity}, in the limit of low contrast $C$, the readout fidelity $\mathcal{F}$ scales as the square root of the mean number of photons collected per NV$^\text{-}$ per measurement. That is, $\mathcal{F} \propto \sqrt{\mathcal{N}/N} = \sqrt{n_\text{avg}}$; see Eqn.~\ref{eqn:ramseyshot}.  In this regime, the sensitivity can be enhanced by improving the geometric collection efficiency $\eta_\text{geo}$, defined as $\mathcal{N}/\mathcal{N}_\text{max}$, where $\mathcal{N}$ and $\mathcal{N}_\text{max}$ are the numbers of photons collected from and emitted by the NV$^\text{-}$ ensemble in a single measurement, respectively~\cite{barry2020sensitivity}. 

\begin{figure}[t]
\centering
\includegraphics[width=\columnwidth]{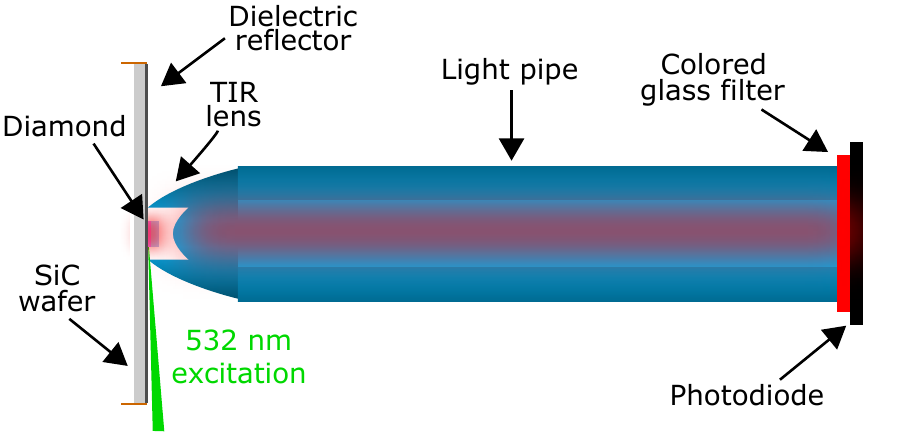}
\caption{Cross section of the fluorescence collection optics employed in this work. The purple-pink area of the diamond denotes the NV-doped portion. The 532 nm laser light passes through a small triangular notch (not shown) in the total internal reflection lens before entering the diamond.  }\label{fig:lightcollection}
\end{figure}

Several approaches have demonstrated high values of $\eta_\text{geo}$ (i.e. $\eta_\text{geo} \gtrsim 0.1$) for cubic-mm-scale NV ensemble volumes~\cite{Lesage2012efficient,Wolf2015subpicotesla,ma2018magnetometry,Barry2016optical,barry2020sensitivity}. We employ a scheme shown in Fig.~\ref{fig:lightcollection} that collects order unity of the fluorescence light emitted from the diamond and provides a higher value of $\eta_\text{geo}$ than existing schemes. The approach introduces two principle innovations. First, the diamond mounting substrate is coated with a multi-layer dielectric reflective coating (Thorlabs E02), which reflects the vast majority of fluorescence light emitted away from the collection optics. Second, the diamond is surrounded by a total-internal-reflection (TIR) lens, which collimates the initial fluorescence into a much smaller angle but over a larger area, as etendue is conserved. Together, these techniques ensure efficient collection from nearly all $4 \pi$ steradian solid-angle of fluorescence. Advantages of this approach are compatibility with unaltered cuboid diamond shapes and the light trapping diamond waveguide method~\cite{Clevenson2015broadband}. Additionally, only a single photodiode is required for the signal channel, and the photodiode location may be well-removed from the location of the diamond.

As there are few routes for light that exits the diamond to escape collection, the value of $\eta_{geo}$ is expected to be close to unity. It is not trivial to reliably quantify the value of $\eta_\text{geo}$ at the $5\%$ level or better. However, Zemax simulations of the light collection suggest 92$\%$ of light emitted from the diamond is delivered to the end of the light pipe. To verify this simulation experimentally, a small spherical Lambertian scatterer (Spectralon) was employed to simulate the quasi-isotropic fluorescence emission of the diamond. When this scatterer was placed at the diamond location and irradiated with 532 nm light, $0.90^{+0.05}_{-0.15}$ of the input 532 nm photons were measured at the output of the light pipe. The light collection system is further detailed in Section \ref{sec:smlight}.

\subsubsection{Integrating balanced photodetector}
\label{sec:integratingcircuit}
In ensemble NV$^\text{-}$ magnetometers, large fluorescence powers, $\gtrsim\!10$~mW, may be produced~\cite{barry2020sensitivity,Schloss2018Simultaneous}. Shot-noise-limited measurement of high-photon-flux fluorescence faces two primary technical difficulties. First, the dynamic range of available analog-to-digital converters is insufficient to digitize the full-scale NV fluorescence signal while keeping additive digitizer noise below optical shot noise. Second, the excitation light's residual intensity noise (RIN) nearly always translates to noise in the collected fluorescence above the shot-noise limit~\cite{hobbs1991shot,hobbs1991reaching,haller1991double,hobbs1997ultrasensitive,hobbs2011building}. 

\begin{figure}[t]
\centering
\includegraphics[width=\columnwidth]{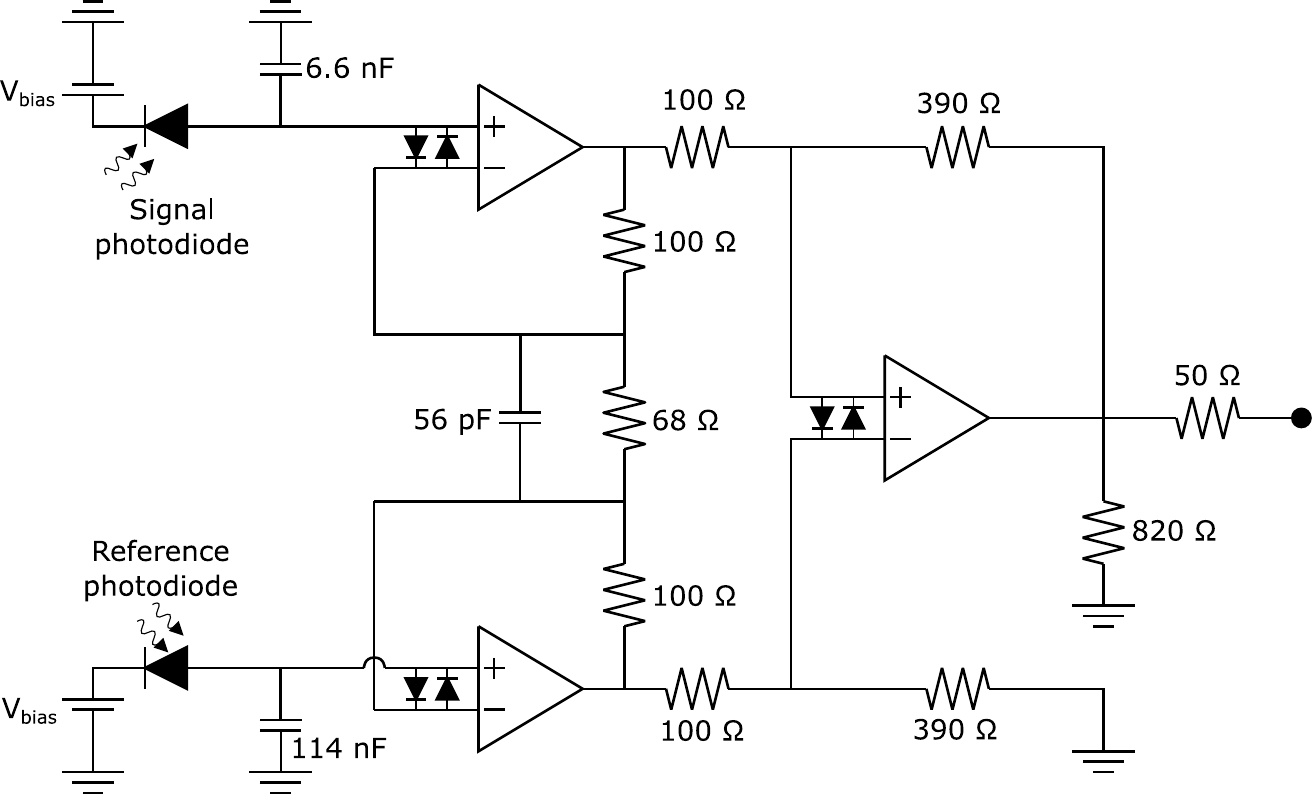}
\caption{\textbf{Integrating balanced photodetector electronics.} Incoming photocurrent in each of the signal and reference arms is integrated on a capacitor. Thereafter, an instrumentation amplifier differences and amplifies the voltages on each integration capacitor. Digitization of the circuit output only occurs after the readout light pulse ends, in a sample-and-hold fashion. Finally, both integration capacitors are discharged using TTL-controlled switches (not shown) so the process can be repeated.}\label{fig:autobalanced_photodiode_diagram}
\end{figure}

Both problems are mitigated by referencing the NV fluorescence photocurrent to that of an identical photodiode sampling the 532 nm excitation light, an approach inspired by Refs.~\cite{hobbs1991shot,hobbs1991reaching,haller1991double,hobbs1997ultrasensitive,hobbs2011building}. For simplicity, we refer to the two photocurrents as the signal photocurrent and reference photocurrent, respectively. The dynamic range problem is circumvented because the digitizer's range must now only cover the \textit{difference} between the signal photocurrent and the reference photocurrent, rather than the signal photocurrent's full scale range. The laser intensity noise problem is greatly reduced, as the vast majority of laser intensity noise will be common mode to both the signal and reference photocurrents. We developed an analog circuit, termed the balancing circuit and shown in Fig.~\ref{fig:autobalanced_photodiode_diagram}, which outputs a voltage proportional to the integrated signal photocurrent normalized to the reference photocurrent. Further details are provided in Section~\ref{App:BalancingCircuit}.

\cleardoublepage

\subsection{Bias magnetic field}\label{sec:staticfield}

In the absence of external fields, the $|m_s=\pm 1\rangle$ Zeeman states of the NV$^\text{-}$ electronic ground state spin triplet are degenerate. Application of a permanent bias magnetic field $\mathbf{B}_0$ breaks this degeneracy, making the \mbox{$|m_s \!=\! 0\rangle \leftrightarrow |m_s \!=\! 1\rangle$} and \mbox{$|m_s \!=\! 0\rangle \leftrightarrow |m_s \!=\! -1\rangle$} transitions spectroscopically resolvable~\footnote{We define the quantization axis for each of the four crystallographic NV orientations so that the quantization axis is along the NV axis, with the direction of the quantization axis set to that which gives a positive projection of the magnetic bias field.}.  We apply a bias magnetic field of $B_0$ = 2.23 gauss oriented normal to the diamond's two large $\{100\}$ faces to ensure Zeeman transitions of NVs aligned along the four different crystallographic directions can be addressed with the same MW frequency (see Section~\ref{sec:ODMR}). All 16 allowed transitions from $|m_s=0\rangle$ to $|m_s=\pm 1\rangle$, addressing both nuclear spin states, can thus be accessed with four MW frequencies. However, in this scheme, precise alignment of the bias field is critical, as the bias field must be uniform across the diamond and directed so that NVs along all four crystallographic orientations are exactly degenerate. Satisfying both criteria will minimize variation of the ensemble's Zeeman resonance frequencies, which can otherwise limit the observed value of $T_2^*$ and degrade sensitivity for Ramsey magnetometry. 

To minimize bias field variation across the diamond and optimize field alignment, the bias magnetic field is created using two large custom-made magnet arrays mounted on motorized actuators. This approach allows \textit{in-situ} adjustment of the field alignment while monitoring dephasing in a precession time sweep. Additional details on the design, construction, and alignment procedure for the bias magnetic field are given in Supplementary Section~\ref{app:staticfielddetails}, while Supplementary Section~\ref{sec:biasmagneticfieldgradients} gives example calculations of uniformity tolerance for the bias magnetic field.

To minimize the influence of external magnetic fields, the sensor is placed in a triple-layer cylindrical magnetic shield made from 1.6-mm-thick $\mu$-metal. The innermost layer is 610 mm in diameter and 2134 mm long, the middle layer is 762 mm in diameter and 2286 mm long, and the outermost layer is 914 mm in diameter and 2438 mm long. While one end of the shield is permanently closed, the other end is left open during testing. Earth's field is attenuated to values of $\sim 10$ nT or less inside the shield.

\subsection{Microwave}

\subsubsection{Dielectric resonator and microwave delivery}\label{sec:DRandMWdelivery}

\begin{figure}[t]
\!\!\!\!\!\!\!
\centering
\includegraphics[width=\columnwidth]{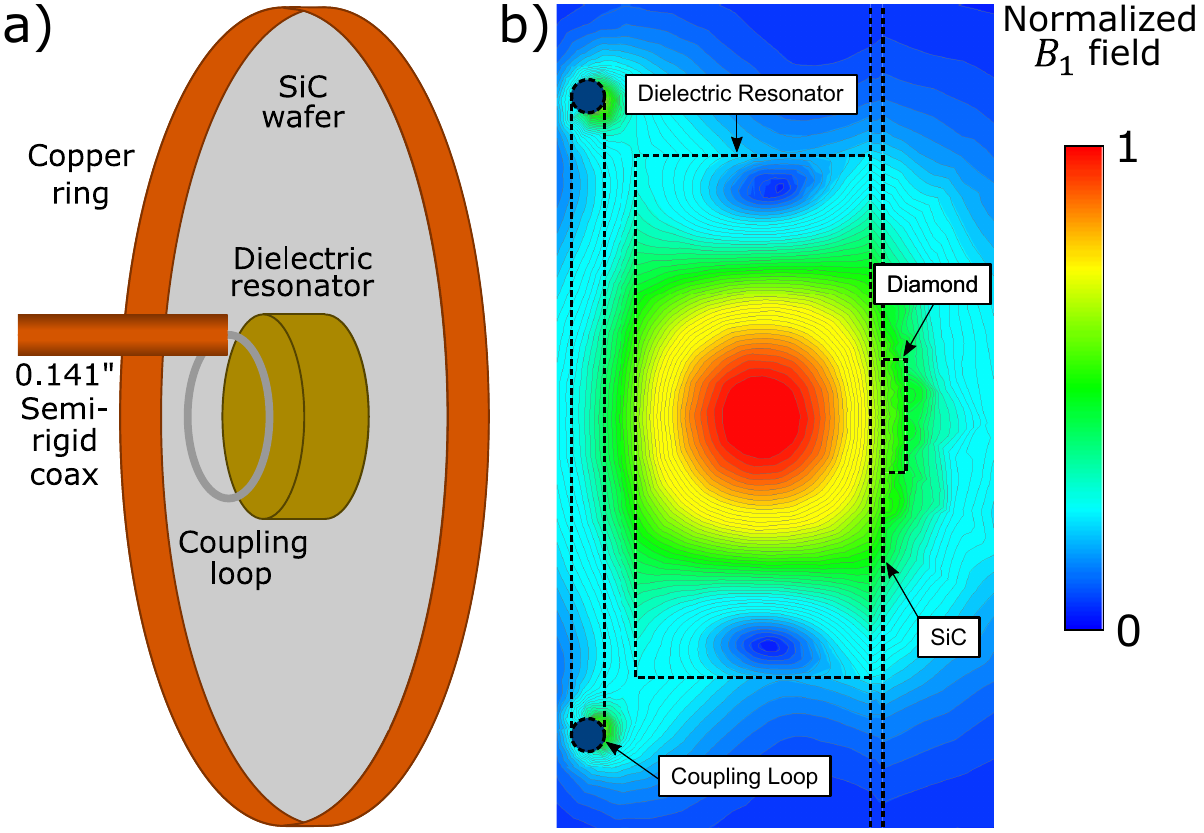}
\caption{\textbf{Dielectric resonator microwave delivery}. (a) The dielectric resonator is placed against the center of the SiC wafer. The diamond is adhered to the opposite side of the SiC wafer along the symmetry axis of the dielectric resonator. A semi-rigid coaxial cable couples the MW excitation signal into the dielectric resonator's TE$_{01\delta}$ mode. (b) Finite element simulation of the normalized $B_1$ field intensity from the dielectric resonator. The resonator, SiC wafer, coupling loop, and diamond are outlined with black dashed lines.}\label{fig:dielectricresonator}
\end{figure}
 
Population in the NV$^\text{-}$ ground-state Zeeman sublevels is manipulated using MW magnetic fields at frequencies near the zero-field splitting $\sim 2.87$ GHz. Uniform amplitude of this MW magnetic field, denoted $B_1$, is desirable so that NVs across the ensemble experience equal Bloch vector rotations. However, with non-resonant MW delivery solutions such as a microstrip or wire loop, achieving uniform Rabi frequencies $\gtrsim 2\pi \times 1$~MHz over the 3 mm $\times$ 3 mm $\times$ 0.07 mm NV-doped volume can require MW powers well above 10 W. However, since excitation is needed only over a $\sim \! 10$~MHz bandwidth near 2.87 GHz, a resonant MW structure can greatly decrease the required mhs relative to non-resonant approaches. For a resonator with a loaded quality factor $Q_L$, the MW magnetic field $B_1$ scales as $\sqrt{Q_L}$, while the full-width half-maximum (FWHM) bandwidth scales as $1/Q_L$.

We therefore employ a cylindrical dielectric resonator in proximity to the diamond to generate the $B_1$ magnetic field, as shown in Fig.~\ref{fig:dielectricresonator}. When installed, the dielectric resonator exhibits a loaded quality factor of $Q_L\sim 260$, resulting in an approximately 11 MHz bandwidth centered at $2.868$~GHz. This solution provides Rabi frequencies of $\sim 2\pi \times 10$~MHz over the 3 mm $\times$ 3 mm $\times$ 0.5 mm diamond volume using a 2~W amplifier (Minicircuits ZVE-8G+).

Additional experimental complications result if the MW drive intended to address a given transition also drives adjacent transitions. This unwanted cross-talk reduces the ability to independently adjust the Bloch vector rotation applied to different NV Zeeman sublevels~\cite{vandersypen2005nmr}. Cross-talk is particularly relevant to this device where the bias magnetic field $B_0 \approx 2.23$~gauss, with projection $1.29$~gauss on each NV axis, produces relatively small Zeeman shifts  $ \approx \pm  3.6$~MHz. The similar scale of the $^{15}$NV hyperfine splitting, approximately 3.0~MHz, makes avoiding such cross-talk even more difficult. 

We find that the Fourier broadening inherent to square-enveloped MW pulses of duration $\sim 1$ $\mu$s produces unacceptable levels of cross-talk. Therefore, Gaussian-enveloped MW pulses, $0.6$~$\upmu$s FWHM in $B_1$ field strength, are employed~\cite{vandersypen2005nmr}. These Gaussian-enveloped MW pulses are created by applying a Gaussian waveform from an arbitrary waveform generator to the IF port of a mixer and the MW signal to the LO port. For additional details on the dielectric resonator and MW delivery, see Supplementary Section~\ref{App:MWDelivery}.

\subsection{Double-quantum magnetometry}\label{sec:dq}

The sensor uses the double-quantum (DQ) measurement protocol~\cite{Fang2013high,Mamin2014multipulse,bauch2018ultralong,barry2020sensitivity} which offers two performance advantages relative to the more conventional single-quantum (SQ) protocol. First, the DQ protocol mitigates strain- and temperature-induced resonance shifts and broadening, thereby increasing the dephasing time and allowing use of longer free-precession times $\tau$. Second, the DQ protocol doubles the rate of magnetic-field-dependent phase accumulation, so that angular precession occurs at $2\gamma_e = 2 \times 2\pi \times 2.8$ MHz/gauss rather than at $\gamma_e = 2\pi \times 2.8$ MHz/gauss. 

The performance benefits of the DQ protocol can be quantitatively evaluated using Eqn.~\ref{eqn:ramseyshot}, where the doubled phase accumulation rate is accounted for by setting $\Delta m_s = 2$ for the DQ protocol rather than $\Delta m_s = 1$ for the SQ protocol, along with each protocol's measured dephasing time. Typical measured dephasing times are $T_{2,\text{SQ}}^* \approx 8.7$~$\upmu$s with the SQ protocol and $T_{2,\text{DQ}}^* \approx 14$~$\upmu$s with the DQ protocol. As the DQ protocol exactly doubles the phase accumulation rate, and adds approximately 50\% to this diamond's observed dephasing time, use of the DQ protocol with the given diamond is expected to grant an approximately three-fold enhancement in sensitivity over the SQ protocol.

\subsection{P$_1$ driving}

In this work, double quantum (DQ) measurement schemes are used to suppress a number of dephasing and decoherence mechanisms~\cite{bauch2018ultralong,barry2020sensitivity}; see Sec.~\ref{sec:dq}. With the diamond used here, the majority of residual NV$^\text{-}$ dephasing and decoherence observed in DQ measurements arises from dipolar coupling to substitutional paramagnetic nitrogen, also called P1 centers. When this coupling is mitigated, the value of $T_2^*$ and $T_2$ can be extended, as first shown in Refs.~\cite{DeLange2012controlling,Knowles2014observing,bauch2018ultralong}. 

To decouple the NV$^\text{-}$ spins from the P1 spin bath, continuous-wave (CW) radio-frequency (RF) fields are applied to near-resonantly drive the P1 spins during the NV$^\text{-}$ free precession time. These RF fields causes the P1 centers to undergo Rabi oscillations; if the P1 spin centers are driven through at least several rotations during the bare NV$^\text{-}$ dephasing time in a Ramsey measurement, the majority of the P1-induced broadening is removed. Ideally therefore the RF magnetic field intensity $B_1^\text{P1}$ driving the P1 centers should satisfy $\gamma_e B_1^\text{P1} \gg 2/T_2^*$. Details on the functional form and scaling of the broadening suppression with $B_1^\text{P1}$ field strength are fully discussed in Ref.~\cite{bauch2018ultralong}. A similar analysis applies for Hahn echo sequences, where several Rabi oscillations would be required during the bare decoherence time.

For a $B_0 = 2.23$ gauss field oriented along the $\langle 100 \rangle$ direction, four distinct dipole-allowed P1 transitions can be observed. These transitions can be driven using 22, 25, 135, and 139 MHz RF magnetic fields. To achieve the necessary RF field strength over the diamond volume, the device employs two multi-turn coils driven by resonant tank circuits, which address the 22 and 25 MHz resonances and the 135 and 139 MHz resonances respectively.  Using these two coils, applying P1 driving leads to an approximately $2\times$ increase of $T_\text{2,DQ}^*$ in double quantum Ramsey sequences (see Sec.~\ref{sec:T2starextension}), whereas an approximately $2.4\times$ extension of $T_\text{2,DQ}$ is found for double-quantum Hahn echo when adding P1 driving (see Sec.~\ref{sec:T2extension}). See Supplementary Section~\ref{Appsec:P1Driving} for a discussion of  P1 resonance frequency calculation, resonant coil loop construction, and RF delivery scheme.

\subsection{Pulse Sequences and Noise Reduction}\label{sec:pulsesequenceandNRS}

 \subsubsection{Resonant driving and phase modulation}
\label{sec:Digitialphaseshiftermaintext}

In a typical Ramsey scheme for an NV system, the MW frequency is tuned off-resonance so that free induction decay (FID) fringes occur as the precession time is varied~\cite{hart2021nv}. This off-resonant excitation prevents optimal population transfer, requires higher power to compensate for the detuning, and may result in increased off-resonant cross-excitation between resonances. Additionally, use of a single MW tone to excite multiple transitions, such as those created by the NV hyperfine structure, often results in multiple FID oscillation frequencies, which may be inconvenient.

To avoid the above non-idealities, the device here instead applies resonant MW pulses to each resolved Zeeman transition. However, no phase accumulates under such resonant drive, and FID fringes are not observed. To counter this problem, the final MW pulse is instead phase-shifted from the first MW pulse by a variable amount for each precession time $\tau$. For single-quantum Ramsey FID, the second MW pulse is phase-shifted by $\omega_m \tau$, producing a single-tone trace with fringe frequency $\omega_m$ under resonant drive.  The fringe frequency can be set as desired by adjusting $\omega_m$.

More generally, when applying a MW tone detuned by $\Delta \omega$ from the target transition, the accumulated phase for a single-quantum Ramsey protocol is
\begin{equation}
    \phi_m = \left(\Delta \omega + \omega_{m} \right) \tau,
\end{equation}
resulting in fringes at angular frequency $\Delta \omega + \omega_m$. To optimize the MW pulse frequency and amplitude, the MW frequencies are first tuned until the observed fringe frequency matches $\omega_m$, and the MW amplitude is subsequently varied to maximize the observed FID fringe amplitude.

In double-quantum Ramsey schemes~\cite{bauch2018ultralong}, fringes arise from differential phase accumulation between the \mbox{$|m_s = 1 \rangle$} and \mbox{$|m_s = - 1 \rangle$} states. Here, additive phases of  $\pm \omega_m \tau$ are applied to the second Ramsey pulse addressing the $|m_s = \pm 1 \rangle$ states, and fringes at angular frequency $2 \omega_m$ are observed for resonant MW drive. In a double-quantum Hahn echo scheme, additive phases of $\pm \omega_m\tau$ are added to the final MW pulse, also giving rise to fringes at angular frequency $2\omega_m$.

Finally, phase shifts can be added to ensure the magnetometer operates at the point of maximum fringe slope. This flexibility is important because the applied MW frequencies are constrained to multiples of the sequence repetition rate (see Supplementary Material Sec.~\ref{App:DQPulseOptimization}), preventing the fringe slopes from being being perfectly maximized by tuning the MW frequencies.

We found that digital phase shifters provide superior performance to analog phase modulation, as well as allowing convenient, precise, and instantaneous modification of the applied phase shift parameters; see Section~\ref{App:DPS}.

\subsubsection{Noise subtraction schemes}\label{sec:NoiseSubtraction}
 
Pulse-sequence-based noise subtraction can improve sensitivity by removing noise correlated between successive sequences. For instance, vibration, low-frequency laser intensity variation, or MW phase and amplitude noise may all produce noise which varies slowly compared to the pulse sequence repetition rate. Noise subtraction schemes typically operate by arranging the magnetic signal to be differential and the noise to be common-mode between consecutive sequences; subtraction of consecutive sequences then retains magnetic signals while suppressing noise.

Differential encoding of the magnetic signal between consecutive sequences is achieved by varying the phase shifts applied to the final MW pulse. For example in single-quantum (SQ) sequences, a $\pi$ phase shift inverts the final population difference between the interferometry states, thereby inverting the magnetic signal~\cite{BarGill2013solid}. Similarly, in double-quantum (DQ) sequences, a $\pi$ phase shift on either resonance involved in the double-quantum superposition inverts the signal, while $\pi$ phase shifts on both resonances restores the signal's original polarity~\cite{Mamin2014multipulse}. In some cases, these noise subtraction schemes can also protect against imperfect pulse sequences, for example removing residual single-quantum content due to imperfect $\pi$ pulses in DQ magnetometry~\cite{hart2021nv}.

Figure~\ref{fig:DPSphases} shows the phases applied for each noise subtraction scheme used in this work. For precession time sweeps, each step in the sweep is repeated four times, and a ``4-state'' progression of phase shifts is applied to provide maximal protection against imperfect MW pulses (Fig.~\ref{fig:DPSphases}b). For Ramsey magnetometry, a ``2-state'' noise subtraction scheme is employed as shown in Fig.~\ref{fig:DPSphases}a. The signal is thereby encoded at higher frequency than in a ``4-state'' scheme, reducing the effect of $1/f$ noise. In both these schemes, any additional phases applied to the final MW pulse (for example to observe on-resonance fringes, see \ref{sec:Digitialphaseshiftermaintext}) are added to the noise subtraction phase shifts. In Hahn echo magnetometry (Fig.~\ref{fig:DPSphases}c), the first sequence has a $\pi/2$ phase shift applied between the DQ pairs, and the second sequence has a $-\pi/2$ phase shift applied; this ensures operation at maximum fringe slope for small test fields, and no additional phase shifts are required. For all schemes,  identical phase shifts are applied for each nuclear spin state of a given electron spin transition.

Depending on the dominant noise source, the pulse-sequence-based noise subtraction can increase the device noise floor. Typically this noise subtraction increases the noise floor by $\lesssim 50$\%, but this effect is not fully understood. It is likely that the increase is related to MW noise, as the pulse-sequence-based noise subtraction increases the noise floor by $\lesssim 1$\% when all MWs are turned off.

\section{Experimental Results}
\label{sec:expresults}
\subsection{Experimental Overview}
Light from a 532~nm, 12~W laser is on/off-modulated by an acousto-optic modulator (AOM) and delivered through a photonic crystal fiber to the sensor head. Upon exiting the fiber, the light passes through a polarizer and a focusing lens. The focused light enters a faceted corner of the diamond in a light-trapping diamond-waveguide configuration. The resulting NV fluorescence is collected by a borosilicate total internal reflection (TIR) lens attached to a hexagonal borosilicate light pipe; upon exiting the light pipe, the light passes through a colored glass filter to a large area photodiode. In order to provide a reference signal for laser noise cancellation, the 532 nm light is sampled prior to entering the diamond, with the sampled light directed to a thin Teflon sheet acting as a diffuser in front of another large area photodiode. Photocurrent from each photodiode is integrated on a capacitor, and an instrumentation amplifier differences and amplifies the resulting voltages. Finally, the amplified difference voltage is digitized for subsequent software-based processing. See SM Section \ref{App:BalancingCircuit} for additional details on the balanced photodetection hardware.

While the $I=\frac{1}{2}$ nuclear spin of $^{15}$NV results in 16 different dipole-allowed electronic-ground-state Zeeman resonances, judicious orientation of the bias magnetic field along the diamond's $[100]$ direction results in only four resolved resonances. Each of the four resonances is addressed using a separate signal generator synthesizing a single MW tone. The four tones are individually digitally phase modulated and then combined. The resulting combined signal is mixed with a Gaussian pulse to create Gaussian-enveloped pulse waveforms, which are then amplified and directed to a wire loop at the end of a semi-rigid nonmagnetic coaxial cable. The MW field from the wire loop is coupled into a cylindrical dielectric resonator, with resonant frequency centered at $\approx 2.865$~GHz, located near the diamond.

In order to perform spin bath (P1) driving, three signal generators create RF signals at frequencies of 23, 135, and 139 MHz. The 135 and 139 MHz signals are combined, amplified, and coupled into a resonant coil with resonance near 137 MHz. The 23 MHz signal is also amplified and sent to a similar coil resonant near 23 MHz. Both resonant coils are proximate to the diamond.

An arbitrary waveform generator controls the timing of trigger and gating pulses for the 532 nm light, the microwaves, the P1 driving, the balanced photodetection circuit, the digitizer, and the digital phase shifters. For additional discussion of each major component of the sensor, see Section~\ref{sec:advancesoverview}.

\subsection{Pulse Sequences}\label{sec:pulsesequences}
\begin{figure}[!ht]
\centering
\begin{overpic}[width=\columnwidth-1.5cm]{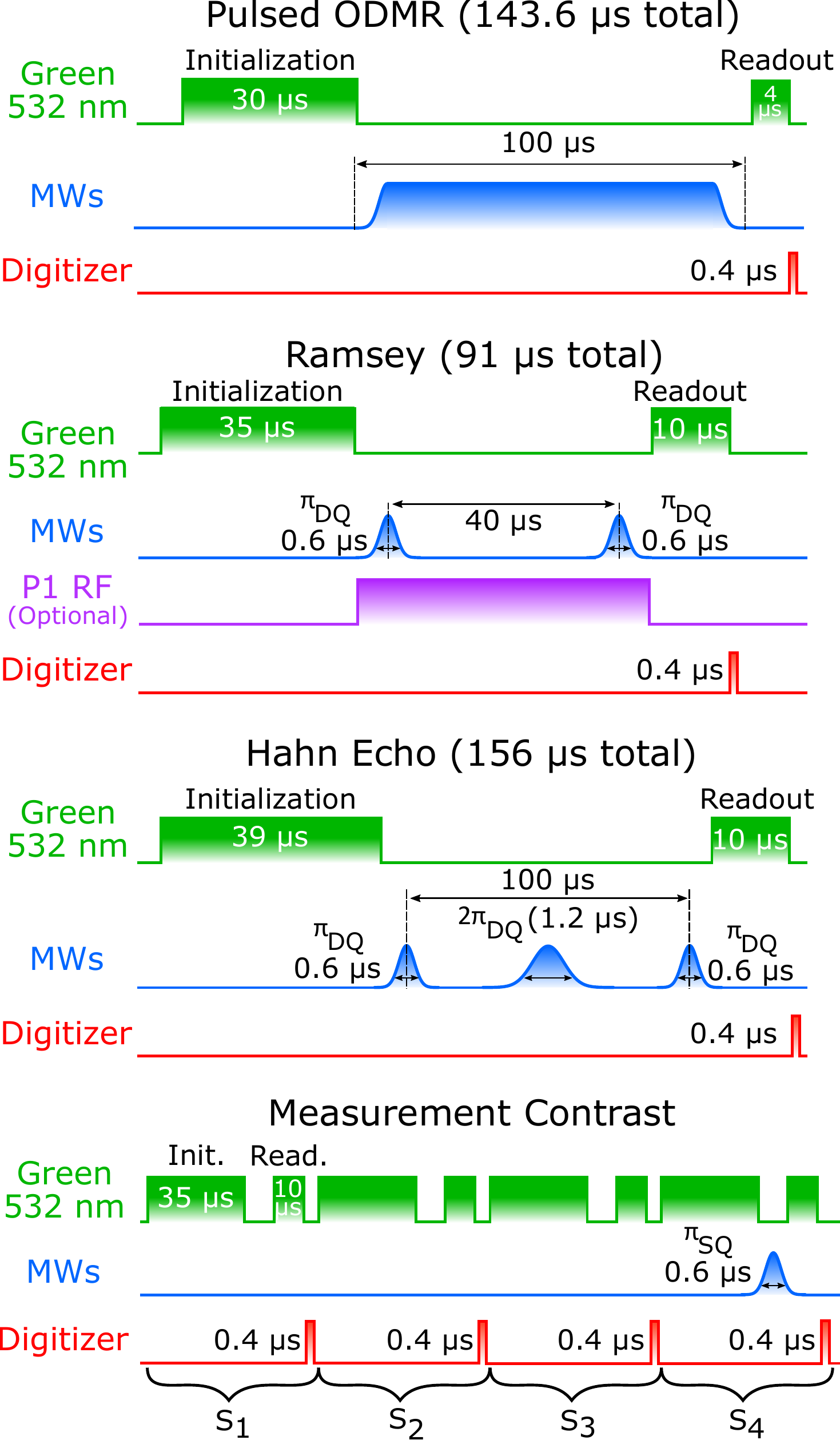}
 \put(3,98){(a)}
 \put(3,75){(b)}
 \put(3,47){(c)}
 \put(3,22){(d)}
\end{overpic}

\caption{\textbf{Selected pulse sequences used in this work.} (a-d) For all sequences, 532-nm laser light is applied for the time shown to initialize NVs. This is followed by a period of MW manipulation and/or free precession of the NV states. Finally, laser light is again applied to read out the ensemble, and once the light has been extinguished, the digitizer is triggered to sample the voltage accumulated on the integrating balanced photodetector. Additional details about the MW manipulation applied in each sequence are given in the text. (b) Optionally, P1 driving can be applied during the precession period to extend the dephasing time. (b-c) For pulse optimization and diagnostics, precession time sweeps can also be performed by fixing the timing of the final MW pulse and varying the timing(s) of the initial pulse(s). (d) To diagnose the initialization efficiency and measurement contrast, a series of four pulse sequences may be performed, with MW manipulation applied only during the final pulse sequence. }\label{fig:pulsesequences}
\end{figure}
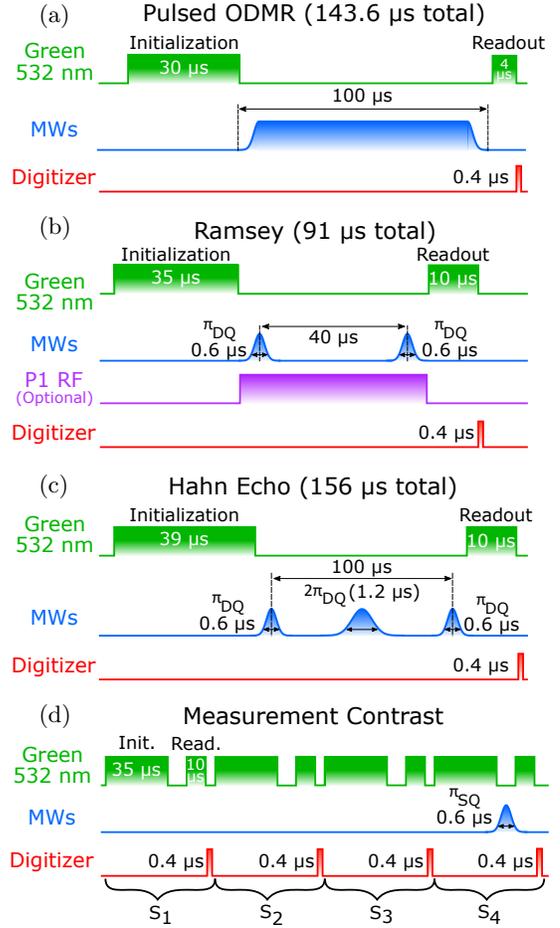

Figure~\ref{fig:pulsesequences} shows selected pulse sequences used in this work. Each sequence begins with a 532-nm laser initialization pulse of duration $t_I$. The middle of each sequence consists of periods of free precession and interspersed MW pulses. Each sequence ends by reading out the NV$^\text{-}$ fluorescence;  a light pulse of duration $t_R$ is applied to the diamond, during which the photocurrent is integrated by the balancing circuit and subsequently digitized (see Sec.~\ref{sec:integratingcircuit}).

The pulsed optically detected magnetic resonance (ODMR) sequence is used to measure NV$^\text{-}$ magnetic resonance spectra~\cite{Dreau2011avoiding}. Between initialization and readout, a 100 $\mu$s Gaussian flat-top SQ microwave pulse transfers population from the $|m_s=0\rangle$ state to one of the $|m_s = \pm 1\rangle$ states, allowing the magnetic resonances to be observed. Ramsey and the related free induction decay (FID) pulse sequences are used for broadband magnetometry and measurement of the dephasing envelope, respectively. In a Ramsey sequence, two Gaussian-shaped DQ $\pi$ pulses, with full width half maximum (FWHM) lengths of $0.6\;\mu$s, define the start and end of the 40 $\upmu$s precession. In order to avoid possible AC Zeeman shifts, the P1 drive was not applied during MW pulses during Ramsey magnetometry. The P1 driving is applied for 35 $\upmu$s during the precession time. 

The FID sequence is similar to the Ramsey sequence, but modified to measure $T_2^*$ dephasing. Although the total length of the sequence remains fixed at 91 $\upmu$s, the first MW pulse is variable-delayed while the second MW pulse remains fixed. The P1 driving is also applied for all times between initialization and readout, which is slightly different from the Ramsey case where the driving does not overlap with the MW pulses. Equivalent Ramsey and FID measurements can also be performed in the single quantum basis.\par

The Hahn echo sequence builds on the Ramsey sequence with a few changes. First, a $1.2\;\mu$s FWHM Gaussian DQ echo pulse is applied at the midpoint of the free precession time to refocus dephasing. Second, the initialization time is increased to $t_I = 39\;\mu$s. Third, the free precession time is increased to $\tau = 100$ $\upmu$s. Similar to the FID sequence, we fix the total length of the sequence at 156 $\upmu$s and measure the $T_2$ decoherence time by variable-delaying the first and second MW pulse while the third is fixed, such that the precession time is varied while maintaining equal spacing between successive MW pulses. If used, the P1 driving is applied for all times between initialization and readout. Optimization of MW pulse frequencies and powers is discussed in Sec.~\ref{App:DQPulseOptimization}.

The measurement contrast sequence is used to determine initialization fidelity and measurement contrast $C$ as a function of initialization and readout laser pulse duration. By observing the change in signal for three adjacent sequences without any MW pulses, the initialization fidelity can be determined. A final, fourth sequence transfers the now well-initialized population to one of the $|m_s = \pm 1\rangle$ states using a single MW pulse, allowing the measurement contrast to be determined; see SM Section~\ref{appsec:readoutfidelityoptimization}.

\subsection{Pulsed ODMR}\label{sec:ODMR}

\begin{figure}
\centering
\includegraphics[width=3.2in]{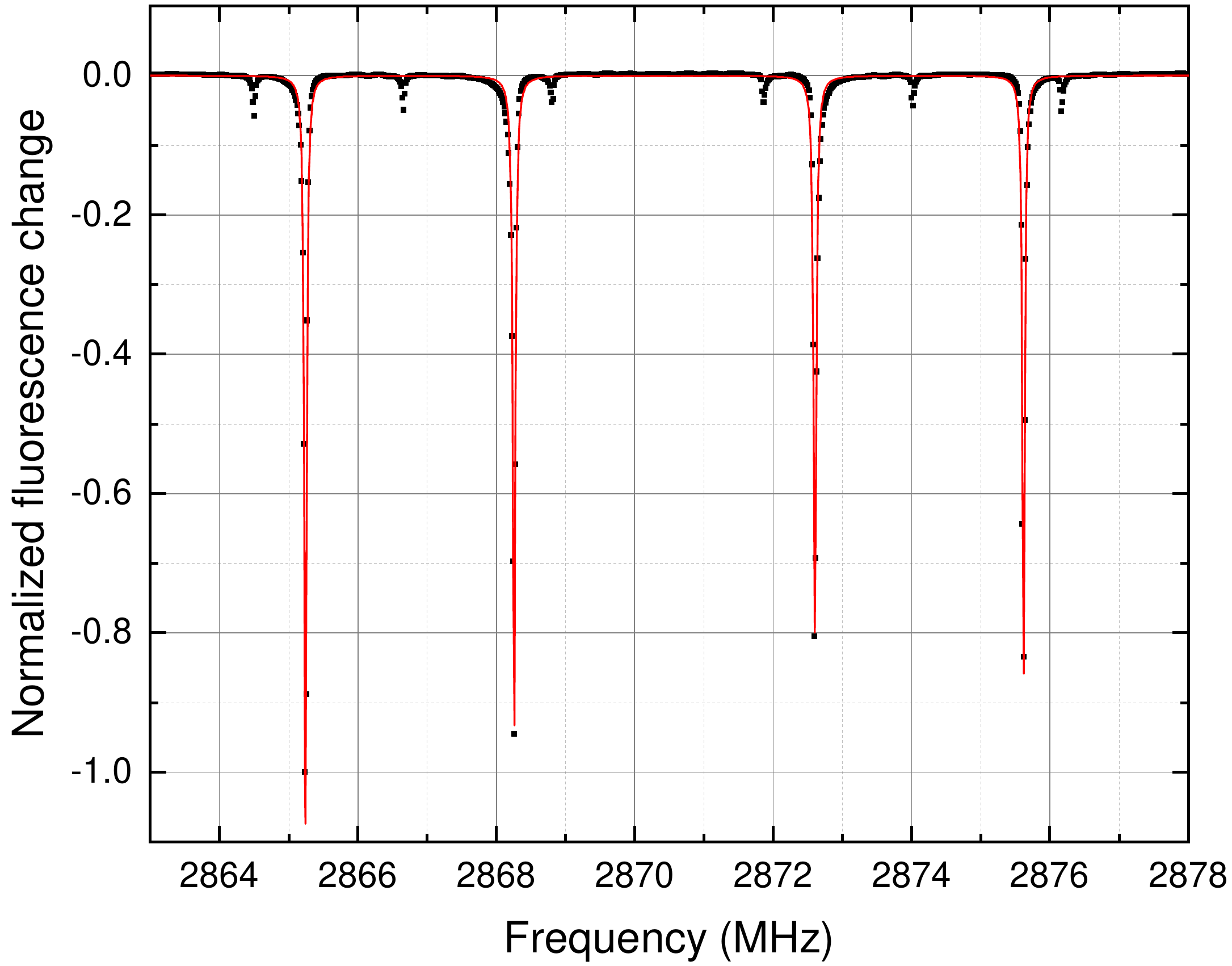}
\caption{\textbf{Pulsed optically detected magnetic resonance of the NV$^\text{-}$ Zeeman resonances.} The four large features arise from $^{15}$NV while the six smaller features arise from unwanted $^{14}$NV in the diamond.  Lorentzian fits to the data reveal FWHM linewidths of 34 to 46 kHz. This data was collected with 1 second of averaging time.}\label{fig:pulsedODMR}
\end{figure}

Accurate measurement of the NV$^\text{-}$ magnetic resonance spectrum allows for Zeeman resonance identification, bias magnetic field alignment, and general diagnostics. However, as the diamond used here exhibits transition linewidths down to 34~kHz, CW-ODMR methods are unsuitable~\cite{Dreau2011avoiding, barry2020sensitivity}. Observing such narrow linewidths with CW-ODMR requires not only MW Rabi frequencies below the transition linewidth but also reduced optical pumping rates, which greatly increase measurement duration~\cite{Dreau2011avoiding}. Alternatively, magnetic resonance spectra can be obtained by the method of Dr\'eau et al.~\cite{Dreau2011avoiding}, termed pulsed ODMR~\cite{barry2020sensitivity}. In this approach, optical excitation and MWs are applied at separate times to the diamond, and the fluorescence is recorded as the MW frequency is varied; see Fig.~\ref{fig:pulsesequences}a. Because the optical excitation and MWs are temporally offset, optically-induced power broadening of the Zeeman resonances is avoided~\cite{Dreau2011avoiding}.\par

Magnetic resonance data collected via pulsed ODMR are shown in Figure~\ref{fig:pulsedODMR}. Because pulsed ODMR is used for diagnostic rather than magnetometetry purposes, avoiding broadening of spectral lines is paramount, while measurement contrast is a secondary concern. Thus, the chosen pulsed ODMR scheme applies a weak $100$~$\mu$s MW pulse, much longer than $T_2^*$, to avoid Fourier broadening the resonance. With optimal alignment of the bias magnetic field, we observe FWHM linewidths varying between 34 and 46 kHz. If Lorentzian lineshapes are assumed, these linewidths correspond to $T_\text{2,SQ}^*$ values of 6.4 to 9.4~$\mu$s, consistent with measurements using Ramsey interferometry; see next section.

\subsection{$T_2^*$ Extension}
\label{sec:T2starextension}

\begin{figure}
\centering
\includegraphics[width=3.2in]{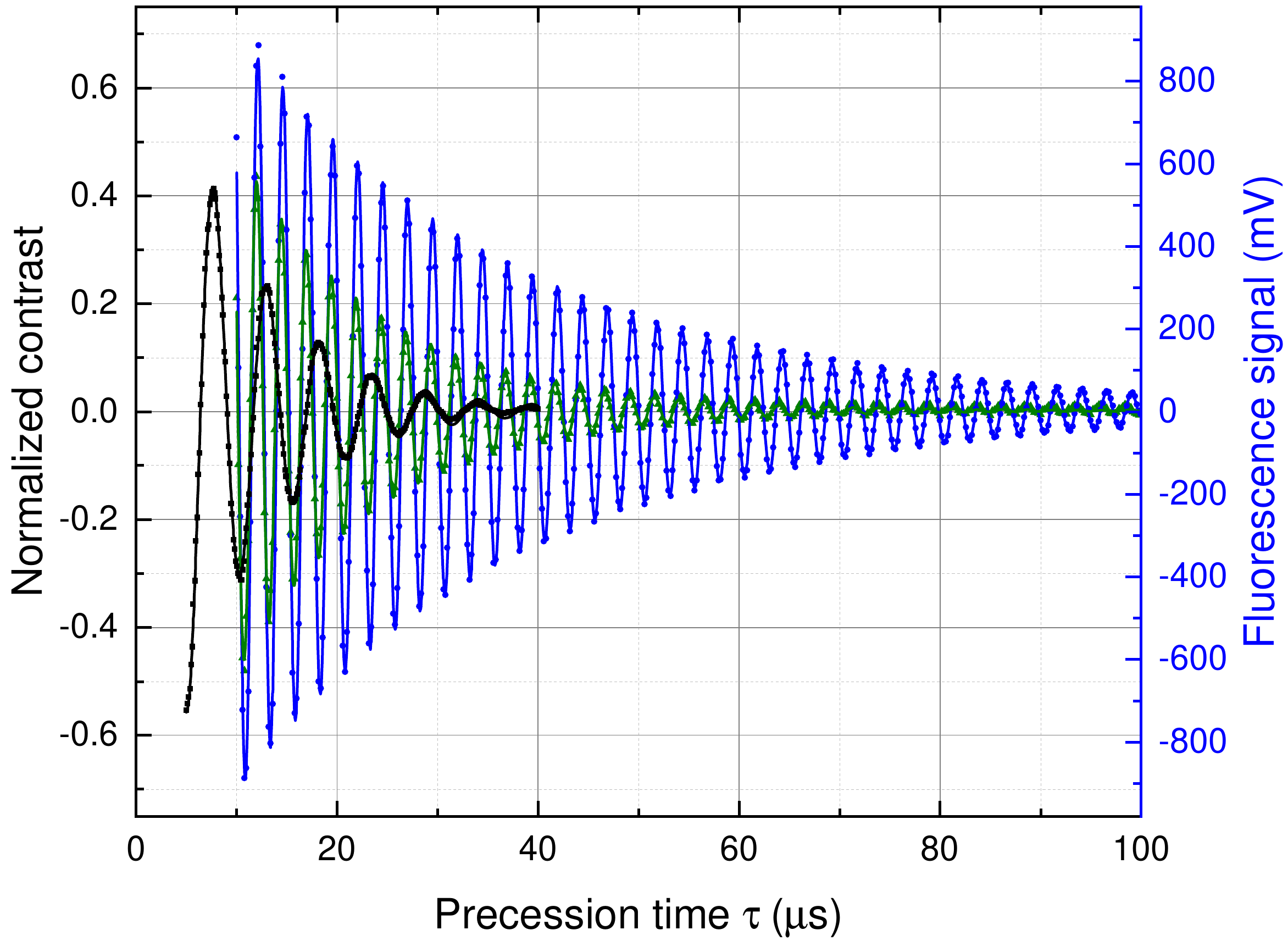}
\caption{\textbf{Extension of $T_2^*$ dephasing time.} Free induction decay fringes are acquired for a total time of 1 second each using the single quantum protocol (\textcolor{black}{\tiny$\blacksquare$}), the double quantum protocol (\textcolor{thegreen}{$\blacktriangle$}), and the double quantum protocol with P1 driving (\textcolor{blue}{\textbullet}). Lines show fits to Eqn.~\ref{eqn:dephasingdecay}, yielding $T_\text{2,SQ}^*= 8.7(1)$ $\mu$s, \mbox{$T_\text{2,DQ}^*=14.0(1)$~$\mu$s}, and $T_\text{2,DQ+P1}^*=28.6(2)$~$\mu$s when the stretched exponential parameter is fixed at $p=1$. The DQ scheme results in an improved dephasing time over the SQ scheme, especially with P1 driving applied. Note the doubled fringe frequency of the two DQ measurements relative to the SQ measurement, depicting the doubled response to magnetic fields as discussed in Section~\ref{sec:dq}. Normalized contrast (left scale) is given as the fluorescence signal at precession time $\tau$ divided by the expected signal at time $\tau = 0$ based on the exponential decay fit. The right scale shows the fluorescence signal in native units for the DQ with P1 driving case.}\label{fig:T2starextension}
\end{figure}

As discussed in Refs.~\cite{bauch2018ultralong,barry2020sensitivity}, combining a double-quantum (DQ) measurement scheme with P1 driving can substantially increase the ensemble dephasing time $T_2^*$. Figure~\ref{fig:T2starextension} shows free induction decay (FID) fringes measured for variable time $\tau$ using the single quantum (SQ) protocol, the DQ protocol, and the DQ protocol with P1 driving. Measurements with the SQ protocol use the $|m_s=0\rangle \leftrightarrow |m_s=-1\rangle$ transition, and drive the two lower-frequency Zeeman resonances shown in Figure~\ref{fig:pulsedODMR} associated with the two $^{15}$NV nuclear spin states. Measurements with the DQ protocol use both the $|m_s=0\rangle \leftrightarrow |m_s=-1\rangle$ and $|m_s=0\rangle \leftrightarrow |m_s=+1\rangle$ transitions, and thus drive all four $^{15}$NV Zeeman resonances shown in the same figure. All four transitions are addressed on-resonance, and the final MW pulse is digitally phase modulated at 200 kHz (SQ) or 400 kHz (DQ) to create FID fringes; see Section~\ref{sec:Digitialphaseshiftermaintext}. To mitigate noise, measurements use the 2-state (SQ) or 4-state (DQ) noise subtraction methods described in Section~\ref{sec:NoiseSubtraction}. 

For all three schemes, FID data are mean-subtracted and fit to a decaying sinusoid $y(\tau)$ of the form
\begin{equation}\label{eqn:dephasingdecay}
   y(\tau) =  A e^{-\left(\tau/T_2^*\right)^p}\sin[2\pi f \tau+\phi],
\end{equation}
where $f$ denotes the fringe frequency, $\phi$ is a phase, $p$ is the stretched exponential parameter, and $A$ is the amplitude~\cite{bauch2018ultralong}. %
For comparison among the three schemes, $p$ is fixed at one so that the $T_2^*$ fit value alone characterizes the temporal dephasing. With $p=1$, the fits yield $T_\text{2,SQ}^*= 8.7(1)$ $\mu$s for the SQ basis, \mbox{$T_\text{2,DQ}^*=14.0(1)$~$\mu$s} for the DQ basis, and $T_\text{2,DQ+P1}^*=28.6(2)$~$\mu$s for the DQ basis with P1 driving. The expected improvement in sensitivity can be determined using Eqn.~\ref{eqn:ramseyshot} under the assumption $p=1$, and finds that employing the DQ scheme instead of the SQ scheme is expected to improve sensitivity by 3.1$\times$. The addition of P1 driving to the DQ scheme is then expected to increase the sensitivity improvement to 5.8$\times$ relative to the SQ scheme. See Supplemental Section \ref{sec:fitparameters} for fit parameters when $p$ is allowed to vary and calculations for this section, and Ref.~\cite{bauch2018ultralong} for the causes of $p\neq 1$.

\subsection{$T_2$ Extension}
\label{sec:T2extension}
\begin{figure}[t]
\centering
\includegraphics[width=3.2in]{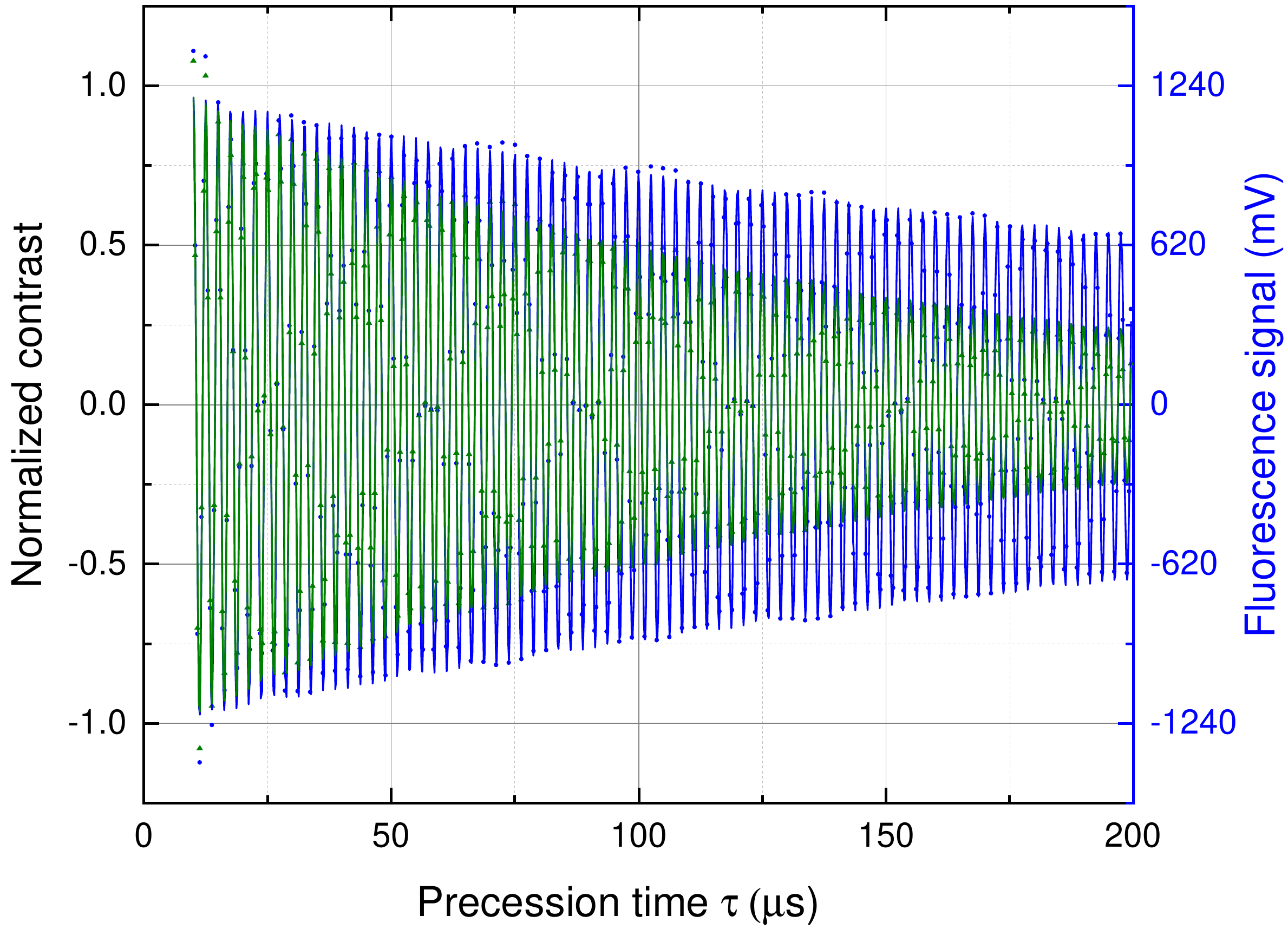}
\caption{\textbf{Extension of Hahn echo $T_2$ decoherence time.} Interference fringes are measured for a total of 1 second each using the double quantum Hahn echo protocol without (\textcolor{thegreen}{$\blacktriangle$}) and with (\textcolor{blue}{\textbullet}) P1 driving. Lines show fits to Eq.~\ref{eqn:dephasingdecay}, yielding $T_\text{2,DQ} = 140(2)$ $\mu$s without P1 driving and $T_\text{2,DQ+P1} = 324(16)$ $\mu$s with P1 driving, both with $p=1$. As with the FID data in Fig.~\ref{fig:T2starextension}, the normalized contrast (left scale) is given as the fluorescence signal at precession time $\tau$ divided by the expected signal at time $\tau = 0$ based on the exponential decay fit. The right scale shows the fluorescence signal in native units for the case with P1 driving.}\label{fig:T2coherencedecay}
\end{figure}

P1 driving is also effective to extend Hahn echo $T_2$ coherence times. $T_2$ coherence time measurements are measured similarly to $T_2^*$, but with an added echo MW pulse applied halfway through the precession. As shown in Fig.~\ref{fig:pulsesequences}, this echo pulse has twice the length of the first and final pulses (see  and refocuses dephasing. Figure~\ref{fig:T2coherencedecay} shows Hahn echo interference fringes measured in the DQ basis with and without P1 driving. Fitting the fringe data to Eqn.~\ref{eqn:dephasingdecay} in the manner described in the previous section yields \mbox{$T_\text{2,DQ}= 136(3)$~$\mu$s} without P1 driving, and $T_\text{2,DQ+P1}=324(16)$~$\mu$s with P1 driving. The addition of P1 driving to the DQ Hahn echo scheme is expected to enhance sensitivity by 1.8$\times$, as detailed in Supplementary Section~\ref{sec:fitparameters}.

In practice, additional technical noise sources and other limitations present during P1 driving can prevent the expected sensitivity gain from being realized. In Hahn echo magnetometry (but not in Ramsey magnetometry), these noise sources could not be sufficiently mitigated in the experiments here, and P1 driving is not used in the Hahn echo magnetometry results presented. These noise sources and mitigations are further discussed in the next section.

\section{Magnetometry Demonstration}\label{sec:magdemo}

\begin{figure}[thp]
\begin{overpic}[width=3.2in]{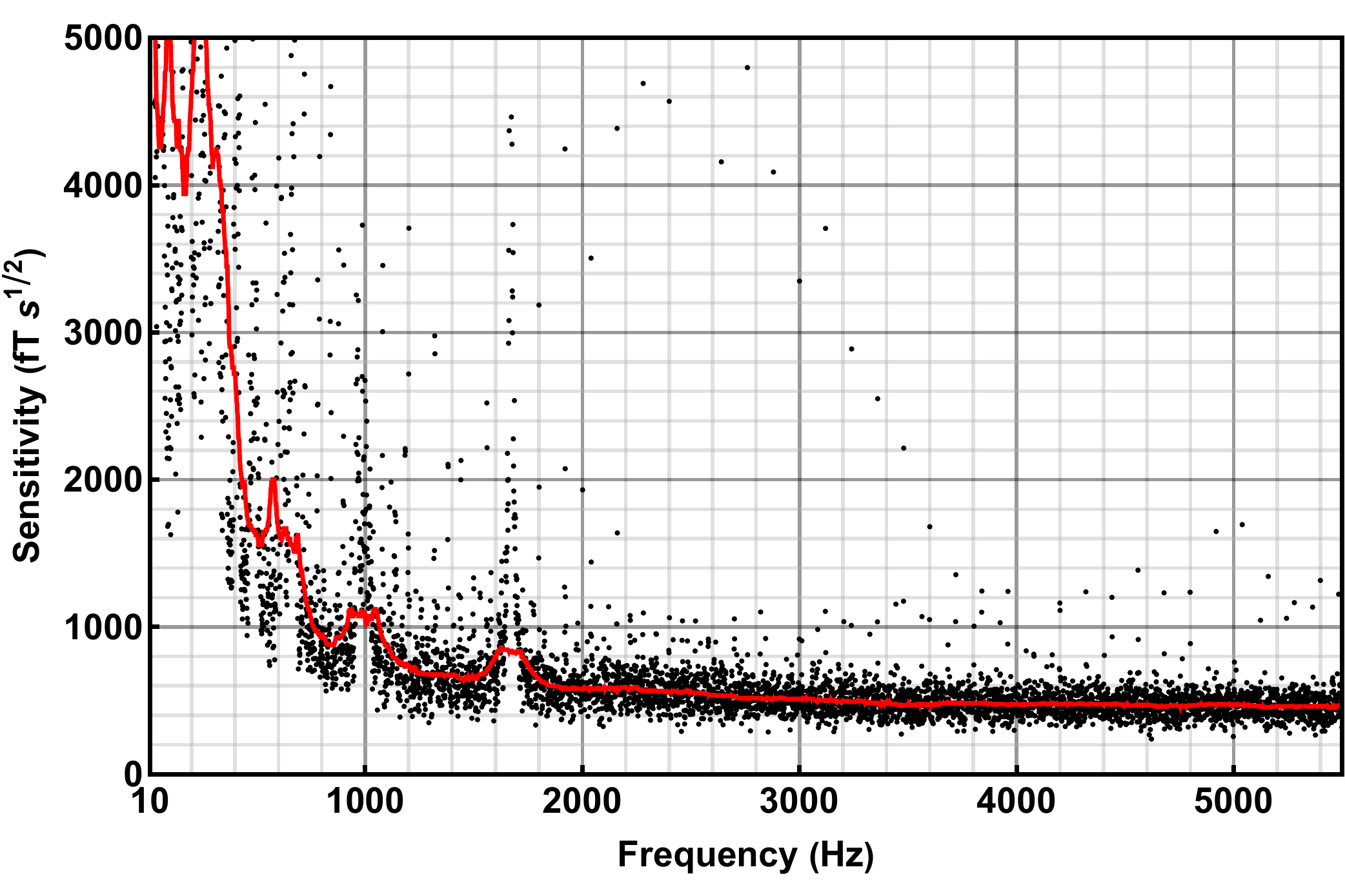}
 \put(-3,63){(a)}
\end{overpic}
\;
\begin{overpic}[width=3.2in]{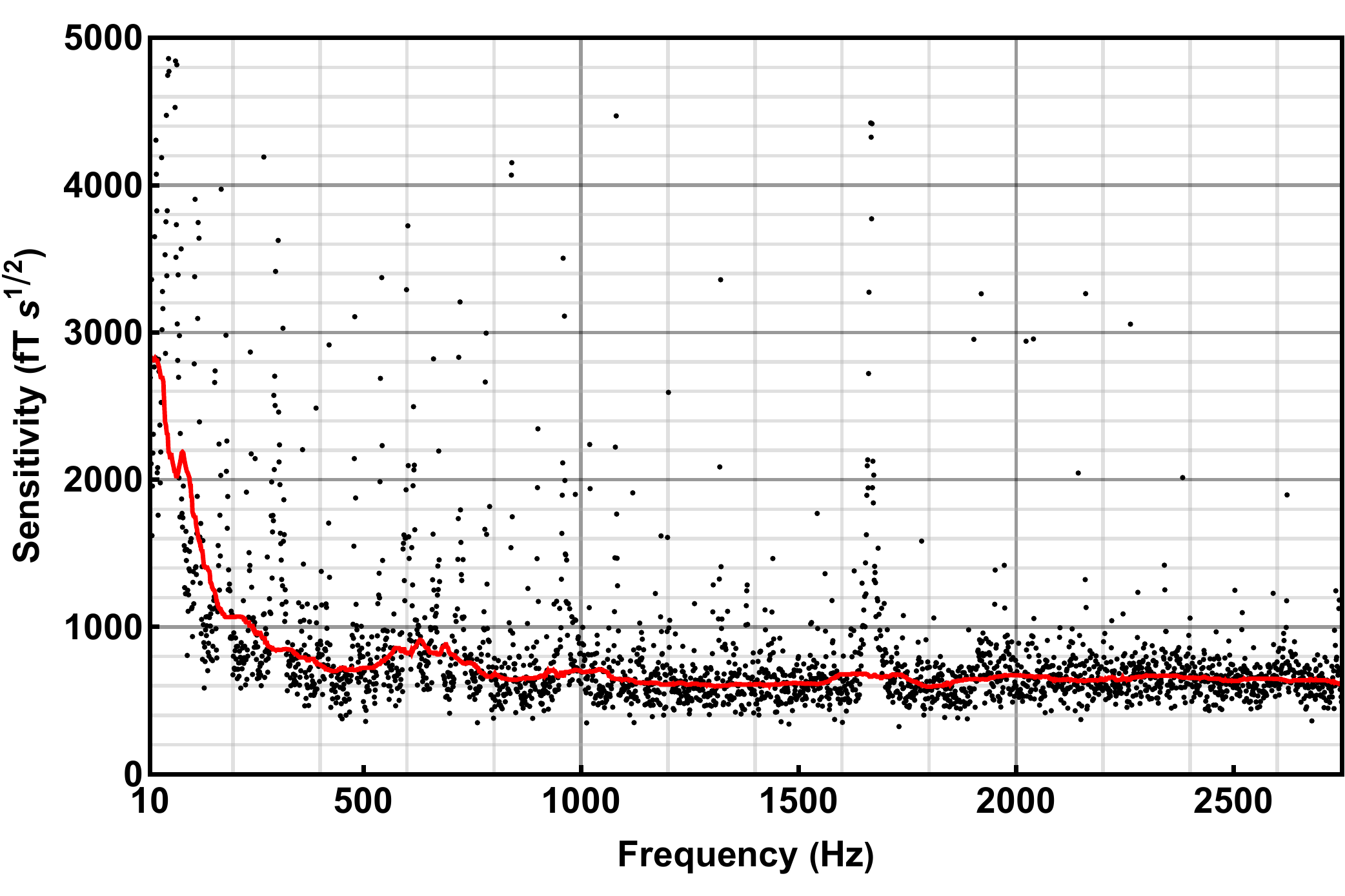}
 \put(-3,63){(b)}
\end{overpic}
\caption{\textbf{\textcolor{mhsnew}{Ramsey magnetometry sensitivity spectra.}} (a) Sensitivity spectrum for double-quantum Ramsey protocol with $40$-$\upmu$s precession time and with P1 driving. No noise subtraction is applied. The device achieves a minimum sensitivity of 460~fT$\cdot$s$^{1/2}$. Device noise is increased by less than 20\% above this value for approximately half the $5.5$~kHz measurement bandwidth. (b) Sensitivity spectrum for double-quantum Ramsey protocol with $40$-$\upmu$s precession time, P1 driving applied, and two-state noise subtraction. The device achieves a minimum sensitivity of 640~fT$\cdot$s$^{1/2}$. Device noise is increased by less than twenty percent for 65$\%$ of the $2750$~Hz measurement bandwidth. For both panels, the raw sensitivity data (\textcolor{black}{$\cdot$}) and 100-point median-filtered data (\textcolor{red}{---}) are calculated by rms averaging ten 1-s long acquisitions. Additional details on the sequence, the P1 driving, and the noise subtraction are given in the main text. To match conventions for sensitivity, data are shown as the positive-frequency half of a double-sided spectrum. See SM Sec.~\ref{sec:calibration} and \ref{App:CalcSensitivity} for additional information on sensitivity measurement.}\label{fig:ramseysensitivity}
\end{figure}

The magnetometer is tested in two configurations. In the first configuration, the device uses a double-quantum Ramsey sequence and is sensitive to broadband magnetic fields at frequencies down to DC. In the second configuration, the device uses a double-quantum Hahn echo sequence and is sensitive to AC magnetic fields around a narrow, pre-selected frequency dependent on the MW pulse spacing. 

In both configurations, the magnetometer response is measured by applying a small, calibrated test magnetic field parallel to the device bias field. This test magnetic field is created by applying a known voltage from a function generator to a multi-turn coil in series with a known resistance. Determining the applied test field requires calculating or measuring the proportionality constant $\kappa$ between the test field amplitude and the function generator voltage output. The value of $\kappa$ is determined in three different ways as described in Supplemental Section~\ref{sec:calibration}: calculating the field using the known geometry and current, measuring the field with a commercial magnetometer, and, finally, calibrating the field using the fixed NV gyromagnetic ratio.

The final method determines the field and associated voltage required to shift the Ramsey or Hahn echo interferometry fringes by one period. As the field value producing $2\pi$ phase accumulation depends only on fundamental constants and the known precession time, this method is expected to be the most accurate. Using a Ramsey magnetometry sequence, we find $\kappa = 201$ nT/V for a DC test field, while the Hahn echo procedure yields $\kappa = 198$ nT/V for a 6.4~kHz test field, a difference of only 1.4\%; see Section~\ref{sec:hahnechosensitivity}.

\subsection{Ramsey magnetometry}\label{sec:ramseysensitivity}

In this work, a single double-quantum Ramsey sequence consists of a $t_\text{I} =35$~$\upmu$s optical initialization, a $\tau = 40$~$\upmu$s free-precession time, a $t_\text{R} = 10$~$\upmu$s optical readout, and $t_\text{D}= 6$~$\upmu$s of dead time. Initialization and readout times are chosen to approximately maximize readout fidelity; see SM Section~\ref{appsec:readoutfidelityoptimization}. While the value of $T_\text{2,DQ+P1}^*$ reported in Section~\ref{sec:T2starextension} suggests an optimum free-precession time $\tau \sim 24\;\upmu$s, somewhat longer precession times are found to produce lower noise floors and improve sensitivity; thus $\tau=40\;\upmu$s is employed instead.

Ramsey measurements are performed at repetition rate $f_\text{rep}^{\text{Ram}} = 11$~kHz. Resulting magnetometer data are recorded in 1-second-long segments, which are then Fourier transformed with rectangular windowing to produce an amplitude spectral density with 1-Hz wide frequency bins. A test field of known size is applied at 10~Hz, allowing the voltage amplitude spectral density to be converted into magnetic field units. P1 driving is applied for 35 $\upmu$s during the precession and is not on during the Gaussian pulses. This duration of P1 driving allows nearly all of the $T_2^*$ extension to be retained while eliminating the excess noise observed when P1 driving is active during readout and digitization.

When operated without any noise subtraction schemes, the device achieves a minimum sensitivity of 460~fT$\cdot$ s$^{1/2}$ within its 5.5 kHz measurement bandwidth, as shown in Fig.~\ref{fig:ramseysensitivity}a and further detailed in SM Sec.~\ref{App:CalcSensitivity}. However, without any noise subtraction schemes, the device exhibits distinct low-frequency noise below $\sim\!2$ kHz. As shown in Fig.~\ref{fig:ramseysensitivity}b, applying 2-state noise subtraction mitigates some of this low-frequency noise but at a cost in bandwidth and minimum sensitivity, now reduced to $f_\text{rep}/4 = 2.75$ kHz and increased to 640~fT$\cdot$s$^{1/2}$ respectively. The mechanism responsible for this noise increase is not fully understood, but with no MWs applied the noise is approximately the same, e.g. to within $\sim 1\%$, with and without the 2-state subtraction. This latter observation suggests the responsible mechanism is likely related to MW noise rather than laser noise alone.

Operating the device using a DQ protocol with 4-state noise subtraction, as described in Sec.~\ref{sec:NoiseSubtraction}, rejects residual SQ signal content from imperfect state transfer~\cite{hart2021nv,kazi2021widefield} and therefore produces the most accurate measurement of $T_\text{2,DQ}^*$. However, this 4-state noise subtraction is not used for magnetometry, as bandwidth is further reduced to $f_\text{rep}/8 = 1.375$ kHz, and the noise floor increases further relative to 2-state noise subtraction.

To check the observed sensitivity is consistent with that expected from the measured contrast and photon shot noise, we calculate the shot-noise-limited sensitivity and find $\eta^\text{ens,sho}_\text{Ram}$ $=  260\;  \text{fT}\cdot$s$^{1/2}$; see SM Sect.~\ref{sec:ramseysensitivity}. With the MW pulses off, the observed noise is approximately $5\%$ above shot noise on the fluorescence photocurrent; see SM Sect.~\ref{sec:shotnoise}. Turning MW pulses on increases the noise to $\sim\!60\%$ over shot noise in the 2 kHz to 5.5 kHz band. This analysis therefore suggests the device should be capable of exhibiting a sensitivity near 420 fT$\cdot$s$^{1/2}$ for frequencies above 2 kHz, in agreement with the minimum wideband sensitivity depicted in Figure~\ref{fig:ramseysensitivity}a.

\subsection{Hahn echo magnetometry}\label{sec:hahnechosensitivity}

\begin{figure}
\centering
\includegraphics[width=\columnwidth]{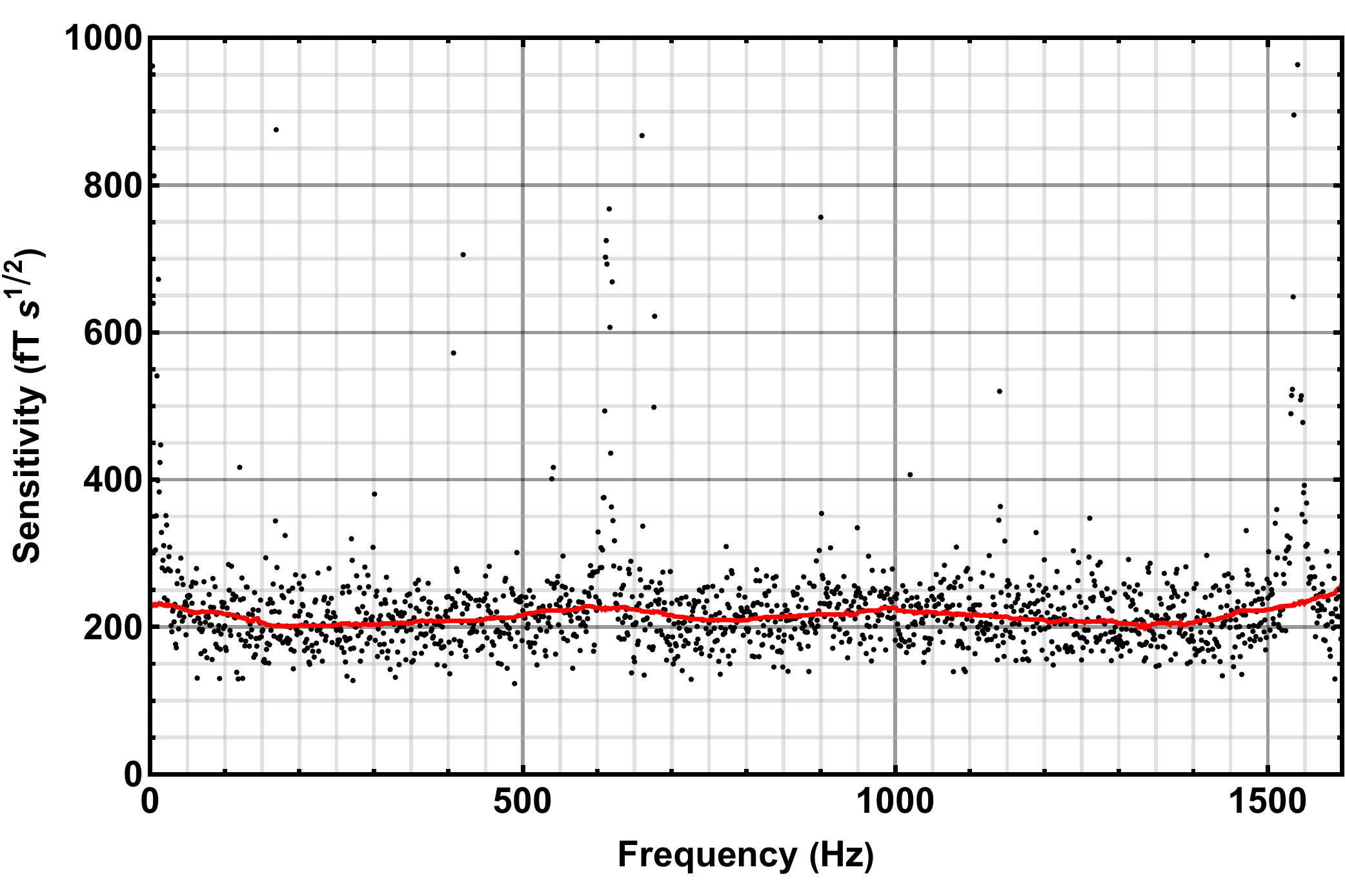}
\caption{\textbf{Hahn echo magnetometry sensitivity spectrum.} Sensitivity spectrum for double-quantum Hahn echo protocol with 100~$\upmu$s total precession time, shown as raw values ($\cdot$) and with a 100-point-median-filter applied (\textcolor{red}{\textbf{---}}). The device achieves a sensitivity noise floor $\approx 210$~fT$\cdot$s$^{1/2}$; device noise is increased by less than twenty percent for 80$\%$ of the $1.6$~kHz measurement bandwidth. No P1 driving is applied. As in Fig.~\ref{fig:ramseysensitivity}, the data are calculated by rms averaging ten 1-s long acquisitions. Additional details on the sequence and the noise subtraction are given in the main text. To match conventions, data are shown as the positive-frequency half of a double-sided spectrum. See SM Sec.~\ref{App:HahnEchoSensitivity} for additional details on sensitivity determination.}\label{fig:HahnEchoData}
\end{figure}

A single double-quantum Hahn echo (HE) sequence consists of a $t_\text{I} =39$~$\upmu$s optical initialization, a $\tau = 100$~$\upmu$s precession time, a $t_\text{R} = 10$~$\upmu$s optical readout, and $t_\text{D}= 7$~$\upmu$s of dead time. Individual Hahn echo sequences are run at repetition rate $f_\text{rep}^{\text{HE}} = 6.4$ kHz. The resulting data is again continuously recorded in 1-second lengths and Fourier transformed and filtered as described in Section~\ref{sec:ramseysensitivity}. The magnetometer is expected to exhibit maximal response to fields with frequency equal to $f_\text{rep}^{\text{HE}}$ and a phase such that the field zero coincides with the midpoint of the refocusing pulse.

To mitigate low frequency noise in the Hahn-echo-based magnetometer, two-state noise subtraction is used; see Section~\ref{sec:NoiseSubtraction}. The first state consists of a sequence where the final pulse has a $\pi/2$ phase difference applied (symmetrically) between the upper and lower transition frequencies (including both hyperfine components), while the second state instead has a $-\pi/2$ phase difference applied. See SM Fig.~\ref{fig:DPSphases} for a graphical depiction of the phases applied in Hahn echo magnetometry. This scheme produces maximum slope at zero applied test field, with the second state's response inverted relative to the first, so that signal from the two states can be subtracted to reject noise and preserve magnetometer response.

Magnetometer sensitivity using Hahn echo sequences is measured as follows. A square wave test magnetic field with frequency $f_\text{rep}^{\text{HE}}$ is applied, and the field's phase is adjusted so the zero-crossing occurs halfway through the precession time. Magnetometer data is recorded as the test field's magnitude is randomized over a range of values, giving the drive response in native signal units per tesla. Magnetometer data is concomitantly recorded with no test field applied, and Fourier transformed, yielding an amplitude spectral density in native signal units times root seconds. Combining these two measurements yields a sensitivity spectrum, with units of $\text{T}\!\cdot\!\text{s}^{1/2}$, as shown in Figure~\ref{fig:HahnEchoData} and detailed in SM Sec.~\ref{App:HahnEchoSensitivity}. The measurement shows a minimum sensitivity $\approx 210\;\text{fT}\cdot$s$^{1/2}$ to amplitude modulation of a 6.4 kHz AC field, as shown in Fig.~\ref{fig:HahnEchoData}. This value is consistent with the calculated photon-shot-noise-limited sensitivity of $\eta^\text{ens,sho}_\text{HE} =  90\;  \text{fT}\cdot$s$^{1/2}$; see SM Sec.~\ref{sec:hahntheorysensitivity}.

As discussed in Section~\ref{sec:T2extension}, P1 driving is not applied for Hahn echo magnetometry since the gain in sensitivity from a longer decoherence time is offset by an increase in electronic noise associated with P1 driving. Similar to the Ramsey case, application of 4-state scheme is therefore only used to determine $T_\text{2,DQ}$.

\section{Discussion and Outlook}
\label{sec:outlook}

This paper represents a concerted effort to implement the approaches suggested in Ref.~\cite{barry2020sensitivity} to improve sensitivity of ensemble-based NV magnetometers. As in Ref.~\cite{barry2020sensitivity}, methods to increase dephasing time are a central focus. With longer dephasing times, phase accrual for a given measurement is increased, which translates to better magnetometer sensitivity. Beyond using a custom-grown diamond with favorable strain and dipolar broadening properties targeted to this application, the $T_2^*$ dephasing time is further augmented by double quantum protocols and P1 driving \cite{bauch2018ultralong}. As a result, the device exhibits the best sensitivity reported to date for an NV magnetometer without a flux concentrator, in both broadband and narrowband magnetometry modes. However, construction and operation of an NV-diamond magnetometer in this new regime of dephasing time and sensitivity produces unanticipated challenges, discussed below.

\subsection{Obstacles encountered}

Major technical obstacles encountered in this work are summarized below. Additional detail on these technical obstacles is provided in SM~Sec.~\ref{sec:obstaclessupplement}.

The need for an extremely low noise MW signal chain constitutes a major challenge of this work, requiring careful selection of components, measurement of the MW noise achieved, and iteration of the signal chain design. Despite this effort, MW noise is still believed to limit device performance.

To minimize noise, as well as thermally-induced drift in MW resonance frequencies, the alignment of the laser light into the fiber and subsequently into the diamond requires careful control. The laser power reaching the diamond is monitored during device optimization and magnetometry, and alignment of laser light into the fiber optic cable is adjusted whenever the power falls by $\gtrsim 1$\%.

The high RF power applied during P1 driving represents another sizable heat load on the diamond, resulting in significant equilibration times despite a high thermal conductivity path from the P1 coils out of the sensor head.

The P1 driving also produces electromagnetic interference (EMI) that required mitigation and limited the power applied. For example, the P1 driving duration could not be extended to the final $\pi$ pulse in DQ Ramsey without increasing device noise; this effect was traced to coupling into the photodiode balancing circuit of RF EMI present during coil ring-down.

Despite using noise subtraction schemes, the device remains sensitive to certain sources of low-frequency non-magnetic noise. The factors limiting performance of these noise subtraction schemes require further investigation to determine whether noise rejection can be improved.

Finally, future applications could introduce additional complexities for magnetometers of this design. Compared to other NV ensemble magnetometers, the reliance here on relatively long dephasing times for Ramsey magnetometry results in increased sensitivity to DC magnetic field gradients. In addition, by applying $\textbf{B}_0$ exactly normal to the diamond's $\{100\}$ front facet, the existing device reduces the number of non-degenerate Zeeman transitions from sixteen in the arbitrary case to four. While these features are useful in a static laboratory device, they require careful tuning of the applied bias field; modification to the scheme will be needed for a portable device subject to a dynamically-oriented Earth magnetic field. One solution might null out Earth's field using coils, but the current sources driving the coils could create excess noise. Alternatively a larger bias magnetic field could be applied along a different axis to resolve all sixteen transitions, which could be interrogated sequentially or simultaneously~\cite{Schloss2018Simultaneous}, but the MW control scheme would be considerably more complex. In both cases, calculations suggest best Ramsey magnetometry performance is only possible with field gradients $\ll 200$ nT over the NV$^\text{-}$ sensing volume; see SM Section~\ref{sec:biasmagneticfieldgradients}.

\subsection{Prospects for improvement}

Several approaches may improve future device sensitivity. For example, by focusing incoming magnetic field lines using flux concentrators~\cite{Fescenko2020diamond,xie2021hybrid,zhang2022pulsed} other NV magnetometers exhibit device response beyond that set by the electron gyromagnetic ratio. However, such devices face limitations. For example, the geometry-dependent flux focusing prevents device response from being tied to fundamental constants. And more critically, the concentrated field lines of the flux concentrator are often accompanied by large field gradients, which are likely to hinder use of long intrinsic dephasing times.

Better readout fidelity would also improve sensitivity. For example, ancilla-assisted repetitive readout has demonstrated readout fidelity approaching one for single NVs~\cite{jiang2009repetitive,neumann2010single,Lovchinsky2016nuclear} and has recently been implemented with NV ensembles~\cite{arunkumar2022quantum}. However, this readout method requires high bias magnetic fields, and the associated larger gradient may also make leveraging long dephasing times difficult. Cavity-enhanced MW-only readout methods \cite{Eisenach2021cavity,ebel2021dispersive} may offer another avenue to improve readout fidelity.

Another worthwhile avenue might focus on bettering the diamond material. As the diamond used in this work is not irradiated, it is likely its brightness could be improved. The diamond here offers a measurement contrast $C = 0.0334$ whereas values as high as 0.06~\cite{edmonds2021characterization} have been achieved.

\subsection{Extension to other sensing protocols}

Advances described in this work extend beyond Ramsey and Hahn echo magnetometry. For example, this apparatus has simultaneously demonstrated dramatically improved sensitivity for Rabi magnetometry~\cite{alsid2022solidstate} relative to existing work. The device is compatible with more complex pulse sequences, such as CPMG~\cite{BarGill2013solid,Pham2012enhancedsolid,Ryan2010robust}, XY~\cite{Maudsley1986modified,Gullion1990new} and others~\cite{Glenn2018high,Boss2017quantum,Schmitt2017submillihertz}. Both CPMG- and XY-based pulse sequences were briefly attempted but produced sensitivities inferior to Hahn echo, despite leveraging longer coherence times. The cause of the reduced performance was not fully investigated, but we presume that MW noise was a primary contributor. To access a far greater range of magnetic field frequencies, work is currently underway to implement the recently-developed quantum frequency mixing technique~\cite{guoqing2022sensing}.

\section{Acknowledgements}
The authors acknowledge Reginald Wilcox for designing the bias magnet array and Chuck Wurio for growing the diamond used in this work. This research was developed with funding from the Defense Advanced Research Projects Agency (DARPA) and the Under Secretary of Defense for Research and Engineering under Air Force Contract No. FA8702-15-D-0001. S.T.A. was supported by the National Science Foundation (NSF) through the NSF Graduate Research Fellowships Program. The views, opinions, and/or findings expressed are those of the authors and should not be interpreted as representing the official views or policies of the Department of Defense or the U.S. Government.

\bibliography{NVBibliography.bib}

\cleardoublepage

\section*{Supplemental Material for ``Sensitive AC and DC Magnetometry with Nitrogen-Vacancy Center Ensembles in Diamond''}

\begin{table*}[h]  
\centering
\caption{Partial list of symbols} %
\centering %
\begin{tabular}{l l c c } %
\hline\hline   
Name & Symbol & Approx. value & Units  \\
\hline   
Gyromagnetic ratio & $g_e$ & $\approx 2$ & unitless \\
Bohr magneton & $\mu_B$ & $9.274\times10^{-24}$ & J/T \\
Vacuum permeability & $\mu_0$ & $1.257\times 10^{-16}$ & H/m \\
Gyromagnetic ratio & $\gamma_e$ & $\approx 2\pi \times 28\times 10^9$ & rad$\cdot$s$^{\text{-}1}$T$^{\text{-}1}$ \\
Static magnetic field& $B_0$ & 2.23 & gauss\\
Signal integration capacitor & $C_\text{sig}$ & $6.6\times10^{-9}$ & Farads \\ 
Reference integration capacitor & $C_\text{ref}$ & $114\times10^{-9}$ & Farads \\ 
Average signal photocurrent & $\bar{I}_\text{sig}$ & $4.8\times10^{-3}$ & Amperes \\ 
Average reference photocurrent & $\bar{I}_\text{ref}$ & $82.9\times10^{-3}$ & Amperes \\ 
Read noise (rms) of digitizer & $\sigma_\text{dig}$ & $0.023\times 10^{-3}$ & Volts \\
Initialization time & $t_I$ & $35\times10^{-6}$ (Ramsey) & s \\
 & & $39\times10^{-6}$ (Hahn echo) & s \\
Free precession time & $\tau$ & $40\times10^{-6}$ (Ramsey) & s \\
 & & $100\times10^{-6}$ (Hahn echo) & s \\
Readout time & $t_R$ & $10^{-5}$ (Ramsey) & s \\
 &  & $10^{-5}$ (Hahn echo) & s \\
Additional dead time & $t_D$ & $6\times10^{-6}$ (Ramsey) & s \\
 & & $7\times10^{-6}$ (Hahn echo) & s \\
Overhead time & $t_O$ & $5.1\times10^{-5}$ (Ramsey) & s \\
 & & $5.6\times10^{-5}$ (Hahn echo) & s \\
Total sequence time & $\tau+t_O$ & $9.1\times 10^{-5}$ (Ramsey)& s\\
 &  & $1.56\times 10^{-4}$ (Hahn echo) &  s\\
Sequence rep. rate & $f_\text{rep}^\text{Ram}$ & $1.1\times10^4$ & Hz \\
 & $f_\text{rep}^\text{HE}$ & $6.4\times10^3$ & Hz \\

Relative initialization efficiency & $\kappa_I$ & 0.98 & unitless\\
\hline 
\end{tabular}\label{tab:varnames}
\end{table*}

\begin{figure*}[thp]
    \centering
    \includegraphics[height = 0.9\textheight]{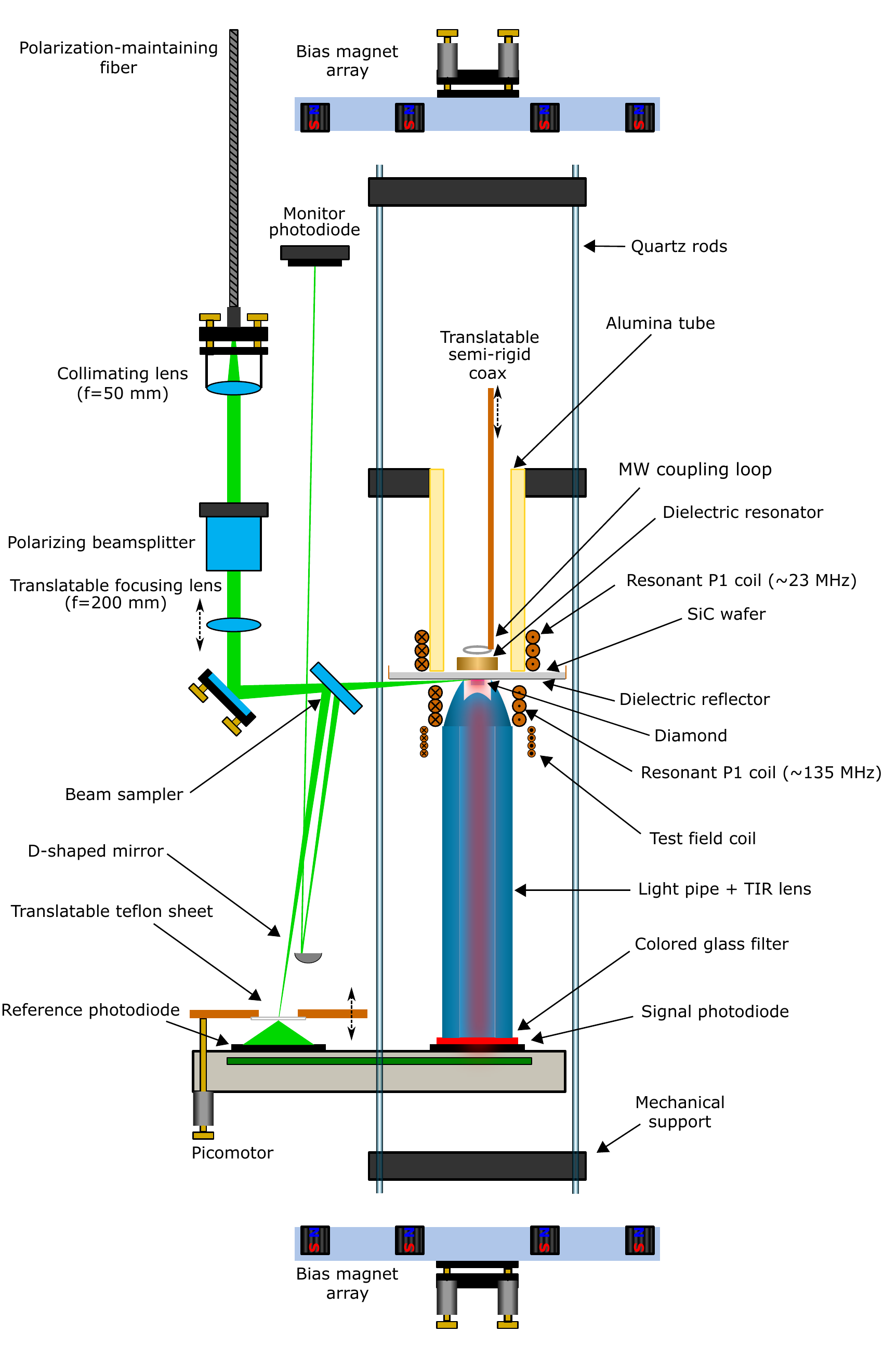}
    \caption{\textbf{Sensor head diagram.} Part numbers are omitted for clarity but can be found in the main text.}
    \label{fig:MasterSetupDiagram}
\end{figure*}

\section{Optical and mechanical design}

\subsection{Laser, acoustic optical modulator, fiber coupling, and light delivery to diamond}\label{sec:laserandAOM}

\begin{figure*}[!ptbh]
    \includegraphics[width= \textwidth]{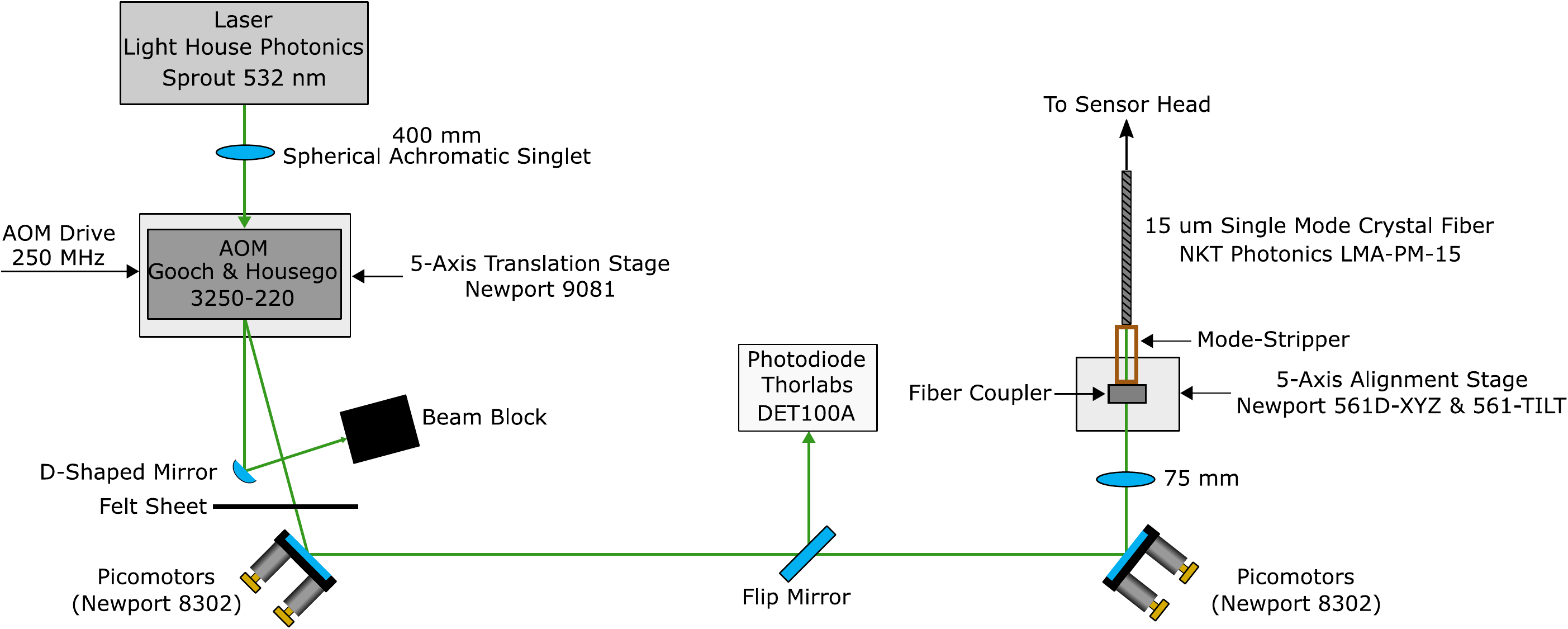}
     \caption{\small\textbf{Laser source optics.} Laser light is sent through an AOM, after which the first-order diffracted beam is aligned into a photonic-crystal fiber. The photonic-crystal fiber delivers light to the sensor head. For safety and to reduce the influence of air currents, these optics are enclosed during normal operation. Fiber alignment is remotely adjusted using two piezo-slip-stick-actuated mirrors.}
    \label{fig:LaserBox}
\end{figure*}

As discussed in the main text, Ramsey and Hahn echo sequences rely on pulsed optical excitation to both initialize and read out the NV$^\text{-}$ quantum states. The required pulsed light is created by gating the output of a continuous-wave laser with an acousto-optic modulator (AOM). The pulsed light is then delivered to the diamond through an optical fiber. The beam path of the continuous-wave laser through the AOM and into the fiber is shown in Figure~\ref{fig:LaserBox}.

In more detail, 12~W of 532 nm green laser light is generated (Lighthouse Photonics Sprout H) and focused into an AOM with a 400-mm-focal-length achromatic lens (Thorlabs AC254-400-A). The AOM (Gooch \& Housego 3250-220) is selected for its high diffraction efficiency \mbox{(up to $\approx$ 93$\%$) depending on focusing}, fast rise time (down to $\sim 10$ ns depending on focusing), and ability to handle high optical powers without damage (as the AOM crystal is quartz rather than tellurium dioxide). The AOM is bolted into a hollowed aluminum block affixed to a large passive heat sink to prevent the applied RF power from thermally damaging the AOM. Even with this heat sinking, duty cycles are limited to $\lesssim 70\%$ to avoid thermal damage. The hollowed aluminum block, AOM, and heatsink are mounted on a five-axis alignment stage (Newport 9081), which allows adjustment of the AOM position and angle relative to the incoming laser light. This adjustment allows the AOM diffraction efficiency, and therefore the amount of light delivered to the optical fiber, to be maximized. We typically realize a diffraction efficiency of 87$\%$ to 90$\%$ with the 400-mm-focal-length lens.

\begin{figure*}[!ptbh]
    \centering
    \includegraphics[width= .6 \textwidth]{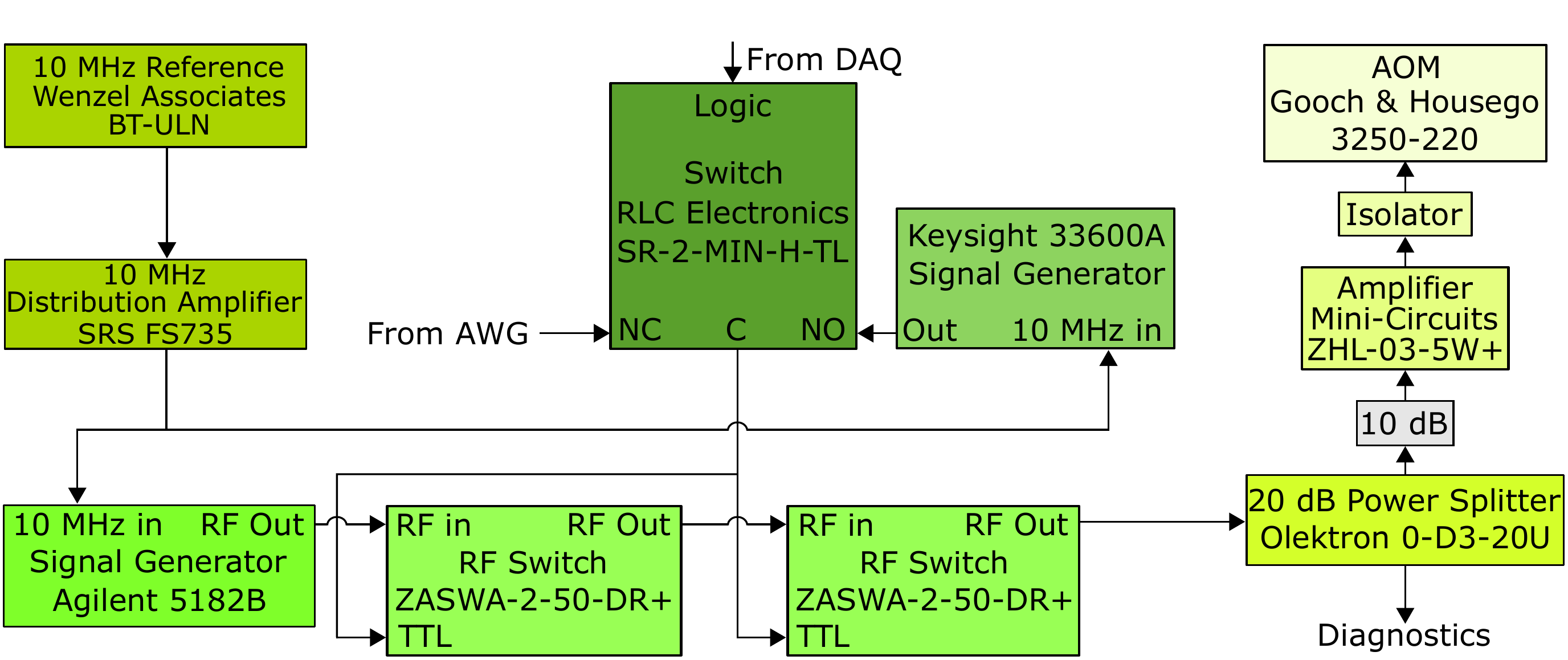}
    \caption{\textbf{AOM signal chain.}  A 250~MHz signal is gated by two RF switches before being amplified and sent to the AOM.}
    \label{fig:AOMElectronics}
\end{figure*}

The electronics chain driving the RF input of the AOM is shown in Figure~\ref{fig:AOMElectronics}. A signal generator (Agilent 5182B) synthesizes the initial 250~MHz RF signal. This RF signal is then gated by two switches in series (Mini-Circuits ZASWA-2-50-DR+), sampled by a 20~dB directional coupler for diagnostics (Olektron 0-D3-20U), attenuated by 10~dB, amplified (Mini-Circuits ZHL-03-5W+), and isolated before driving the AOM itself. 

The AOM's first-order diffracted beam is focused by a 75 mm best form lens (Thorlabs LBF254-075-A) into a single-mode, polarization-maintaining photonic-crystal fiber (NKT Photonics LMA-PM-15), mounted on a five-axis fiber alignment stage (Newport 561D-XYZ with Newport 561-TILT). Two mirrors prior to the 75 mm lens allow beam walking, which may be adjusted by stick-slip piezo actuators (Newport 8302). After the initial fiber alignment is set up, thereafter only the two mirror mounts are adjusted rather than the five-axis fiber alignment stage. 

During experiment sequences, the AOM gating is directly controlled by a digital channel on the arbitrary waveform generator (Tektronix 5014C). However, the arbitary waveform generator cannot output waveforms while a new sequence is being loaded into memory (e.g., in order to switch from precession time sweep mode to magnetometry mode). Therefore, during sequence loading, a computer-controlled switch (RLC Electronics SR-2-MIN-H-TL) temporarily reroutes the signal from a separate waveform generator (Keysight 33600A) to gate the AOM as shown in Figure~\ref{fig:AOMElectronics}. This waveform generator outputs a square wave with the same AOM gate duty cycle used in the experiment sequences. As a result, while a sequence is loaded into memory, the time-averaged RF power applied to the AOM and time-averaged the optical power applied to the diamond remain unchanged. As the AOM, fiber, and diamond exhibit thermal hystersis and lengthy response times to changing thermal loads, use of the substitute AOM gate signal improves the sensor's stability and avoids effects from thermal transients after changing sequences.


The photonic crystal fiber delivers the pulsed optical excitation light to the sensor head inside the magnetic shields (see Section~\ref{sec:staticfield}). At the sensor head an $f= 50$~mm lens (Thorlabs AC254-050-A) collimates the light exiting the photonic crystal fiber. The light then passes through a polarizing beam splitter oriented to transmit only vertical polarization, thereby transforming polarization noise into intensity noise. Converting polarization noise into intensity noise is prudent; while both noise sources induce changes in diamond fluorescence, the balancing circuit corrects for fluorescence variation from intensity noise but does not correct for fluorescence variation from polarization noise. After the polarizing beam splitter, the light is focused into the diamond by an $f = 200$~mm lens (Thorlabs LBF254-200-A). The excitation light is spatially aligned into the diamond using tilt-tilt controls on two non-magnetic mirror mounts (Radiant Dyes MDI-3-3025 non-magnetic), one controlling the photonic crystal fiber output and the other controlling a mirror after the $f=200$~mm focusing lens. The intensity of the excitation light inside the diamond may be varied by translating this $f=200$~mm lens along the beam path.

Immediately prior to entering the diamond, the pulsed optical excitation light is sampled; the reflection off the front facet of a beam sampler (Thorlabs BSF10-A) is sent to the reference photodiode (see Figure~\ref{fig:MasterSetupDiagram}) while the reflection off the beam sampler's back facet is picked off with a D-shaped mirror (Thorlabs PFD05-03-P01) and monitored on a diagnostic photodiode. The diagnostic photodiode allows continuous monitoring of the light delivered to the sensor head and therefore of the coupling efficiency through the photonic crystal fiber. If the diagnostic photodiode signal drops below a certain threshold, the coupling into the photonic crystal fiber is re-optimized using remote control of the two alignment mirrors prior to the photonic crystal fiber (see Figure~\ref{fig:LaserBox}). Ensuring maximal coupling efficiency into the photonic crystal fiber mitigates multiple difficulties; not only does the light then exert a consistent thermal load on the sensor head, but alignment-related intensity noise on the pulsed optical excitation is also minimized.

\subsection{Additional diamond details}\label{sec:diamondappendix}

\begin{figure}[t]
\centering
\includegraphics[width=80mm]{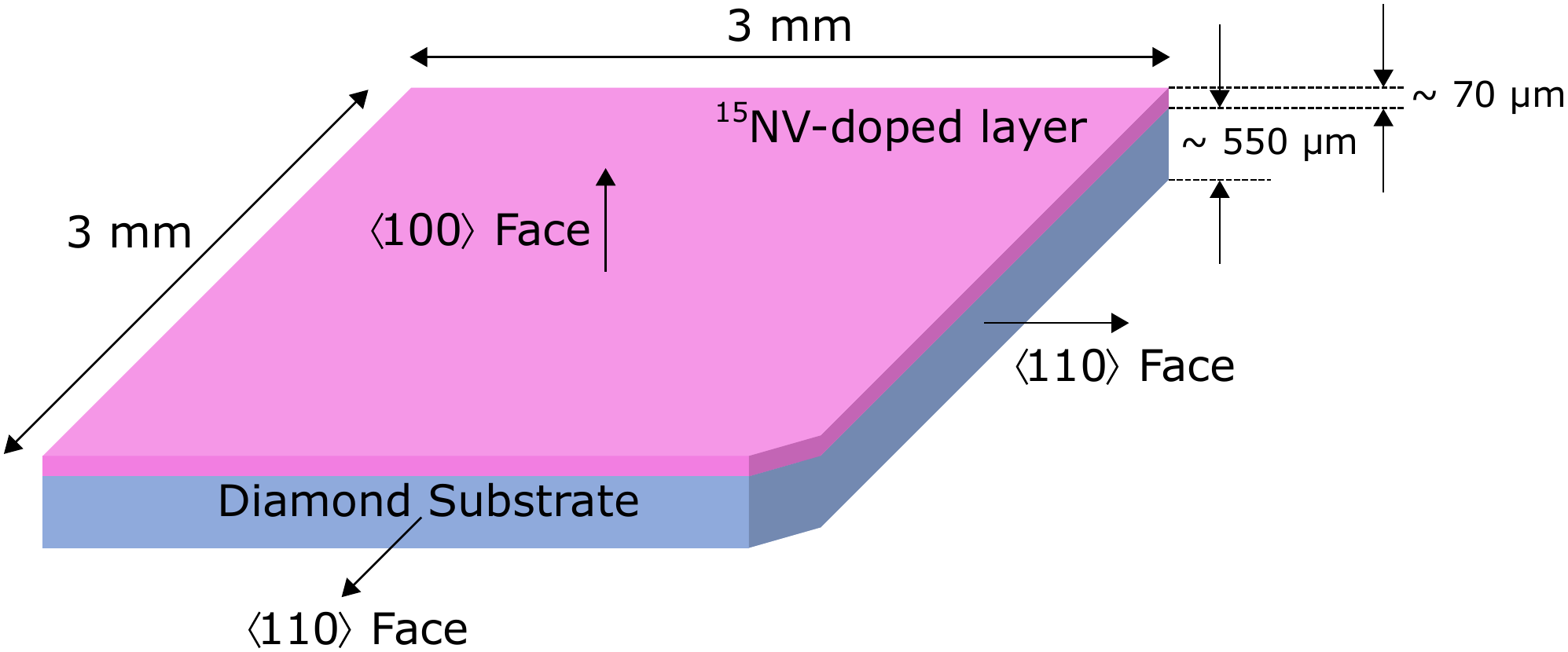}
\caption{Diamond geometry used in this work. The $^{15}\text{NV}^\text{-}$ layer is approximately 70~$\mu$m thick and is grown atop an approximately 550~$\mu$m thick substrate. One corner is faceted to allow use of the diamond light trapping waveguide technique.}\label{fig:diamondface}
\end{figure}

The diamond crystal is a $3\times3\times0.62$ mm$^3$ chip with $\langle 110 \rangle$ sides and a $\{100\}$ front facet. To enable the light-trapping diamond waveguide technique~\cite{Clevenson2015broadband}, a 250 $\upmu$m wide notch is faceted on an edge at 45 degrees relative to the adjacent sides as shown in Fig.~\ref{fig:diamondface}. The diamond is not irradiated; consequently, the N-to-NV$^\text{\text{tot}}$ conversion efficiency is low, and estimated to be $\sim\!10^{-2}$. With low-power 532 nm excitation, we estimate \mbox{[NV$^\text{-}$]/[NV$^\text{T}$]$\sim \frac{2}{3}$} using the setup and method outlined in Ref.~\cite{Alsid2019photoluminescence}.

\subsection{Choice of adhesive for diamond mounting}\label{sec:adhesion}

The type of adhesive adhering the diamond to the SiC heatsink was found, under some conditions, to deleteriously affect the linewidth of the magnetic resonances. We speculate that the broadening observed with several adhesives occurs because the differential coefficients of thermal expansion among the diamond, SiC, and adhesive produce substantial stress on the diamond upon heating with the initialization/readout laser. In addition, expansion or contraction of the adhesive during curing may produce stresses even in the absence of temperature changes.  Regardless of the precise mechanism producing stress, we hypothesized that an adhesive with a low Young's modulus would be least likely to induce broadening. In addition, the ideal adhesive should withstand temperatures up to about 100$^{\circ}$C, be transparent to visible light, exhibit high thermal conductivity and low dielectric loss (to not absorb microwaves applied by the dielectric resonator), and be removable without extraordinary effort (e.g., removal via sonication in acetone is preferable to removal requiring immersion in heated sulfuric acid).

To evaluate the strain imparted by the adhesive, resonance linewidths were first measured with the diamond adhered to the SiC wafer with double-sided tape and interrogated with pulsed ODMR~\cite{Dreau2011avoiding}; this approach produced consistent, narrow linewidths which are assumed to not be broadened. These linewidths were then compared to the linewidths achieved with adhesive mounting to the SiC wafer. Several adhesives were tested; adhesives which resulted in notable ($\gtrsim 20$ kHz) line broadening were Loctite 454 and Loctite 495 (cyanoacrylates), Devcon 20845 5-Minute epoxy, and Clear Gorilla Glue (polyurethane). We were unable to achieve consistently narrow magnetic resonances with any of these adhesives. Broadening was observed regardless of whether the diamond was mounted on SiC or sapphire and regardless of whether the substrate was coated with the reflective dielectric mirror coating.

Ultimately, the diamond was adhered to the SiC wafer with Sylgard 184 (polydimethylsiloxane, also known as PDMS), which resulted in no excess broadening. PDMS was chosen due to its unusually low Young's modulus ($\sim 2$ MPa) and because PDMS satisfies most of the criteria listed above for the ideal adhesive.

\subsection{Light trapping diamond waveguide details}\label{App:LTDWDetails}

The light-trapping diamond waveguide technique is expected to be sensitive to the polarization of the incoming light relative to the plane of the diamond~\cite{Clevenson2015broadband}, and we find that the amount of fluorescence produced depends strongly on the angle of the laser propagation axis relative to the diamond input facet as well. Though light is delivered to the sensor head via a polarization-maintaining fiber, a subsequent polarizer ensures consistent laser polarization is applied to the diamond. In addition, the custom mount for the diamond and SiC assembly includes the ability to reproducibly make small changes in the angle of the diamond facet relative to the incoming laser light. The mount is composed of several parts, including a cylindrical insert on which the diamond/SiC assembly is clamped.  The insert includes lever arms, and the stationary part of the mount has push screws impinging on either side of the lever arms. Adjusting these screws allows rotation of the insert,and hence diamond/SiC, while maintaining otherwise approximately fixed positioning of the diamond.

The light trapping diamond waveguide technique does however introduce additional considerations to the magnetometer design. Because the technique excites NV centers throughout the NV layer, the bias magnetic field and the MW magnetic field must be sufficiently uniform, and the MW and P1 fields must be sufficiently strong, over this entire volume.

\subsection{Light collection details}\label{sec:smlight}

The TIR lens is joined to the 125 mm long hexagonal BK7 light pipe (Edmund Optics $\#84-535$) with optical epoxy (Epotek 302-3M). This light pipe is followed by a 3 mm thick Schott RG645 colored glass filter (Edmund Optics $ \# $66-097) and then a 28 mm $\times$ 28 mm photodiode (Hamamatsu S3584-09). No index-matching oil or adhesive is used between the light pipe-filter or filter-photodiode interfaces.

Two TIR lenses were employed: a plastic TIR lens (Carclo Optics, $ \# 10193 $) and a custom borosiliate TIR lens of similar geometry to the plastic TIR lens.  While all collection efficiency measurements were taken with the plastic TIR lens, the borosilicate TIR lens was used for all magnetometry measurements. The two TIR lenses exhibited similar light collection efficiency, but the borosilicate lens does not burn or melt when inadvertently exposed to focused laser light. A small ($\sim$ 1 mm$^2$) triangular notch in the TIR lens allows the 532 nm excitation light to reach the diamond.

One non-negligible loss mechanism is expected to arise from imperfect reflection of p-polarized light by the dielectric reflector stack at certain angles for NV emission wavelengths $\gtrsim 700$ nm~\cite{thorlabs2021private}. We estimate this mechanism causes loss of between $1\%$ and $10\%$ of NV$^\text{-}$ fluorescence. This loss could be reduced or eliminated in the future by applying a custom coating to the SiC wafer rather than the employed stock broadband reflective coating (Thorlabs E02). Absorption of fluorescence light by nitrogen or NVs in the diamond is expected to be quite small, order 1$\%$ or less~\cite{Fraczek2017laser,barry2020sensitivity}.

This light collection scheme does not required altering the diamond geometry from the native diamond plate cuboid shape, a possible advantage not exhibited by other enhanced light collection schemes~\cite{Wolf2015subpicotesla}. However, ray-tracing-based modelling (Zemax) of the collection system indicates that for a cuboid-shaped diamond, closed paths within the diamond volume exist where the light cannot leave the diamond. This hypothesis was tested by roughening one of the large faces of a 3 mm $\times$ 3 mm $\times$ 0.5 mm diamond. We found the roughening of the surface with diamond grit sandpaper (Thorlabs LF30D) increased collected light by 30$\%$. Although the roughening technique was not implemented for any data collected in this paper, the approach may be valuable in future iterations of this device. 

A bright, brilliant-cut diamond was also employed to evaluate the system's light collection efficiency. A calibrated amount of green power is applied to the brilliant-cut diamond, and the amount of red fluorescence collected is measured. Applying a 532 nm excitation power corresponding to 116 mA of direct photocurrent results in 28.6 mA of photocurrent due to collected red fluorescence light, where these photocurrents have been scaled by the quantum efficiency (which differs between 532~nm and $\sim 700$~nm) so that one unit of photocurrent corresponds to the same flux of photons for each wavelength. Correcting for the 17$\%$ loss at the input facet of the diamond, this corresponds to a green-input-to-red-collected-photon conversion efficiency of 0.30. The diamond and NV system both exhibit a number of loss mechanisms (absorption of 532 nm by nitrogen, closed light paths that do not exit the diamond, absorption and subsequent fluorescence by NV$^0$ defects at filtered wavelengths, and non-unity radiative efficiency for the NV system), so that this number cannot directly serve as an estimate of $\eta_\text{geo}$. However, the near-order-unity conversion of input photons into collected fluorescence photons with the given test diamond provides a notable demonstration of the unusually high efficiency of this collection system.

While the geometric collection efficiency is high for this system, other inefficiencies in the optical readout exist. Quantum efficiency of the photodiodes is 95$\%$ at 700 nm and above 90$\%$ over the range 600-825 nm. As detailed above, not all emitted light exits \textcolor{mhs}{the} cuboid-shaped diamond. Finally, the chosen dielectric reflective coating may be slightly sub-optimal for p-polarized NV$^\text{-}$ emission, with a portion of the angular distribution of light of this polarization escaping through the dielectric coating.

\subsection{Additional details on balancing circuit}\label{App:BalancingCircuit}

As discussed in the main text, shot-noise-limited readout of the NV fluorescence photocurrent is complicated by both laser intensity noise and limited digitizer dynamic range.

A digitizer's dynamic range is defined as the ratio of the rms full-scale range to the rms additive noise imparted during digitization. Ideally this additive noise should be smaller than shot noise on the signal photocurrent, so that the magnetometer's signal-to-noise (SNR) is not degraded. At the same time, the digitizer needs to digitize the full-scale range of the photocurrent pulse. Combining these two criteria sets the dynamic range requirement for the digitizer. Unfortunately, the expected photocurrent ($\sim$ 10 mA) results in a dynamic range requirement equal to or beyond that available with current digitizers. In CW-ODMR experiments such as those in Refs.~\cite{Barry2016optical,Schloss2018Simultaneous} this difficulty is circumvented by AC-coupling the photodiode signal, so that the digitizer range instead only needs to cover the contrast on the photocurrent, a few percent of the DC value, rather than the full photocurrent. For a pulsed Ramsey magnetometer however, this workaround is not obviously applicable and some other method is required.

Moreover, fractional intensity fluctuations on the 532 nm excitation light typically translate to similar fractional fluctuations on NV fluorescence emission. At kHz-scale frequencies and below, these fractional intensity fluctuations are often many orders of magnitude above shot noise. For example, the noise on a multi-watt high-quality green laser (e.g. Coherent Verdi V or Lighthouse Photonics Sprout, the one used in this work) is specified to total about $0.02\%$ integrated from 10 Hz to 1 GHz. Empirically, this laser intensity noise results in noise on the collected NV photocurrent in the vicinity of 30$\times$-100$\times$ above shot noise. Without mitigation, this increased noise will translate directly to worse sensitivity by the same factor. As described in the main text, our device employs a balancing circuit where the fluorescence photocurrent light is referenced to the green excitation light. This approach, which largely resolves both problems outlined above, will now be described in detail.

During state readout, NV fluorescence light is directed to a photodiode, termed the signal photodiode, which converts incoming light into photocurrent during the $t_\text{R} = 10$~$\upmu$s readout duration. This signal photocurrent, with mean value $\bar{I}_\text{sig}$, is integrated on a capacitor with value $\mathcal{C}_\text{sig} = 6.6$~nF. Simultaneously, the 532~nm excitation light is sampled and directed to an identical photodiode, termed the reference photodiode, which produces mean photocurrent $\bar{I}_\text{ref}$ and  is integrated on a capacitor of size $\mathcal{C}_\text{ref}=114$~nF. To decrease the noise contribution from this referencing step, the reference light is made larger than the signal light and hence requires a larger capacitor, as detailed in Section \ref{sec:shotnoise}. After the readout light pulse ends, the voltages on the signal and reference capacitors are proportional to the integrated photocurrent seen by each photodiode during the readout period. The voltages on the signal and reference capacitors are then differenced and amplified by an instrumentation amplifier. After digitization, an analog switch (Vishay Si1900DL) controlled by a TTL signal grounds both integration capacitors until just before the next readout laser pulse, and the process repeats.

The instrumentation amplifier's output is passed to a 50~$\Omega$ distribution amplifier (SRS FS730), whose four output channels are input into a 4-channel digitizer (Gage Applied Razor Express CSE1632) and averaged in software. The digitizer's contribution to noise can be decreased by allowing a longer digitization period. We find that a 0.4~$\mu$s digitization period at a sampling rate of 200~MHz is sufficient to limit the digitizer's additive noise to $\approx 2.4\%$ of shot noise during operation, as discussed in Section~\ref{sec:shotnoise}. As detailed further in Section~\ref{sec:shotnoise}, the performance of the complete balancing circuit and digitization system is near-optimal, achieving a noise only $\approx 5\%$ larger than the shot noise limit on the NV$^\text{-}$ fluorescence photocurrent alone.

For troubleshooting purposes, a non-magnetic mechanical switch also allows both signal and reference photodiodes to be terminated into 50~$\Omega$ resistors included on the balancing circuit board. The voltages across both 50~$\Omega$ terminations are available via two SMA connectors. These signals can be connected to high-impedance inputs of an oscilloscope. This mode is useful to perform initial laser alignment, to check the signal and reference photocurrent values, and to give a coarse check of stability/noise. Photodiode rise time can also be evaluated but requires a low-impedance termination.  

\subsubsection{Balancing circuit electrical design considerations}

\begin{figure}
\centering
\includegraphics[width=80mm]{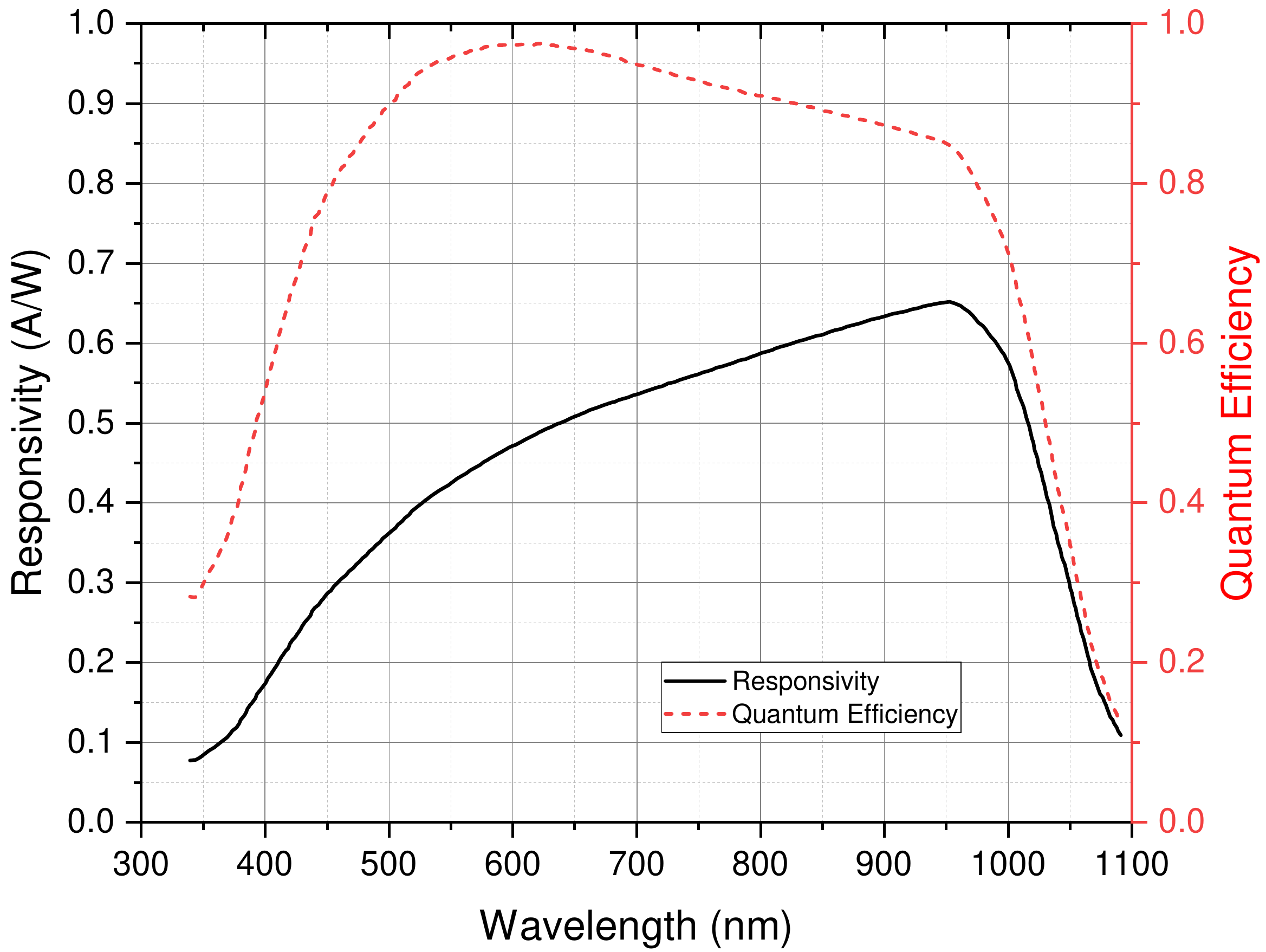}
\caption{\textbf{Photodiode quantum efficiency and responsivity.} Data are repoduced from figures provided by the manufacturer, Hamamatsu.}\label{fig:photodiodequantum}
\end{figure}

Further design considerations in the system warrant mentioning. First, large photodiodes (Hamamatsu S3584-09) with active areas of 28~mm $\times$ 28~mm are used. As the etendue of the collection system is a conserved quantity, large-area photodiodes allow light emitted over nearly all $4\pi$ solid angle from the diamond to be collected as discussed in Section~\ref{sec:lightcollection}. The chosen photodiodes are AR-coated for maximum efficiency near 600~nm, as shown in Fig.~\ref{fig:photodiodequantum}.  The manufacturer-specified quantum efficiency is $>$ 90$\%$ from 500~nm to 825~nm, which overlaps well with the NV$^\text{-}$ emission spectrum ($\sim$ 600 to 850~nm). In earlier prototypes, smaller photodiodes were used (Hamamatsu S3590-09) but were later replaced with the current photodiodes to allow increased light collection.

Unfortunately, large photodiode areas are necessarily accompanied by large capacitances, which can easily result in unacceptably-long rise times. Photodiode capacitance can be decreased by increasing the reverse bias voltage ~\cite{hobbs2011building}. The photodiodes are therefore reverse biased by +32 V, which results in a capacitance of about 330 pF, a 3 dB bandwidth of about 10 MHz, and an expected rise time of 35 ns. This rise time is fast enough for most desired diagnostics and is much faster than required for magnetometry operation, where all photocurrents are integrated during a 10~$\upmu$s readout period. Ideally, the NV signal rise time would be limited by the $^3$E excited-state lifetime of approximately 13~ns~\cite{doherty2013nitrogen}, rather than the photodiode capacitance in order to preserve information about the temporal shape of fluorescence from the diamond. If required, the balancing circuit could be replaced by a smaller-area and faster photodiode for such diagnostics, but this was not deemed necessary for the present work.

Noise on the reverse bias voltage for each photodiode could possibly result in noise on the collected photocurrent. The reverse bias voltage for each photodiode is filtered by two capacitance multipliers in series~\cite{hobbs2011building}, followed by three 33~$\upmu$F non-magnetic tantalum polymers capacitors (Kemet T543X336M050AHE040) in parallel and located in close proximity to each photodiode. The voltage filtering is extensive enough that the circuit continues to operate for several seconds after power to the photodiode bias has been cut.

As discussed in Section~\ref{App:BalancingCircuit} and in the main text, the balancing circuit employs an instrumentation amplifier architecture. This architecture ensures good common mode rejection and amplifies the difference in voltage on the signal and reference integration capacitors.  The op-amps used to make the instrumentation amplifier are Analog Devices, ADA4898-1, chosen because of their low $1/f$ noise corner of approximately 10~Hz and excellent dynamic range; the part exhibits an input voltage noise density of $e_n = 0.85 \;\text{nV}/\sqrt{\text{Hz}}$ and a common-mode voltage swing of up to $\pm$11.4~V. Each op-amp is protected from differential over-voltage of either polarity by two protection diodes (1N4148). In addition, both inputs to the instrumentation amplifier are protected from common-mode over-voltage by two transient voltage suppression (TVS) diodes connected in series (Littelfuse SLVU2.8HTG and Texas Instruments TPD1E05U06DPYT).

Given the resistor values in Figure \ref{fig:autobalanced_photodiode_diagram}, the instrumentation amplifier should exhibit a gain of 15.36. However, to impedance match the output to the 50 ohm coaxial line connected the SRS distribution amplifier, a 50 $\Omega$ resistor is inserted in series with the coaxial line, thereby decreasing the effective gain to 15.36/2 = 7.68.

Proper selection of the type of capacitors employed as the integration capacitors is critical for the circuit to perform as designed. The chosen integration capacitors are non-magnetic (determined empirically) film capacitors (Kemet R82MC2100DQ50K or similar). Film capacitors are used because they are less likely to exhibit acoustic electromechanical resonances or other non-ideal behavior compared to other commonly-used capacitor types. We note that many other capacitors were tried, including C0G, Y5V, Z5U, and X7R ceramic capacitors.  The Y5V, Z5U, X7R capacitors resulted in ringing, which was attributed to piezoelectric resonances of the capacitors. Problematic ringing was not observed with either film or C0G ceramic capacitors. Ultimately film capacitors were chosen over C0G capacitors since it was easier to procure a variety of values in an empirically non-magnetic package. 
The values of the integration capacitors are chosen for a given integration time and signal photocurrent so that the integrated charge on each capacitor results in a voltage in the vicinity of 8~V, slightly below the threshold for activating the TVS diodes protecting the op amps. The use of lower integrated signal voltages is not desirable because of the fixed additive read noise in both the instrumentation amplifier and the digitizer.

It is sometimes necessary to adjust the integrated voltage on either the signal or reference arm of the balancing circuit. In theory, best noise cancellation performance is expected when the two arms are exactly balanced. Additionally, in practice, if the light levels on the two arms are too far from producing balance, the balancing circuit will fail to operate; the instrumentation amplifier following the balancing circuit requires the voltage on each integration capacitor following the readout light pulse to be similar enough not to rail the amplifier output. 

There are two methods of obtaining balance between the arms. Coarse adjustment of the balance is performed by replacing the integration capacitors. Fine adjustment of the balance, as is necessary in day-to-day operation, is realized by translating a thin piece of Teflon mounted a variable distance in front of the reference photodiode. This translation is performed remotely using a non-magnetic actuator (Newport 8301-UHV-NM Picomotor). Several other approaches were tried but were found to be less satisfactory, including biasing one of the capacitors with a DC voltage and heating one of the capacitors. However, balance adjustment through capacitor voltage bias required use of high-$\kappa$ dielectric capacitors (X7R, Y5V), which produced ringing, presumably from piezoelectric resonances.  Adjustment via temperature was unacceptably slow and also required the use of similarly non-ideal capacitors. On the other hand, the mechanical balance adjustment was found to be straightforward to implement and troubleshoot.

\section{Bias magnetic field}
\subsection{Additional details on bias magnetic field design, construction, and alignment}\label{app:staticfielddetails}
The bias magnetic field $\mathbf{B_0}$ is created by two circular arrays of magnets positioned on opposite sides of the diamond's large $\{100\}$ faces. Each array consists of eight axially-magnetized cylindrical Sm$_2$Co$_{17}$ permanent magnets, each 12.7~mm in diameter and 12.7~mm in length. The eight magnets within each array are symmetrically spaced on the circumference of a 173.2-mm-diameter circular layout; magnets are oriented with their axial magnetization normal to the plane of the circular layout. The magnets within each array are held in the circumference of a monolithic wheel-and-spoke structure 3D-printed from a ceramic-plastic composite (Somos PerFORM) and clamped in place using nylon screws. Sm$_2$Co$_{17}$ magnets are chosen instead of conventional NdFeB magnets due to the lower temperature coefficient of the former, approximately $-0.03\%$/K. Both arrays are placed so that their centers each lie $\approx 590$~mm along the normal to the diamond's largest $\{100\}$ faces.

As discussed in the main text, bias field alignment is critical to maximize the magnetometer's performance due to two considerations. First, gradients across the diamond should be minimized, as such gradients decrease the observed dephasing time. Second, for schemes such as the one in this work, the bias magnetic field should project exactly equally onto all four NV orientations, so that transition resonances for different orientations are excited by the same MW frequency. To aid in satisfying these two criteria, the hub of each wheel-and-spoke structure is mounted on a non-magnetic mirror mount which allows tip, tilt, and axial translation (towards or away from the diamond) adjustment of the circular magnet array. 

The arrays are tuned as follows: First, a single array is placed approximately 295 mm from the diamond, and the array orientation is manually adjusted while monitoring the magnetic resonance spectra using pulsed ODMR~\cite{Dreau2011avoiding}. Once the NV orientations appear degenerate and the linewidth of the overlapped transitions has been roughly minimized, the alignment is fine-tuned to further reduce the observed linewidth using remote control of piezo slip-stick actuators (Newport 8301-UHV-NM Picomotor) on the non-magnet mirror mounts. The second array is then added to the apparatus, and the same process is repeated. Finally, the magnetometer is configured in double-quantum Ramsey precession time sweep mode to measure the free induction decay, and the piezo slip-stick actuators are iteratively adjusted to maximize the observed $T_2^*$.

In order to estimate $T_2^*$ in real time during alignment, the free induction decay measurement is mean-subtracted and Hilbert-transformed to produce a complex-valued signal. The magnitude of this complex signal yields the envelope of the FID. The envelope is linearized assuming $p=1$ and fit, providing an estimate of $T_2^*$.

\subsection{Bias magnetic field gradients}\label{sec:biasmagneticfieldgradients}

Spatial variation of the bias magnetic field applied to the diamond can broaden the NV$^\text{-}$ Zeeman resonances, thereby decreasing sensitivity for Ramsey magnetometry. To minimize the detrimental effects of magnetic field gradients on dephasing time, the gradient-induced linewidth should be small compared to the total linewidth. In other words, in the SQ basis where angular precession occurs at $\gamma_e$, the gradient $\Delta B$ over the ensemble should ideally satisfy
\begin{equation}
\gamma_e \Delta B \ll \frac{2}{T^*_{2,\text{SQ}}}.
\end{equation}
In the DQ basis the doubled precession rate $2\gamma_e$ must be accounted for, and gradients will be problematic unless
\begin{equation}
\gamma_e \Delta B \ll \frac{1}{T^*_\text{2,\text{DQ}}}.
\end{equation}
For a typical DQ dephasing time measured in our experiment, $T^*_{2,\text{DQ}}$ = 30 $\upmu$s, the magnetic field gradients  will decrease the measured dephasing time unless $\Delta B \ll 200$ nT over the volume of NV$^\text{-}$ centers used for sensing. %

\section{Microwaves}

\subsection{Additional details on microwave delivery and dielectric resonator}
\label{App:MWDelivery}

\begin{figure*}[t]
    \centering
    \includegraphics[width = \textwidth]{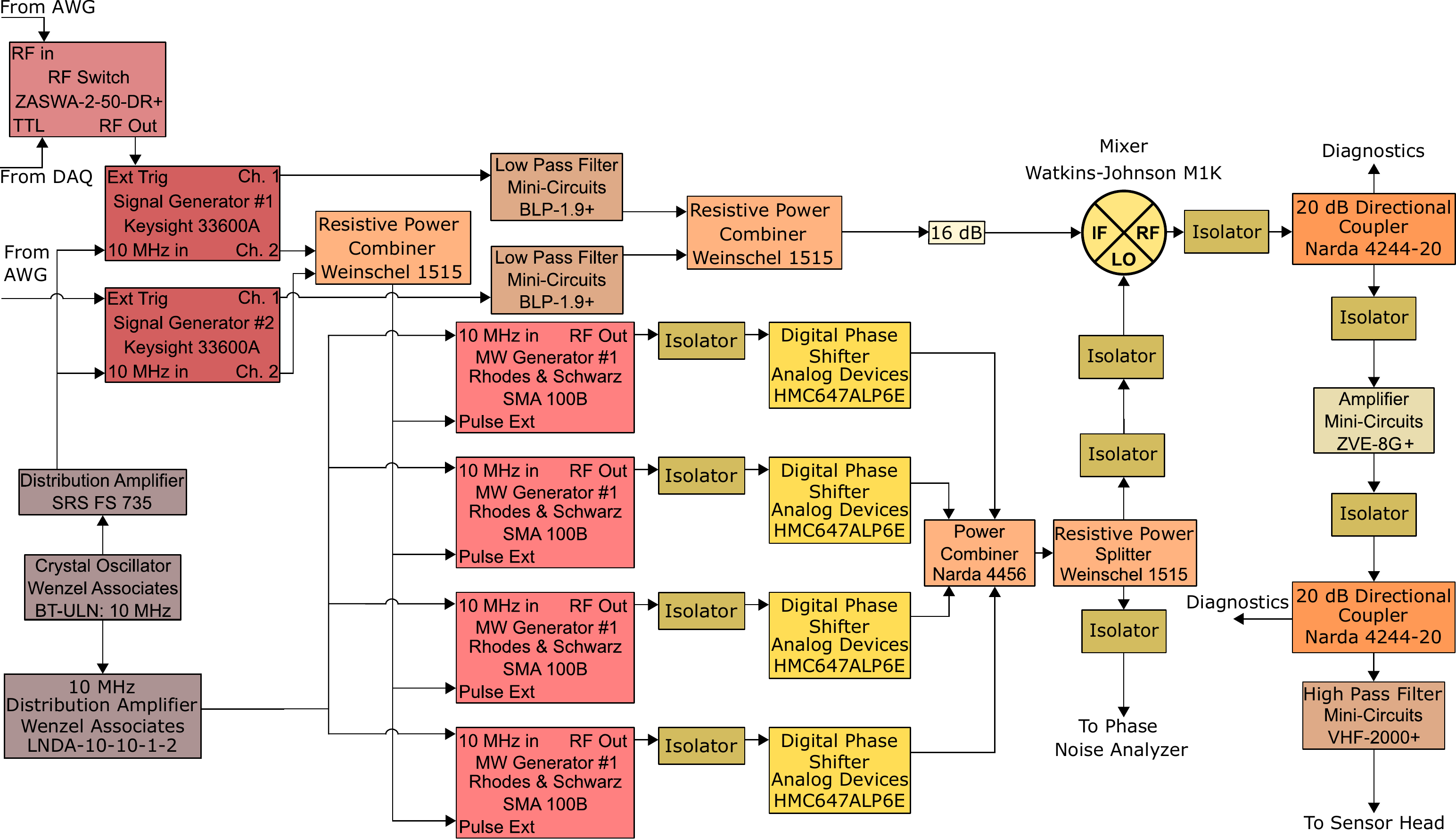}
    \caption{\small \textbf{Microwave signal generation and control.}  Generated MWs pass through an isolator and a digital phase shifter before being combined and sent to the LO port of a mixer. A separate Gaussian signal is generated, passed through a low-pass filter, and sent to the mixer's IF port, resulting in Gaussian-enveloped MW pulses. The resulting signal is amplified, high-pass filtered, and sent to the sensor head.}
    \label{fig:MWElectronics}
\end{figure*}

As discussed in the main text, Ramsey and Hahn echo pulse sequences~\cite{bauch2018ultralong,hart2021nv} rely on resonant MW pulses to manipulate the Zeeman sublevels of the NV$^\text{-}$ ensemble. This task can be broken down into four principle parts: generation of the MW signals, Gaussian envelope shaping of the MW signals, power amplification, and, finally, application of the MW field to the NV$^\text{-}$ ensemble. The full signal chain, including MW isolators, is shown in Figure~\ref{fig:MWElectronics}.

With the chosen bias magnetic field alignment (see Section~\ref{sec:staticfield}), four distinct MW frequencies are needed in the double quantum protocol to address all Zeeman transitions for both nuclear spin states. The four MW signals are generated separately (Rohde \& Schwarz SMA 100B) and then independently digitally phase-shifted (Analog Devices HMC647ALP6E; see~Appendix~\ref{App:DPS}) before being combined (Narda 4456). This implementation allows individual frequency, amplitude, and phase control of the four distinct MW signals. The aggregate MW signal is thereafter split (Weinschel 1515), with half the power either sent to a phase noise analyzer (Keysight E5052B or Rohde \& Schwarz FSWP8) for diagnostics or terminated, while the other half of the power is sent to the LO port of a mixer for pulse-shaping, discussed next.

To mitigate cross-excitation errors, Gaussian-enveloped MW pulses are employed rather than square-enveloped MW pulses~\cite{freeman1998shaped,vandersypen2005nmr,Fuchs2009gigahertz}. The two Gaussian envelopes required for a Ramsey sequence are generated using a dedicated channel on an arbitrary waveform generator (Keysight 33622A), which is triggered for each envelope at the appropriate time (Tektronix AWG5014C). For a Hahn echo sequence, an additional arbitrary waveform generator (Keysight 33622A) creates the Gaussian envelope for the echo pulse, and the length of the echo pulse is set to double that of the $\pi$ (Ramsey) pulses. All Gaussian envelopes are low-pass-filtered (Mini-Circuits BLP-1.9+), combined (Weinschel 1515), and attenuated by 16 dB before driving the IF port of a mixer (Watkins Johnson M1K). The aggregate MW signal described in the previous paragraph drives the mixer LO port, so that the desired Gaussian-enveloped MW pulses are generated at the mixer's RF port. The Gaussian-enveloped MW pulses have a full-width-half-maximum (FWHM) of 0.6~$\upmu$s for the Ramsey pulses and 1.2~$\upmu$s for the Hahn echo pulse, measured in voltage rather than power.

The signal exiting the mixer's RF port is amplified (Mini-Circuits ZVE-8G+) and high-pass filtered (Mini-Circuits VHF-2000+) before delivery to the sensor head through a low-loss coaxial cable (Carlisle UTiFLEX UFB311A). A pair of 20-dB directional couplers are placed immediately before and after the amplifier; the coupled ports of both directional couplers are sent to a 40 GS/s oscilloscope for diagnostics. Viewing the actual Gaussian-enveloped MW pulses in real time is helpful to diagnose synchronization, reference-clock, or other timing-related problems.

To reduce required MW power and improve $B_1$ uniformity over the diamond, the sensor head employs a cylindrical dielectric resonator in proximity to the diamond to generate the $B_1$ magnetic field. The dielectric resonator is 13.85 mm in diameter and 6.22 mm in height, with a relative dielectric of $\epsilon_r \approx 45$, and operates using the TE$_{01\delta}$ mode. The dielectric resonator is mounted on a non-magnetic, non-conductive translation mechanism and is pressed against the SiC wafer (which decreases the resonant frequency due to SiC's relative permittivity of $\epsilon_r \approx 10$) on the side opposite the diamond. As the dielectric resonator is non-conductive, it should not interfere with sub-MHz AC magnetic fields applied to the sensor. Microwave radiation is coupled into the dielectric resonator with a loop antenna fabricated from 0.141" semi-rigid non-magnetic coaxial cable (Tek-Stock UT-141C). The position of the coupling loop relative to the dielectric resonator is adjusted to achieve critical coupling using a vector network analyzer, and microwave absorbing foam is added near the dielectric resonator. This foam broadens the dielectric resonator's resonance to cover all necessary NV$^\text{-}$ resonances. A 56-mm-diameter ring of 60-micron-thick copper foil provides a partial shield for the dielectric resonator, without which the resonance could not be observed. Numerical modeling indicates that this foil has a negligible effect on the applied AC test magnetic fields.

\subsection{Double-quantum microwave pulse optimization}\label{App:DQPulseOptimization}
As described in the main text, the sensor uses a double-quantum (DQ) protocol, which offers performance advantages over a single-quantum (SQ) protocol. Here we detail the optimization procedure for MW pulses applied in the DQ protocol.

To ensure consistency among experiment repetitions, the MW frequencies $f_1,f_2,f_3,f_4$ are constrained to integer multiples of $f_\text{rep}$, the rate at which measurement sequences are executed. For instance, when Ramsey interferometry sequences are executed at $f_\text{rep}=5500$ Hz, all MW frequencies are constrained to an integer multiple of 5500 Hz. This constraint ensures the phase of all MW signals is the same sequence-to-sequence. When this constraint is ignored, strong spurious signals occur at frequencies $f_i$ mod $f_\text{rep}$, $f_i-f_j$ mod $f_\text{rep}$, as well as higher order products. As the lengths of the MW state preparation pulses are similar to the $1/[f_i-f_j]$ beat note period between MW carrier frequencies, the varying levels of constructive and destructive interference between these pulses might explain the presence of the spurious signals when $f_1,f_2,f_3,f_4$ are not constrained to integer multiples of $f_\text{rep}$.

For a Ramsey sequence the frequency and power of each of the four MW sources are set as follows. First, the device is loaded into a precession time sweep mode, and each of the four distinct frequencies is tuned to maximize the fringe amplitude in a SQ protocol free induction decay (FID). Here the power and frequency of each of the four MW signals are optimized independently. Next, the powers of all four MW sources are increased by 3 dB to achieve a $\pi$ pulse in the DQ protocol, denoted $\pi_\text{DQ}$. At this point, the generators corresponding to one $^{15}$NV nuclear spin state are switched off, and the FID fringe frequency is checked to ensure that the two-photon transition is as close as possible to resonance. This process is then repeated for the generators corresponding to the other  $^{15}$NV nuclear spin state, and all four generators are then switched on. We note that the FID fringes from the two pairs of DQ resonances may interfere constructively or destructively at a given precession time in the Ramsey sweep sequence. Because the applied MW frequencies are constrained to multiples of the experiment repetition rate, it is not possible to exactly match the two-photon detunings for the two pairs of DQ resonances. The phase shift applied to each generator during the final Ramsey pulse can be independently varied in order to ensure maximal constructive interference of the FID fringes at a given precession time. In addition, adjusting these four phases allows the slope of the FID signal to be maximized at a chosen precession time (i.e., to produce a zero-crossing in the FID curve at a given precession time).

The device is then loaded into the Ramsey magnetometry mode with the chosen precession time $\tau$, and a small test magnetic field is applied. The final-pulse phase shift applied to each generator is varied to observe maximum test field signal. However, the parameters of the four MW generators cannot be fully optimized independently of each other. To correct for amplifier saturation, off-resonant cross-excitation, and interference between MW frequencies, the power of each MW generator is again varied to maximize the test field signal size. Further optimization is achieved by iteratively adjusting each generator's frequency, power, and, if necessary, final-pulse phase. Joint changes in the frequencies of all four generators are used to account for small changes in the temperature of the diamond. Because the generators do not maintain consistent relative MW phase upon changes in frequency, even joint changes in the frequencies of the generators require adjustment of the applied powers to account for differing levels of constructive and destructive interference during the Ramsey pulses. However, additional changes to the final-pulse phase shift are generally required only when the generator frequencies are changed in a manner that alters the two-photon detuning for a DQ resonance pair.

For a Hahn echo sequence, the optimization proceeds as in Ramsey mode, with the echo pulse disabled. Just prior to switching from the precession time sweep mode to the magnetometry mode, the echo pulse is enabled. The echo pulse length is set to twice that of the Ramsey pulses \cite{Mamin2014multipulse}, and this pulse length is confirmed to provide the optimal observed decoherence time as expected.

\subsection{Digital phase shifters}\label{App:DPS}

Phase control of the applied MW signals is necessary for phase-modulation-based noise-subtraction schemes, observation of resonant FID oscillations, and sensor operation at the maximum interferometer fringe slope (see Sec.~\ref{sec:pulsesequenceandNRS}). However, the additive noise and non-linearities inherent to analog phase shifters were found to create unwanted microwave modulation sufficient to become the device's dominant noise contributor. Consequently, phase control in this work is implemented using digital phase shifters.

A 6-bit digital phase shifter (Analog Devices EV1HMC647ALP6) is placed on the output of each microwave generator as shown in Figure~\ref{fig:MWElectronics}. Each phase shifter is individually-controllable, and the desired phase shifts are loaded into the buffer of a USB digital output device, and trigger signals from the arbitrary waveform generator advance the output through the desired phases. This solution allows flexible, nearly instantaneous re-programming of phase shifts, for example to enable or disable the phase modulation that produces on-resonance free induction decay, or to switch between alternate noise subtraction schemes.

\textcolor{mhsnew}{Although we find the performance of digital phase shifters to be superior to that of analog implementations, we note two imperfections of these devices. First, the phase shifters produce a slight shift in amplitude along with the intended phase shift. Second, the devices have limited resolution, and the phase shifts applied are not perfectly independent of frequency. Therefore, the phase shift applied may vary slightly from the intended shift.}

\begin{figure*}[t]
    \centering
    \begin{overpic}[width=\textwidth]{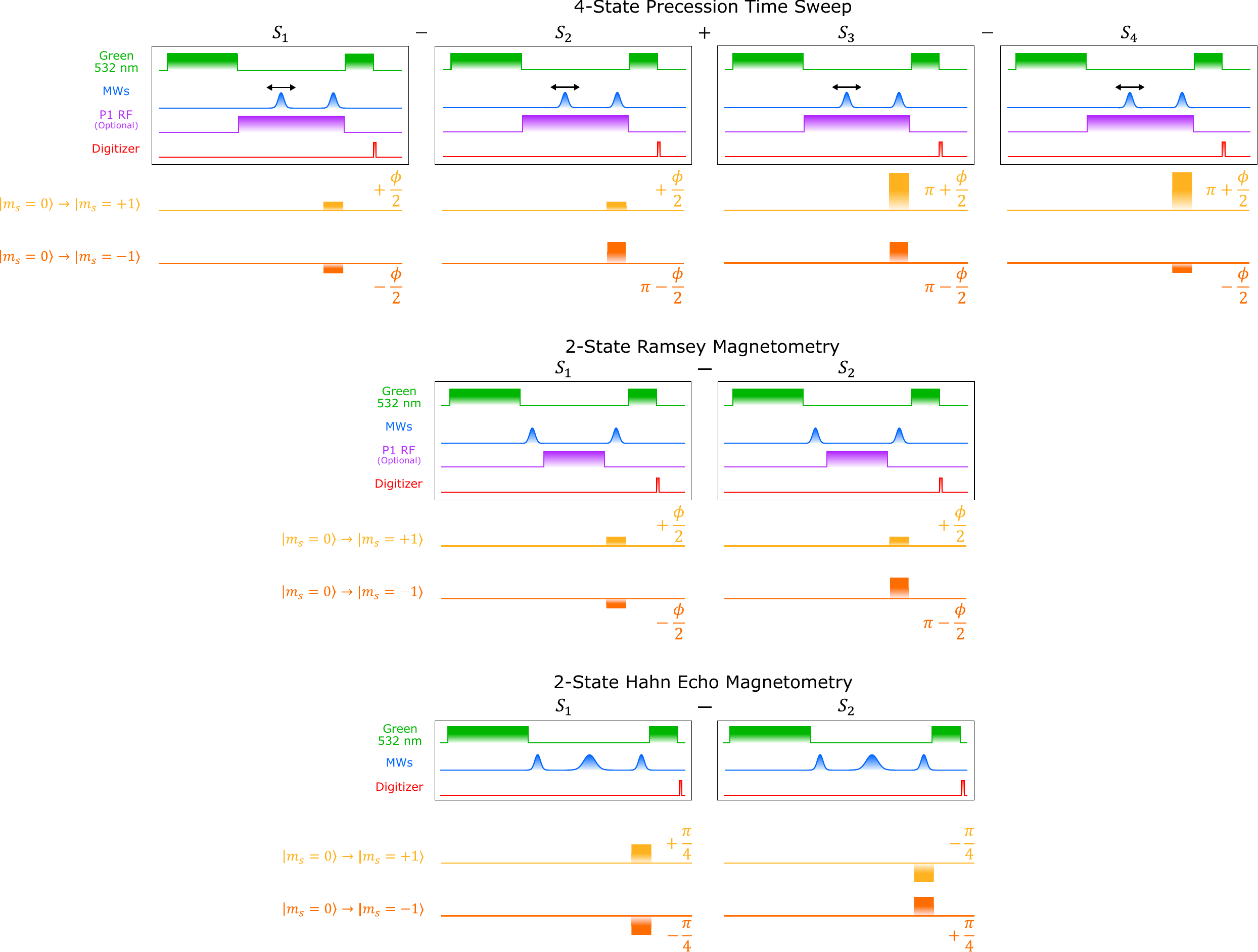}
    \put(0,74){\textbf{(a)}}
    \put(22,47){\textbf{(b)}}
    \put(22,21){\textbf{(c)}}
    
    \end{overpic}
    \caption{\small \textbf{Phases applied in magnetometry and FID pulse sequences.}   (a) In DQ measurements with 4-state noise subtraction, pulse sequences are completed with the phases shown, and signals are added and subtracted as shown to provide noise subtraction and rejection of imperfect DQ $\pi$ pulses. The phase shift $\phi$ varies with the precession time of a given sequence and allows observation of fringes even with MWs applied on resonance. While the panel depicts a Ramsey precession time sweep, addition of an echo pulse allows application to Hahn echo sweeps. (b) In DQ Ramsey magnetometry with 2-state noise subtraction, pulse sequences are completed with $\phi$ relative phase shift and then with $\phi-\pi$ relative phase shift between the MWs addressing the upper- and lower-frequency spin transitions; the signal from these two pulse sequences are subtracted. The phase shift $\phi$ allows operation at the maximum fringe slope, and the phase shift $\pi$ inverts the signal for noise subtraction. (c) In DQ Hahn echo magnetometry, a pulse sequence is completed with relative phase shift $\pi/2$, followed by a pulse sequence with relative phase shift $-\pi/2$. Because Hahn echo magnetometry is insensitive to static detunings, these phase shifts ensure operation at maximum fringe slope (for small test fields), and the signal of the second sequence is inverted to allow noise subtraction. For all panels, a non-zero phase shift is applied to only the final pulse in each sequence. Details of the magnetometry pulse sequence timings are shown in Fig.~\ref{fig:pulsesequences}.}
    \label{fig:DPSphases}
\end{figure*}

\section{P1 Driving}\label{Appsec:P1Driving}

\subsection{P1 resonance frequency calculation}
\label{ref:App:P1-Hamiltonian}
Near-resonant driving of P1 centers requires determining their transition frequencies under the static bias field $B_0$. These frequencies can be calculated from the eigenenergies of the spin Hamiltonian $H_{P1}$ for a substitutional $^{15}$N defect. Neglecting strain and the nuclear Zeeman interaction, the full spin Hamiltonian is: 
\begin{equation}
\frac{H_{P1}}{h} = \frac{\mu_B}{h} \mathbf{B}\cdot\mathbf{g}\cdot\mathbf{S} + \mathbf{S}\cdot \mathbf{A}\cdot\mathbf{I}, 
\end{equation}
where $\mu_B$ is the Bohr magneton, $h$ is Planck's constant, $\mathbf{B}$ is the magnetic field, $\mathbf{g}$ is the electronic g-factor tensor, $\mathbf{S}$ is the electronic spin vector $(S = 1/2)$, $\mathbf{A}$ is the hyperfine tensor, $\mathbf{I}$ is the nuclear spin vector $(I = 1/2)$~\cite{Bauch2018thesis}, and no electronic quadrupole effect is present since $I = 1/2$ for $^{15}$N.  The dynamic Jahn-Teller effect~\cite{Loubser1978electron,Ammerlaan1981reorientation} displaces the P1 defect along any of the $\langle 111 \rangle$ crystal axes, which leads to a trigonal symmetry. For a specific $[111]$ direction, the Hamiltonian can be simplified by transforming to the principle coordinate system, where $\mathbf{g} = \textrm{diag}\{g_{\perp},g_{\perp},g_\parallel\}$ and $\mathbf{A} = \textrm{diag}\{A_\perp,A_\perp, A_\parallel\}$. The axial and transverse hyperfine couplings are  $A_\perp = -113.83$ MHz and $ A_\parallel = -159.7$ MHz, respectively~\cite{Cox1994ENDOR}, and for simplicity the g-factor is assumed to be isotropic so that $g_\perp = g_\parallel = g$~\cite{Smith1959electron}.\par

With these simplifications, the Hamiltonian for each Jahn-Teller orientation under a static field $\mathbf{B}_0$ is
\begin{equation}
\frac{H_{P1}}{h} = \frac{g\mu_B}{h} \mathbf{B}_0\cdot\mathbf{S} + A_\parallel \mathbf{S}_z\cdot \mathbf{I}_z +A_\perp \left(\mathbf{S}_x\cdot \mathbf{I}_x + \mathbf{S}_y\cdot \mathbf{I}_y\right). 
\end{equation}
The P1 resonances are calculated as follows: using the values $g\mu_B/h = 2.8 $ MHz/G and $B_0 = 2.23$~gauss, the Hamiltonian is diagonalized, and the 12 possible transition frequencies are calculated from the differences between eigenenergies. To determine which transitions are dipole-allowed, a transition matrix element between each eigenstate is calculated using an electron spin operator representing the P1 drive field direction. We find that there are dipole-allowed transitions at approximately 22, 25, 135, and 139~MHz, and we target these frequencies for P1 driving.

\subsection{P1 coil design and implementation}\label{App:P1-Coil}
Spin-bath driving implementations in the literature ~\cite{DeLange2012controlling,Knowles2014observing,bauch2018ultralong} typically combine all RF drive signals onto a single broadband delivery structure such as a wire loop or coplanar waveguide. While this approach is sufficient for interrogating small volumes of NVs, the light-trapping diamond waveguide configuration (see Sec.~\ref{sec:LTDW}) requires a high P1 Rabi frequency over the entire NV-doped layer.  We therefore employ two multi-turn coils driven by  resonant  tank  circuits. The first coil addresses the low-frequency P1 resonances at $\omega^\text{P1}_1 \approx 2\pi \times 22$~MHz and $\omega^\text{P1}_2 \approx 2\pi \times 25$~MHz; the second coil addresses the higher-frequency resonances at $\omega^\text{P1}_3 \approx 2\pi \times 135$~MHz and $\omega^\text{P1}_4 \approx 2\pi \times 139$~MHz. The low-frequency P1 coil is a single-layer solenoid with close winding pitch, constructed using 6 turns of 2.05-mm diameter copper wire on a 33 mm diameter form, whereas the high-frequency P1 coil consists of two planar turns of 2.05-mm diameter wire with mean coil diameter $\approx\!21$ mm. Because up to 30 and 100~W may be dissipated in the low- and high-frequency coils respectively (at full drive power and 100\% duty cycle), the coils are each heat sunk to an aluminum nitride plate with thermal adhesive (Arctic Alumina). These plates in turn are clamped to separate mounting blocks (Thorlabs TS240) with thermal compound (Arctic Silver 5). The mounting blocks are affixed to a large aluminum breadboard beneath the sensor, providing a high thermal conductivity path to a large surface area for passive air cooling.

\begin{figure*}[t]
\begin{overpic}[width=15cm]{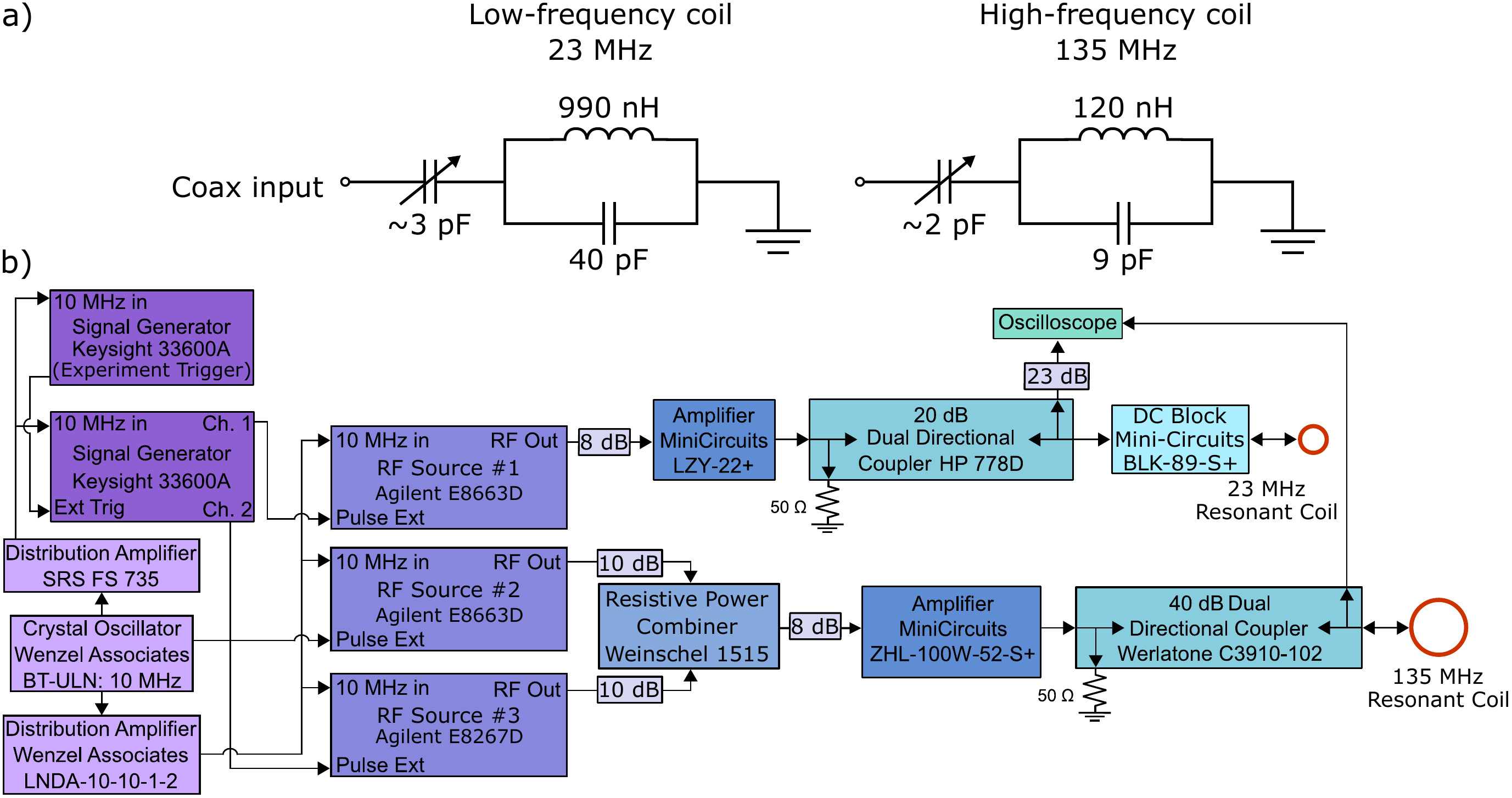}
\end{overpic}
\caption{\textbf{RF components and setup for P1 driving.} (a) Resonant tank circuit design for both the low-frequency and high-frequency P1 coils, as described in the main text. (b) Signal chain for the P1 driving.}\label{fig:P1Electronics}
\end{figure*}

Figure~\ref{fig:P1Electronics}(a) shows the circuit topology of the network used to match to the two coils. The low-frequency P1 coil is connected in parallel with four high-voltage $10$~pF capacitors, and these elements are connected in series with a home-made adjustable capacitor constructed from strips of copper shim stock covered in Kapton tape. The entire circuit is connected across the end of a semi-rigid coaxial cable. The bare coil inductance is measured using an HP 4192A LF impedance analyzer to be $990$~nH.  The high-frequency P1 coil is constructed similarly, with an \mbox{$\approx\!9$~pF} capacitor in parallel with the coil and a similar adjustable copper-shim-stock-and-kapton-tape capacitor in series; the coil inductance is 120~nH.

For both circuits, the resonant frequency $f_0$ and loaded quality factor $Q_L$ are determined by measuring  S$_{11}$ on a vector network analyzer (Agilent FieldFox N9923A).  Measured \textit{in situ} with the same near-critical coupling used during device operation, the low-frequency circuit resonance is centered at $\approx\!23$~MHz, with loaded quality factor $Q_\text{L}^\text{low} \approx\!60$ and FWHM $\approx\!0.4$~MHz, while the high-frequency circuit resonance is centered at $\approx\!135$~MHz, with $Q_\text{L}^\text{high} \approx\!20 $ and FWHM $\approx\!8$~MHz.  For a given $Q_L$ and resonant frequency $f_0$, the $1/e$ ring down time for the transverse field $B^{P1}_1$ exciting the P1 centers is given by $\tau_{\textrm{ring}} = Q_L/(\pi f_0)$~\cite{Eisenach2018broadband}. The calculated ring down times are $\tau_1 ~\sim 800$ ns and $\tau_2 \sim 50$ ns for the low-frequency and high-frequency resonant tank circuits, respectively.

\subsection{P1 drive delivery}
Three RF signal generators are used to drive the P1 spin bath, as shown in Fig.~\ref{fig:P1Electronics}(b). For the high-frequency P1 drive, two generators (Agilent E8663D and Agilent E8267D) create signals that are first attenuated by 10~dB and then combined (Wienschel 1515). The combined output is further attenuated by 8~dB and then amplified (Mini-Circuits ZHL-100W-52-S+) before passing through a 40~dB dual directional coupler (Werlatone C3910-102), with the forward-direction coupled port terminated into 50~$\Omega$.  The signal from the through port is capacitively coupled into the high-frequency resonant tank circuit, and any reflected signal travels back through the directional coupler. The backward-direction coupled port is sent into an oscilloscope; monitoring the reflected signal allows detection of any changes in the high-frequency tank circuit's resonance.

For the low-frequency P1 drive, the single signal generator's output (Agilent E8663D) is first attenuated by 8~dB, sent through a high-power amplifier (Mini-Circuits LZY-22+) and then to a 20~dB dual-directional coupler (HP 778D). The coupled port for the forward direction is terminated into 50~$\Omega$, while the signal from the through port passes through a DC block (Mini-Circuits BLK-89-S+) before being capacitively coupled into the low-frequency P1 coil resonant tank circuit. Any reflections return through the directional coupler, with the backward-direction coupled port directed through a 23~dB attenuator and into an oscilloscope.  Only one generator is used due to the limited bandwidth of the low-frequency P1 coil; the generator's frequency is set to 23.2~MHz.

All three RF generators are gated by a dual channel arbitary waveform generator (Keysight 33622A), with one channel gating the low-frequency P1 coil's generator and the other channel gating the two generators for the high-frequency P1 coil. In this experiment only continuous P1 driving is employed, but to reduce heat load and electronic noise, the P1 driving is applied only between the first and final MW pulses of both Ramsey and Hahn echo sequences. We find it critical to switch the P1 driving off well before the laser readout pulse and digitization; RF radiation from the P1 coils was found to greatly interfere with the photodiode balancing circuit, and the coils must be switched off well in advance of laser readout to account for the tank circuit ring-down time.

The drive frequencies applied to P1 coils are tuned by maximizing the $T_2^*$ value in real time during an FID measurement (see Sec.~\ref{app:staticfielddetails}). For variable precession time experiments, as used for FID diagnostics, the P1 drive is set to be on for a fixed length of time corresponding to the longest precession time in the experiment. Due to the substantial heat load on the P1 coils and their close proximity to the diamond, it is important that the diamond be allowed time to reach a steady-state temperature before final tuning of the applied MW pulse frequencies, as described in SM Sect.~\ref{App:DQPulseOptimization}.

\subsection{$T_2^*$ and $T_2$ extension details}\label{sec:fitparameters}

\begin{table*}[t]  
\centering
\centering %
\begin{tabular}{l l l l } %
\hline\hline   
Name & SQ & DQ & DQ+P1 driving  \\
\hline
$T_2^*$ with variable $p$ & 10.8(2); $p=0.72(2)$ & 7.5(6); $p = 0.72(2)$ & 24.7(9); $p = 0.90(2)$ \\
$T_2^*$  with fixed $p$ & 8.1(1); $p\equiv 1$ & 14.0(1); $p\equiv 1$ & 28.6(2); $p\equiv 1$ \\
\hline
$T_2$ with variable $p$ & Not measured & 142(3); $p = 1.14(3)$ & 320(20); $p = 1.02(7)$ \\
$T_2$  with fixed $p$ & Not measured & 136(3); $p\equiv 1$ & 324(16); $p\equiv 1$ \\
\hline 
\end{tabular}
\caption{\textbf{Ensemble dephasing and decoherence data.} All values are given in $\upmu$s.} %
\label{tab:fitparameterstable}
\end{table*}  

Extending the NV ensemble dephasing time $T_2^*$  allows for increased free precesssion time $\tau$ in a Ramsey sequence, which will improve sensitivity. The expected sensitivity improvement can be calculated as follows: First, for a given value of $T_2^*$, the value of $\tau$ which minimizes the following factor from Eqn.~\ref{eqn:ramseyshot},
\begin{equation}\label{eq:optimaltaufactor}
\frac{1}{\Delta m_s}\frac{1}{e^{-\tau/T_2^*}} \frac{\sqrt{\tau + t_\text{O}}}{\tau},
\end{equation}
is determined; here $t_\text{O} = t_\text{I}+t_\text{R}+t_\text{D}$ is the overhead time and is the sum of the initialization time $t_\text{I}$, the fluorescence readout time $t_\text{R}$, and any additional dead time $t_\text{D}$; and we have implicitly assumed $p=1$. 
Second, the optimal $\tau$ is plugged into Eqn.~\ref{eq:optimaltaufactor} along with the given value of $T_2^*$. Using this procedure for the $T_2^*$ values obtained using the SQ Ramsey, DQ Ramsey, and DQ Ramsey with P1 driving protocols, the sensitivity enhancements obtained using DQ and P1 driving with DQ can be evaluated. We find DQ Ramsey offers an expected 3.1$\times$ improvement over SQ Ramsey, while adding P1 driving to DQ Ramsey offers a 5.8$\times$ improvement over SQ Ramsey. 

The evaluation process is similar for Hahn echo magnetometry. First, the optimal $\tau$ is determined using Eqn.~\ref{eq:optimaltaufactor} but with $T_2$ substituted for $T_2^*$. Second, the optimal value of $\tau$ is plugged back into  Eqn.~\ref{eq:optimaltaufactor} for the given $T_2$ value, and the results are compared for DQ Hahn echo with and without P1 driving. Based on the extension of $T_2$ alone, the data suggest that adding P1 driving to DQ Hahn echo should give a 1.8$\times$ sensitivity enhancement; this calculation does not include additional techical noise sources introduced by P1 driving as discussed in Section~\ref{sec:T2extension}. 

For ease of comparison among the different protocols, $T_2^*$ and $T_2$ values are determined with the stretched exponential parameter fixed at $p=1$. This approach ensures that the $T_2^*$ or $T_2$ value alone characterizes the temporal dephasing or decoherence. More complex analysis may be possible if $p$ is allowed to vary; see Table~\ref{tab:fitparameterstable} for a list of fitted $T_2^*$ and $T_2$ values for both $p=1$ and variable $p$.

\section{Magnetometry}
\subsection{Readout fidelity optimization}
\label{appsec:readoutfidelityoptimization}
The distinguishability of the $|m_s\!=\!0\rangle$ and $|m_s\!=\!\pm1\rangle$ NV$^\text{-}$ electronic ground states is quantified by the readout fidelity $\mathcal{F}$~\cite{taylor2008high}. The readout fidelity satisfies $0\le \mathcal{F} \le 1$, with the upper bound representing projection-noise-limited spin-state readout~\cite{itano1993quantum} and the lower bound representing null information readout. This quantity directly affects the achievable sensitivity of NV magnetometers. For interferometry-based NV magnetometers, the readout fidelity is given by~\cite{taylor2008high,barry2020sensitivity}
\begin{equation}
\mathcal{F} =  \frac{1}{\sqrt{1+\frac{1}{C^2 n_\text{avg}}}},
\end{equation}
where $C$ is the measurement contrast~\cite{taylor2008high}, representing the asymmetry in photons collected for NVs maximally prepared into the $|m_s\!=\!0\rangle$ state versus the $|m_s\!=\!\pm 1\rangle$ state, and $n_\text{avg}$ is the average number of fluorescence photons detected per NV$^\text{-}$ in a single readout period. The measurement contrast $C$ is sometimes referred to as the fringe visibility.

Although $C$ is fundamentally limited by spin-dependent branching ratio asymmetries of the NV$^\text{-}$ $^3E$ excited state~\cite{Goldman2015state, Goldman2015phonon}, observed measurement contrast also depends on the initial polarization of the NV$^\text{-}$ ensemble into the $|m_s=0\rangle$ spin state, the (laser) readout duration $t_R$, and the fraction of fluorescence from unwanted NV$^0$ defects~\cite{barry2020sensitivity}.

We employ a scheme of four consecutive pulse sequences, referred to as $S_1,S_2,S_3$ and $ S_4$, to determine the measurement contrast $C$ and the efficiency of optical initialization. As shown in Figure~\ref{fig:pulsesequences}d, all four sequences consist of a $t_I = 35$ $\upmu$s optical initialization and a $t_R=10$ $\upmu$s optical readout. The fourth sequence $S_4$, however, introduces a MW $\pi_{SQ}$ pulse between the optical initialization and optical readout; the $\pi_{SQ}$ pulse transfers both hyperfine populations of the $|m_s=0\rangle$ state to the $|m_s\!=\!-1\rangle$ state. With this scheme, the signal from $S_4$ corresponds closely to fluorescence from the $|m_s = -1\rangle$ state while signals from $S_1$, $S_2$ and $S_3$ correspond predominantly to fluorescence from the $|m_s = 0\rangle$ state with increasingly efficient preparation.

Although the photodiode balancing circuit effectively AC-couples collected fluorescence signals by referencing the collected NV$^\text{-}$ fluorescence to the applied 532 nm light, the DC offset common to $S_1$, $S_2$, $S_3$, and $S_4$ can be determined by measuring the signal photocurrent along with the known gain of the balancing circuit. With this common-mode signal included, fluorescence measurements after $S_1, S_2,S_3$ and $S_4$ result in normalized signals of 0.99871, 0.99993, 1.00000, and 0.93416 respectively. The achieved measurement contrast $C$ is given by~\cite{taylor2008high}
\begin{equation}
C = \frac{S_1-S_4}{S_1+S_4} = 0.0334,
\end{equation}
which corresponds approximately to the fractional amplitude (i.e. approximately half the peak-to-peak value) of the interferometry fringes. 

As $S_2 \approx S_3$, we assume signal from $S_3$ corresponds to near-maximal initialization. The initialization efficiency  $\kappa_{I}$ of a single $t_I = 35$ $\upmu$s initialization is then crudely estimated as 
\begin{equation}
    \kappa_I \approx \frac{S_1-S_4}{S_3-S_4} = 0.980.
\end{equation}
We note that $\kappa_I$ represents the observed initialization efficiency relative to that which would be achieved with an infinitely long initialization pulse; this differs from the absolute polarization efficiency due to the continuous $T_1$ decay process.

As $\kappa_I$ is already quite close to one, initialization into the $|m_s = 0\rangle$ state would not be expected to improve substantially by increasing the optical initialization time beyond the 35 $\upmu$s used. On the other hand, as $t_I \lesssim \tau+t_R+t_D$, substantially shorter values of $t_I$ would result in only marginal decreases in the total length of a measurement sequence and offer little if any improvement in sensitivity. 

The duration of the fluorescence readout, $t_R$, influences both $n_\text{avg}$ and $C$. While $n_\text{avg}$ increases with readout duration, observed values of $C$ eventually decrease as NVs repolarize during the readout process. The optimal value of $t_R$ balances these two competing effects to maximize the readout fidelity. We observe optimum readout fidelity for $t_R\approx10$~$\upmu$s and employ $t_R=10~\upmu$s in this device.

\subsection{Test field calibration}\label{sec:calibration}
\begin{figure*}[t]
\centering
\includegraphics[width=\textwidth]{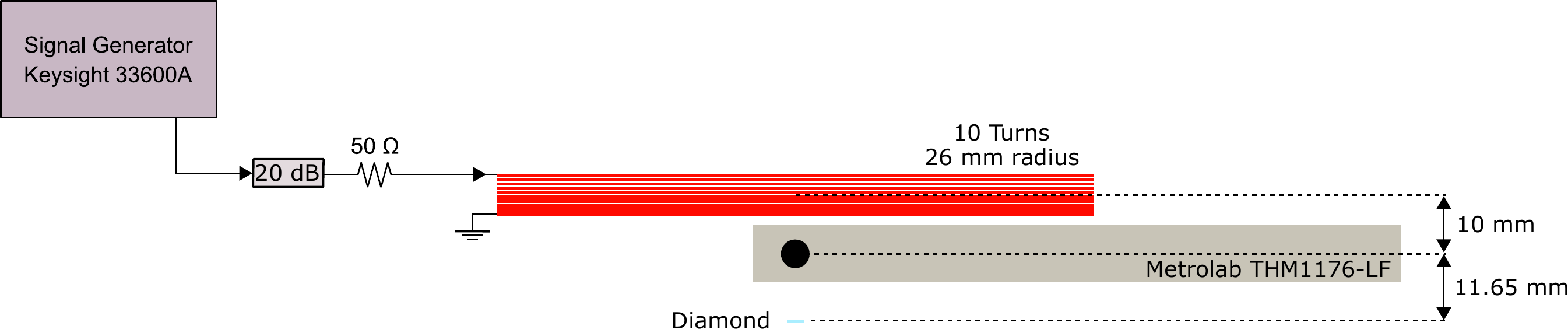}
\caption{\textbf{Test coil signal chain and geometry.} Test magnetic fields are applied using a 10-turn test coil placed a short distance from the diamond. The test field drive voltage is provided by a signal generator, whose output is connected to a series circuit consisting of a 20 dB attenuator, a 50 $\Omega$ resistor, and the test coil itself. A commercial magnetometer placed in the position shown provides one of the multiple methods to calibrate the test magnetic field; this magnetometer is removed during device operation.}\label{fig:testcoil}
\end{figure*}

The coil and electronics to create the test magnetic field are shown in Figure~\ref{fig:testcoil}. Careful calibration of the test field amplitude is important to accurately evaluate the magnetometer's sensitivity. The calibration factor $\kappa$, with units nT/V, characterizes the scaling between the signal generator's voltage and the resulting test magnetic field amplitude. The value of $\kappa$ is determined using four distinct methods.

The first calibration method determines the value of $\kappa$ from the known coil geometry and the applied current. The test coil has radius $r_t= 26$ mm, $N_t = 10$ turns, and is located $z_\text{coil} = 21.7$ mm from the diamond as shown in Fig.~\ref{fig:testcoil}. The test coil is connected in series with a 50~$\Omega$ resistor, and driven by a function generator whose output is first attenuated by 20 dB, reducing the voltage output by 10$\times$. The calculated rms magnetic field is 
\begin{equation}\label{eqn:fieldfromloop}
B_\text{rms}(z) = \frac{\mu_0 I_\text{rms} N_\text{t}}{2}\frac{r_t^2}{\left(r_t^2+z^2\right)^{3/2}},
\end{equation}
where $I_\text{rms} = \frac{V_\text{rms}}{50\Omega}\times \frac{1}{10}$, $V_\text{rms}$ is the rms voltage output of the generator, and $\mu_0$ is the vacuum permeability. With these parameters, Eqn.~\ref{eqn:fieldfromloop} gives $\kappa = 219$ nT/V.

For the second calibration method, the light pipe is removed, and a commercial magnetometer (Metrolab THM1176-LF) is placed near the diamond as shown in Figure~\ref{fig:testcoil}. The size of the commercial magnetometer restricts placement to approximately 10 mm from the test coil and 11.65 mm from the diamond. With the signal generator outputting a sinusoidal signal of 3~V rms at 10 Hz, the commercial magnetometer measures a 1165 nT rms magnetic field. Accounting for the offset between the commercial magnetometer and the diamond and using Eqn.~\ref{eqn:fieldfromloop}, we calculate $\kappa = 230$ nT/V.

The third calibration method leverages the fixed gyromagnetic ratio $\gamma_e$ of the NV$^\text{-}$ and is expected to be more accurate than the two methods described above. In a double quantum Ramsey sequence with free precession time $\tau$, the presence of a constant magnetic field $B$ leads to a phase accrual of
\begin{equation}
\label{eq:accumulatedphase}
\phi = 2\gamma_e B F_\text{pro}\tau.
\end{equation}
Here $\gamma_e = g_e \mu_B/\hbar$, and $F_\text{pro} = 1/\sqrt{3}$ is an angular correction to account for the alignment of the magnetic field, which is applied normal to the diamond's (100) front facet. For a given precession time $\tau$, if the magnetic field is varied to advance the interferometer phase by $2\pi$, i.e. one fringe, the change in field $\Delta B$ is
\begin{equation}\label{eqn:fringespacingfieldcalibration}
\Delta B = \frac{1}{F_\text{pro}} \frac{2\pi}{2 \gamma_e}\frac{1}{\tau}.
\end{equation}
For a Ramsey sequence with a precession time $\tau = 19.8$~$\upmu$s, $\Delta B = \sqrt{3} \times 901$~nT = 1561~nT. Dividing $\Delta B$ by the measured voltage required to advance one fringe, we find $\kappa=201$~nT/V.

Our final calibration method applies the previous technique to double-quantum Hahn echo magnetometry. If a square wave field with amplitude $B$ is present during the free-precession interval and switches sign at the echo pulse, then the accumulated phase is again given by Eqn.~\ref{eq:accumulatedphase}. For a Hahn echo sequence with $\tau = 100$~$\upmu$s, we calculate that a fringe period corresponds to $\Delta B = \sqrt{3} \times 178$~nT = 309~nT. Recording the Hahn echo magnetometry signal versus voltage, as shown in Fig.~\ref{fig:HahnEchoFit}, we find $\kappa = 198$~nT/V; see SM Section~\ref{App:HahnEchoSensitivity} for additional information on the field calibration in Hahn echo magnetometry.
This value is 1.4\% smaller than the value determined by using the Ramsey fringes. This difference is likely due to a combination of several effects that are non-negligible at the percent level. In particular, field attenuation due to metal near the diamond (e.g., P1 drive coils), reduced test coil current due to increased resistance at higher frequencies, and reduced test coil current due to increased inductance at higher frequencies may all contribute to smaller fields for a given voltage amplitude as frequency increases.

While these four distinct methods produce sensor calibrations that agree to within 15\%, we use the calibrations determined from fringe spacing to calculate sensitivity in Ramsey and Hahn echo magnetometry, due to the expectation that these techniques are the most accurate. Calculation of the sensitivity is further discussed in Sec.~\ref{App:CalcSensitivity}.

\section{Sensitivity measurement}
\label{App:CalcSensitivity}
\subsection{General considerations}
The magnetic field sensitivity is defined as
\begin{equation}\label{eqn:taylorsensitivity}
\eta \equiv \delta B \sqrt{T}
\end{equation}
where $\delta B$ is the minimum detectable field, defined as the field giving an SNR of 1 after a measurement duration $T$~\cite{taylor2008high,Lesage2012efficient}. For white Gaussian noise,  the sensitivity can be written equivalently~\cite{Schloss2018Simultaneous} as
\begin{equation}\label{eqn:svmsensitivity}
\eta \equiv \sigma_B\sqrt{T} = \frac{\sigma_B}{\sqrt{2\Delta f}},
\end{equation}
where $\sigma_B$ is the standard deviation of a series of measurements, each with measurement time $T$, and $\Delta f = 1/(2 T)$ is the measurement bandwidth on a single-sided spectrum, by convention.

For white noise, the sensitivity of the device defined by Eqn.~\ref{eqn:taylorsensitivity} is equal to the rms noise in a single 1 Hz bin of the double-sided amplitude spectral density (or the noise in a 0.5 Hz band of the single-sided amplitude spectral density, as implied by Eqn.~\ref{eqn:svmsensitivity}, except at DC). This definition of sensitivity is valid both for DC and AC fields, and implicitly assumes the phase of the signal is known. This definition, with its assumption of known phase, is chosen because it is consistent with theoretical calculations of sensitivity from the spin projection limit, with the prevailing Hahn echo literature where the phase of the AC signal is assumed known, and with the definition chosen in the initial seminal NV magnetometry paper~\cite{taylor2008high}. Unfortunately, this definition is inconsistent with the rms noise in a 1 Hz bandwidth on a single-sided amplitude spectral density, which definition would give a sensitivity $\sqrt{2} \times$ higher. For additional discussion see Ref.~\cite{Fescenko2020diamond}.

In general, in the presence of non-white noise sources, it is more appropriate to consider the noise within a band of interest in the frequency domain. As is common practice, we may Fourier transform the time series of measurements and write the resulting amplitude spectral density in rms magnetic field units by dividing the spectral density in the magnetometer's native voltage units by a calibration factor produced by applying a known (rms) $B$-field, assuming a uniform response across the spectrum (see Section \ref{sec:calibration}). This produces a spectrum of sensitivity as a function of frequency of interest. The minimum sensitivity of the device is then commonly reported as the amplitude spectral density of the device noise over the low-noise band or bands.

However, the usual practice of determining device sensitivity from the noise floor of the amplitude spectral density produces a subtle challenge that is seldom discussed in the magnetometry literature. According to Parseval's theorem, the integral over time of the square of the magnetic field is equal to the integral over frequency of the square of the field spectrum. Thus, the appropriate average of a frequency band of interest in the amplitude spectral density is an rms average (or, equivalently, the mean of the power spectral density in the band of interest), and \textit{not} the mean of the amplitude spectral density. Alternatively, to determine the sensitivity from the average of successive measurements, the rms average of the amplitude spectral density should be taken. In this case, we have
\begin{align}
\eta(f) = \left< B^d (f) \right>_\text{rms},
\end{align}
where $\eta(f)$ denotes the sensitivity amplitude spectral density, $B^d(f)$ is the double-sided magnetic field amplitude spectral density, and $<\cdot>_\text{rms}$ indicates an rms average. We note that we can also write
\begin{align}
\eta(f) = \frac{1}{\sqrt{2}}\left< B^s (f) \right>_\text{rms},
\end{align}
where $B^s(f)$ is the single-sided magnetic field amplitude spectral density, and the prefactor $\alpha_\text{s-d} = 1/\sqrt{2}$ is valid to convert from single-sided to double-sided spectral density except at DC.

Whether an average is taken across frequency bins of a single acquisition's spectrum or an average is taken across the spectra of a number of acquisitions, an rms average must be used to appropriately preserve the total noise power present in the data. In the latter case, once an rms average has been performed over spectra from a sufficiently large number of acquisitions, the mean or median of the appropriate band of the amplitude spectral density may be taken to correctly determine the minimum sensitivity of the device. We have verified in simulation that, for additive white Gaussian noise, rms-averaging of spectra from 5 acquisitions is sufficient to allow the median of amplitude spectral density to produce a minimum sensitivity correct at the $\sim 1$\%-level. \textcolor{mhsnew}{The sensitivity data presented in Fig. \ref{fig:ramseysensitivity} and Fig. \ref{fig:HahnEchoData} are produced from rms-averaging 10 acquisitions per plot.}

In the presence of sizable spurs at particular frequencies, use of the median is preferable to naive calculation of the rms mean of the noise floor, as the latter can be significantly influenced by even a small number of large spurious values. Though frequencies with spurs can be manually excluded or automatically excluded by filtering outliers above a particular value, we have found sensitivity calculation via the median to be more convenient and robust to variations in the experimental conditions, as is especially important during real-time optimization of the device.

Finally, though not relevant to the data presented in this work, we note that subtle complications arise when applying this median-based sensitivity calculation method to the amplitude spectral density of a single data acquisition. In particular, several correction factors must be applied in order to produce an accurate sensitivity using the median-based method with the spectrum of a single data acquisition. Assuming white Gaussian noise, the median amplitude spectral density must be multiplied by a factor
\begin{equation}
\alpha =  \left(\sqrt{\frac{4}{\pi}} \right) \left( \frac{1}{2} \sqrt{\frac{\pi}{\ln(2)}} \right) \approx 1.201,
\end{equation}
where the first factor accounts for the ratio between an rms mean and an ordinary mean, and the second factor accounts for the ratio between an ordinary mean and a median. These factors result from the statistical properties of time-domain (real) additive white Gaussian noise after Fourier transforming into (complex) frequency-domain signals. In particular, the factors may be viewed as resulting from properties of the Rayleigh distribution, which describes the magnitude of the complex frequency-domain noise.

\subsection{Ramsey magnetometry sensitivity measurement}

During Ramsey magnetometry, a 10-Hz calibrated test magnetic field is applied, with rms amplitude \mbox{$B_\text{rms} = \kappa \cdot 50 \text{ mV} = 10$~nT}. Each 1-s long acquisition is Fourier transformed, and the resulting double sided spectrum with 1 Hz bins is scaled so that the magnitude of the +10 Hz bin and -10 Hz bin are both $B_\text{rms} =  \frac{1}{\sqrt{2}} \times 10$~ nT. This scaling ensures that the magnitude in the sole 10 Hz bin on a single-sided amplitude spectral density is $B_\text{rms} = 10$~ nT.

The sensitivity spectra provided in this paper are calculated using an rms average of 10 spectra from successive 1-s long acquisitions. The minimum sensitivity is then determined by calculating the median sensitivity across the appropriate frequency band. Approximately two dozen 1-Hz wide notch filters, mostly at harmonics of 60 Hz, remove spurious signals, but this filtering has no effect on the sensitivity. For Ramsey magnetometry, the minimum sensitivity is calculated as the median over the final 10\% of frequency bins in the spectrum.

\subsection{Hahn echo sensitivity measurement}\label{App:HahnEchoSensitivity}

To calculate the Hahn echo sensitivity, we begin by verifying the calibration of magnetic field versus voltage applied to the test coil, as the presence of residual metal components may partially shield the sensor from the \textcolor{mhs}{6.4}~kHz magnetic field applied during Hahn echo testing. We first roughly optimize the DQ Ramsey microwave parameters in a precession time sweep mode as described in Sec. \ref{App:DQPulseOptimization}. We then enable the echo pulse and switch to the magnetometry mode. A square wave magnetic field is applied using the test coil, with the transition between high and low voltage occurring at the midpoint of the echo pulse. As described in the main text (see Section~\ref{sec:hahnechosensitivity}), phase shifts of $\pm \pi/2$ are applied to the final pulse on successive experiment repetitions. The difference between these two experimental repetitions is the Hahn echo magnetometry signal, and this cycle is repeated indefinitely.

\begin{figure}
\centering
\includegraphics[width=\columnwidth]{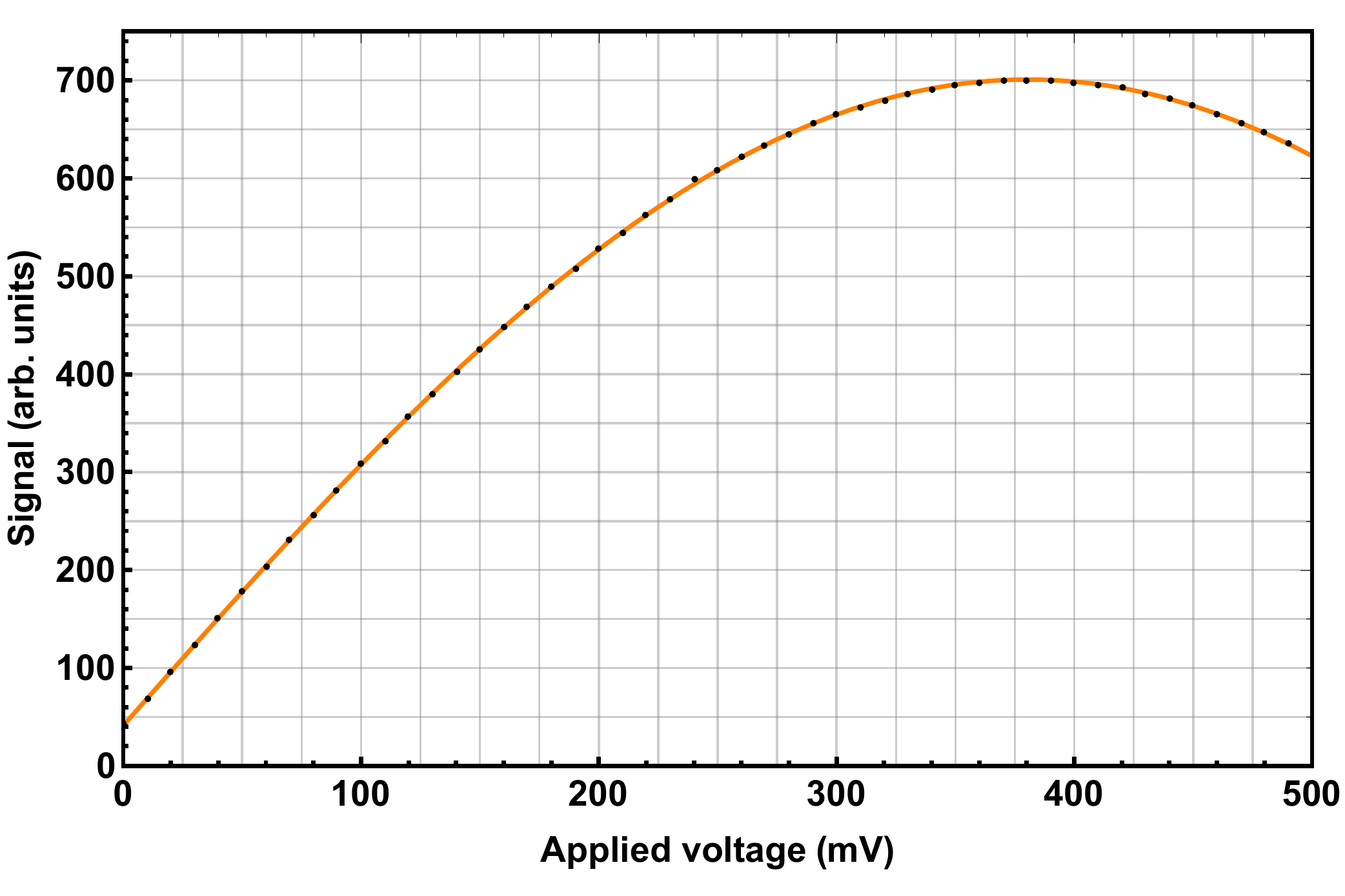}
\caption{Hahn echo magnetometry signal vs. nominal amplitude of applied square wave. Data ($\cdot$) are fit to a sinusoid with additive offset (\textcolor{orange}{\textbf{---}}), as described in the text. The best fit is given by an angular frequency of $4.03\times 10^{-3}$~(mV)$^{-1}$ and amplitude, phase, and offset of 684, 0.0365, and 16.1, respectively.  The slight offset from zero $y$-intercept is attributed to imperfections in the digital phase shifters, in particular the slight dependence of insertion loss on selected phase.}\label{fig:HahnEchoFit}
\end{figure}

The amplitude of the square wave is then scanned, and the average value of the Hahn echo magnetometry signal is recorded. Data are then fit to a sinusoid with additive offset, as shown in Fig. \ref{fig:HahnEchoFit}. Following Eqn. \ref{eqn:fringespacingfieldcalibration}, with total precession time $\tau=100$~$\upmu$s for the Hahn echo sequence, we find that the fringe spacing in magnetic field units is $309$~nT, and this quantity is set equal to the period obtained from the sine wave fit to produce a calibration of magnetic field per applied voltage. As discussed in Sec. \ref{sec:calibration}, the value obtained for the 6.4~kHz test field calibration is consistent with the near-DC calibration to within 1.4\%.

After the initial calibration, further optimization may then be performed iteratively. First, the test field amplitude is set to approximately that of the first signal maximum found in the calibration procedure; MW powers and frequencies are then adjusted to maximize the signal observed. Next, the test field is turned off, and the MW power and frequencies are adjusted to produce minimum noise, and the procedure is repeated as desired.

After finding the MW power and frequency settings that produce the highest signal and the lowest noise, the applied field amplitude is again scanned over a series of values in random order, and magnetometry traces are recorded for each step. As in the calibration procedure, the average signal values for each step of the field amplitude scan are fit to a sinusoid to determine the maximum slope in units of device signal per nT of applied field. The traces at zero applied field, recorded during and immediately after the field amplitude scan, are used to determine the device noise spectrum. In particular, the rms average of 10 spectra from 1-s long acquisitions is performed to produce the noise spectrum. The noise spectrum is then divided by the measured slope to produce a sensitivity spectrum. By calculating the median of this sensitivity spectrum, we find an optimized Hahn echo sensitivity $\eta \approx 210$~fT/$\sqrt{\text{Hz}}$.

\section{Sensitivity Theoretical Limits}

Calculating theoretical limits can confirm that reported results are sensible, and provide insight into avenues for device improvement. 

\subsection{Shot noise, reference noise, digitizer noise}\label{sec:shotnoise}
The number of photoelectrons collected on the signal and reference photodiodes during a single readout of duration $t_R$ are
\begin{equation}
    \mathscr{N}_\text{sig} = \frac{\bar{I}_\text{sig} t_R}{q},\hspace{10 mm}  \mathscr{N}_\text{ref} = \frac{\bar{I}_\text{ref} t_R}{q},
\end{equation}
where $\bar{I}_\text{sig}$ and $\bar{I}_\text{ref}$ are the average photocurrents on the signal and reference photodiodes during the readout period, respectively, and $q$ is the magnitude of the electron charge. Data in this work are taken with $\bar{I}\approx 4.8$~mA, $\bar{I}_\text{ref} \approx 82.9$~mA, and $t_R = 10$~$\upmu$s, so that $\mathscr{N}_\text{sig}\approx 3.0\times10^{11}$ and  $\mathscr{N}_\text{ref}\approx 5.2\times10^{12}$.

The signal and reference photocurrents are integrated on capacitors with values $\mathcal{C}_\text{sig}= 6.6$~nF and $\mathcal{C}_\text{ref}=114$~nF, producing voltages
\begin{equation}
    V_\text{sig} = \frac{\bar{I}_\text{sig} t_R}{\mathcal{C}_\text{sig}} \quad \text{and} \quad V_\text{ref} = \frac{\bar{I}_\text{ref} t_R}{\mathcal{C}_\text{ref}}.
\end{equation}
For a given value of $\bar{I}_\text{sig}$, the values of $\mathcal{C}_\text{sig}$ and $\mathcal{C}_\text{ref}$ are first coarsely adjusted so that $V_\text{sig} \approx V_\text{ref}$. Next, the value of $\bar{I}_\text{ref}$ is adjusted by varying the fraction of light that reaches the reference photodiode, so that $V_\text{ref}$ closely matches $V_\text{sig}$. The given values of $\bar{I}_\text{sig},\bar{I}_\text{ref},\mathcal{C}_\text{sig},\mathcal{C}_\text{ref}$ and $t_\text{R}$ result in $V_\text{sig}\approx V_\text{ref} \approx 7.3$~V at the end of each readout integration period. 

Shot noise on the number of collected photoelectrons is often a fundamental noise source for fluorescence-based measurements. Photoelectron shot noise results in rms variations $\sqrt{\mathscr{N}_\text{sig}}$ and $\sqrt{\mathscr{N}_\text{ref}}$ on the number of photoelectrons observed on the signal and reference photodiodes for a single readout, respectively. This photoelectron shot noise results in an integrated voltage noise on the signal and reference integration capacitors of
\begin{equation}
\sigma_\text{sig} = \frac{\sqrt{q \bar{I}_\text{sig}t_R}}{\mathcal{C}_\text{sig}} \quad \text{and} \quad \sigma_\text{ref} = \frac{\sqrt{q \bar{I}_\text{ref}t_R}}{\mathcal{C}_\text{ref}},
\end{equation}
so that $\sigma_\text{sig} \approx 13$~$\upmu$V and $\sigma_\text{ref} \approx 3.2$~$\upmu$V. As photoelectron shot noise $\sigma_\text{sig}$ represents a fundamental limit, additive noise sources can be evaluated relative to $\sigma_\text{sig}$. For the balancing scheme described in SM Sec.~\ref{App:BalancingCircuit}, subtraction of the reference voltage from the signal voltage increases the noise relative to $\sigma_\text{sig}$ alone by the factor
\begin{equation}
    \kappa_\text{bal} = \sqrt{1+\frac{\bar{I}_\text{ref}\mathcal{C}_\text{sig}^2}{\bar{I}_\text{sig}\mathcal{C}_\text{ref}^2}} \approx \sqrt{1+\frac{\mathcal{C}_\text{sig}}{\mathcal{C}_\text{ref}}},
\end{equation}
which yields $\kappa_\text{bal} = 1.029$. 

Other noise sources present in the balancing circuit contribute negligibly. The input voltage noise ($e_n \approx 0.85$~nV/$\sqrt{\text{Hz}}$) of each ADA4898 op-amp integrates to 190~nV~rms over the $t_\text{R} =10$~$\upmu$s readout time. The ADA4898's input current noise (2.4~pA/$\sqrt{\text{Hz}}$), Johnson-Nyquist noise on the relevant $\sim 100$~$\Omega$ resistors in the instrumentation amplifier, and thermal reset noise on the integration capacitors contribute negligible noise as well. 

The effective gain of the analog balancing circuit is $\mbox{G=7.69}$. The $V_\text{sig}\approx 7.3$~V integrated signal voltage corresponds to a hypothetical output voltage of $GV_\text{sig} \approx\!56$~V. Although subtraction of the reference voltage reduces the balancing circuit output to near zero in practice, this $\approx\!56$~V value can be thought of as the signal which would be observed at the output of the balancing circuit without the reference subtraction. Noise voltages, multiplied by G where appropriate, can be compared with the value of $GV_\text{sig}$ to determine fractional additive noise, and evaluated in comparison to fractional shot noise.

The digitization process introduces an additional rms voltage noise of $\sigma_\text{dig} = 23$~$\upmu$V per readout. This digitization noise adds in quadrature to the noise from the output of the balancing circuit. The total rms voltage noise for a single readout due to signal shot noise, reference shot noise, and digitizer noise is then
\begin{equation}
\sigma_\text{tot} = \sqrt{G^2\sigma_\text{sig}^2 \kappa_\text{bal}^2 + \sigma_{\text{dig}}^2}.
\end{equation}
Inserting the values given above, we find that the digitization noise increases the total noise by $\approx 1.024\times$. We find $\sigma_{\text{tot}} \approx 0.1075$ mV for a single 10 $\upmu$s readout, of which $\approx 95\%$ is attributed to shot noise on the signal photodiode. In other words, the entire balancing circuit and digitization process increases the noise by only $\approx 5\%$ over the shot noise limit on the NV fluorescence photocurrent. \par

\subsection{Ramsey theoretical sensitivity}\label{sec:ramseytheorysensitivity}

The theoretical sensitivity of a photon-shot-noise-limited, Ramsey NV-ensemble broadband magnetometer is~\cite{barry2020sensitivity}
\begin{align}\label{appeqn:ramseyshot}
\eta^\text{ens,sho}_\text{Ram} \approx &\frac{\hbar}{\Delta m_s g_e \mu_B} \frac{1}{C e^{-\left(\tau/T_2^*\right)^p}\sqrt{\mathscr{N}}} \nonumber \\ & \quad \times \frac{\sqrt{t_\text{I}\!+\!\tau\!+\!t_\text{R}\!+\!t_\text{D}}}{\tau} \times \frac{1}{F_\text{pro}}
\end{align}
where $F_\text{pro}$ represents the projection of the applied magnetic field onto the chosen NV axis or axes. When the bias field and test field are both normal to the diamond $\{100\}$ front facet, then $F_\text{pro}=1/\sqrt{3}$.

When employing a double-quantum scheme with P1 driving, we typically observe dephasing times $T_{2,\text{DQ+P1}}^* \approx 30$ $\upmu$s. For $C = 0.0334$, $T_2^*=28.6$ $\upmu$s, $\tau = 40$ $\upmu$s precession time, and $p=1$, we expect that $C e^{-(\tau/T_2^*)^p} \approx$ 0.0079. In practice, we measure $C e^{-(\tau/T_2^*)^p} = 0.0082$, which we employ in the following calculation instead. 

The timing of our Ramsey magnetometry sequence is described by values $t_\text{I} =35$~$\upmu$s, $\tau = 40$~$\upmu$s, $t_\text{R} = 10$~$\upmu$s, and $t_\text{D}= 6$~$\upmu$s. Under typical experiment conditions, we measure a signal photocurrent $I_\text{sig} \approx 4.8$~mA, which results in $\mathcal{N}_\text{sig} \approx 3\times 10^{11}$ signal photoelectrons per measurement. Finally, with \mbox{$\Delta m_s$ =2} for double-quantum magnetometry, we calculate from Eqn.~\ref{appeqn:ramseyshot} that the photon-shot-noise-limited double-quantum Ramsey sensitivity is
\begin{equation}
    \eta^\text{ens,sho}_\text{Ram} \approx 260 \;\text{fT}/\sqrt{\text{Hz}}.
\end{equation}

\subsection{Hahn echo theoretical sensitivity}\label{sec:hahntheorysensitivity}
The theoretical sensitivity of a photon-shot-noise-limited NV-ensemble magnetometer employing Hahn echo to an AC telegraph signal is

\begin{align}\label{eqn:hahnshot}
\eta^\text{ens,sho}_\text{Hahn} \approx &\frac{\hbar}{\Delta m_sg_e \mu_B} \frac{1}{C e^{-\left(\tau/T_2\right)^p}\sqrt{\mathscr{N}}} \nonumber \\ & \quad \times \frac{\sqrt{t_\text{I}\!+\!\tau\!+\!t_\text{R}\!+\!t_\text{D}}}{\tau}
\frac{1}{F_\text{pro}},
\end{align}
where $F_\text{pro} = 1/\sqrt{3}$ is the projection of the applied $[100]$-oriented AC magnetic field onto the chosen NV axis or axes. Order unity differences, e.g. $2/\pi$, in Eqn.~\ref{eqn:hahnshot} relative to other references arise from defining sensitivity relative to rms amplitude rather than amplitude~\cite{taylor2008high} and for a telegraph signal rather than a sinusoid~\cite{barry2020sensitivity}.

Hahn echo is performed without P1 driving, and thus we calculate the approximate expected sensitivity based on $T_\text{2,DQ}$ rather than $T_\text{2,DQ+P1}$. For $C = 0.0334$, $T_\text{2,DQ}=136$ $\upmu$s, $\tau = 100$ $\upmu$s precession time, and $p=1$, we expect that $C e^{-(\tau/T_\text{2,DQ})^p} \approx$ 0.0160. In practice, the value of the term $C e^{-(\tau/T_\text{2,DQ})^p}$ is measured to be 0.0125 for a precession time of $\tau =100$ $\upmu$s. %
With $\Delta m_s =2$ for a measurement using the double quantum scheme, $t_\text{I} =39$~$\upmu$s, $\tau = 100$~$\upmu$s, $t_\text{R} = 10$~$\upmu$s, $t_\text{D}= 7$~$\upmu$s, and a signal photocurrent of $I_\text{sig}=4.8$~mA which results in $\mathcal{N} = 3\times 10^{11}$ photoelectrons per measurement, we calculate from Eqn.~\ref{eqn:hahnshot} a photon-shot-noise-limited double-quantum Hahn echo sensitivity of
\begin{equation}
    \eta^\text{ens,sho}_\text{Hahn} = 90 \;\text{fT}/\sqrt{\text{Hz}}.
\end{equation}

\subsection{Additional detail on obstacles encountered}\label{sec:obstaclessupplement}

Major technical obstacles encountered in this work include electrical noise associated with P1 driving; thermal issues resulting from laser, MW, and RF power applied to the sensor head; and MW noise which limited device performance to above the shot noise limit.

The high RF power applied during P1 driving results in both electrical and thermal challenges. In addition to the limitations on P1 drive duration and power described in the main text, ferrite chokes needed to be installed on some DC power supply lines to allow consistent device operation with P1 driving. While P1 driving improved Ramsey magnetometry sensitivity, the additional noise associated with P1 driving offset the longer achieved coherence time in Hahn echo magnetometry, and better Hahn echo sensitivities were realized without P1 driving than with. In the future the electrical noise associated with P1 driving might be mitigated by better shielding of sensitive components, by turning the P1 driving off during the echo pulse, or by employing pulsed P1 driving. Although such electrical noise is not fundamental as are shot noise or Johnson noise, ensuring that the resonantly-enhanced fields produced by the $\sim\!100$~W applied RF powers produce no increase in device noise may prove difficult.

The heat load of the P1 drive coils located near the diamond, as well as the high optical power impinging on the diamond, necessitates substantial thermal equilibration periods before device operation, as well as care to maintain a consistent heat load from these sources. The MW power applied to the dielectric resonator may also produce a non-negligible heat load, but this was not found to be a primary source of thermal drift.

In contrast, the laser power reaching the diamond was monitored during device optimization and magnetometry, and alignment of laser light into the fiber optic cable was adjusted whenever the power fell by $\gtrsim 1$\%. Alignment of the laser light into the diamond itself was required much less frequently, but for optimum performance this alignment should be tailored to the operating configuration of the device. For example, a change in laser power or sequence repetition rate might necessitate a change in alignment. Future devices could include automated feedback to maintain consistent optical power, for example by adjusting the amount of light through an acousto-optic or electro-optic modulator. Automated fiber alignment tools might also reduce drift by maintaining near-optimal fiber coupling efficiency. Regardless of methods used to maintain consistent heat loads, however, the high powers required for device operation essentially guarantee that the diamond will reach a steady state temperature well above that of the surrounding environment.

The need for an extremely low noise MW signal chain constitutes another major challenge of this work. Even with phase and amplitude noise as the primary criteria for signal generator (Rohde and Schwartz SMA100B with SMAB-B711(N) option) selection, turning on the MWs increased device noise beyond the nearly-shot-noise-limited level observed otherwise. For both Hahn echo and Ramsey magnetometry, the limiting noise source in this device is believed to be MW phase noise. Though amplitude noise may not be entirely negligible, DQ Ramsey and Hahn echo are first-order insensitive to MW amplitude noise, and amplitude noise is additionally suppressed by amplifier saturation. In future devices, lower MW phase noise might be achieved by employing superior fixed-frequency sources mixed with tunable sources operating at relatively low frequencies (e.g., $\sim 100$~MHz).

\end{document}